\documentclass[twocolumn,english,aps,prl,reprint,floatfix]{revtex4-2}
\usepackage[T1]{fontenc}
\usepackage[utf8]{inputenc}
\usepackage{xcolor}
\usepackage{amsmath}
\usepackage{amssymb}
\usepackage{graphicx}
\PassOptionsToPackage{normalem}{ulem}
\usepackage{ulem}
\usepackage{titlesec} %
\titlespacing*{\paragraph} %
  {\parindent}               %
  {0pt}               %
  {0em}               %

\makeatletter

\usepackage{silence}
\usepackage{braket}
\usepackage{babel}
\usepackage[hidelinks]{hyperref}

\makeatother

\begin{document}
\title{Gaussian boson sampling: Benchmarking quantum advantage}
\author{Ned Goodman}
\author{Alexander S. Dellios}
\author{Margaret D. Reid}
\author{Peter D. Drummond}
\email[Contact author: ]{peterddrummond@protonmail.com}
\affiliation{Centre for Quantum Science and Technology Theory, Swinburne University of Technology, Melbourne 3122, Australia}
\begin{abstract} 
Gaussian boson sampling experiments claim quantum advantage: exactly reproducing the largest sample of an 8176-mode experiment would take $10^{449}$ years classically. Yet the experiments only approximate their ideal output. We introduce a fast approximate classical sampler based on the positive-$P$ representation. On the statistical tests used, its samples are closer to ideal predictions than those of all current experiments, and within sampling error for nearly all datasets beyond exact classical simulation. This highlights the need to reduce unitary errors and improve system characterization.

\end{abstract}
\maketitle
\paragraph*{Introduction---}
Quantum advantage experiments aim to perform computational tasks that classically require an exponentially long time~\cite{zhong2020quantum,zhongPhaseProgrammableGaussianBoson2021,deng2023gaussian,liu2025robust,madsenQuantumComputationalAdvantage2022}, but can be performed quickly on a quantum computer (QC). In Gaussian boson sampling (GBS), squeezed light is sent through a large interferometer and the photons at its outputs are counted; the task is to produce samples of these counts~\cite{hamiltonGaussian2017,quesadaGaussian2018,kruseDetailed2019}. In the GBS proposal, multi-mode pure squeezed states are subject to a linear transformation implementing a Haar-random unitary, turning the input state into a highly entangled Gaussian state~\cite{hamiltonGaussian2017,kruseDetailed2019}. Despite its simplicity, exactly sampling the output photon counts of such a state is classically hard~\cite{AaronsonArkhipov2013LV,hamiltonGaussian2017,quesadaGaussian2018,grier2022complexity} and is expected to be hard even for moderate losses~\cite{deshpandeQuantumComputationalAdvantage2022a,qi2020regimes}.

Multiple GBS experiments have been conducted that claim quantum advantage. These include Borealis~\cite{madsenQuantumComputationalAdvantage2022}, which uses photon-number-resolving (PNR) detectors to generate the random counts, and the \textit{Jiuzhang} experiments~\cite{zhongPhaseProgrammableGaussianBoson2021,deng2023gaussian,liu2025robust}, whose threshold detectors output binary clicks. Together with demonstrations at Xanadu~\cite{arrazola2021quantum}, Paderborn~\cite{stefszky2025benchmarking} and elsewhere, these experiments and their data are a large advance in quantum technology. Despite this achievement, as with earlier experiments~\cite{drummondSimulatingComplexNetworks2022,MartinezCifuentes2023classicalmodelsmay,dellios2023validationtestsgbsquantum,dellios2024validation}, the experimental random counts do not match the expected output distribution, or ground truth, for even the lowest-order moments (Supplemental Material~\cite{SupplementalMaterial}, Tables~S3--S8). Linear loss, the dominant source of errors, is already included in the ground truth and thus cannot explain the discrepancies.

The largest experiment, \textit{Jiuzhang~4}, has set a landmark for continuous-variable quantum computing by implementing an impressive 8176-mode optical network~\cite{liu2025robust}, possibly the highest-dimensional QC Hilbert space to date. One of its samples, with 3050 clicks, would take about $10^{449}$ years to generate on the Fugaku supercomputer with current exact classical algorithms~\cite{bulmerBoundaryQuantumAdvantage2022a,madsenQuantumComputationalAdvantage2022} (Supplemental Sec.~S6). Yet such experiments reproduce the ideal distribution only approximately, so the relevant comparison is a fast approximate sampler: can one come closer to the ideal distribution than the experiment? Approximate sampling methods exist~\cite{villalonga2021efficient,oh2024classical,dodd2025fast}. However, the most accurate of them, the matrix-product-state (MPS) method~\cite{oh2024classical}, would require more than $10^{42}$ years at the \textit{Jiuzhang~4} scale~\cite{liu2025robust}.

Here, we introduce a fast, scalable classical approximate algorithm for GBS counts using the positive-$P$ phase-space representation, with timescales of $\sim30$~ms per sample on 18 cores at 8176 modes (Supplemental Sec.~S7). Judged by the largest deviation among 23 commonly used observables, its samples are closer to the ideal distribution than those of every experiment in the quantum-advantage regime.

\paragraph*{Background---}

Our sampler uses the phase-space description of light, which we summarize first. The Glauber--Sudarshan {$P$ distribution}~\cite{Glauber_1963_P-Rep, Sudarshan_1963_P-Rep}, used to describe classical fields, is a diagonal expansion of the density matrix $\hat{\rho}$ of $M$ modes in terms of bosonic coherent states $\left|\boldsymbol{\alpha}\right\rangle$~\cite{walls2008quantum}:
\begin{equation}
 \hat{\rho}=\int P\left(\boldsymbol{\alpha}\right)\left|\boldsymbol{\alpha}\right\rangle \left\langle \boldsymbol{\alpha}\right|d^{2M}\boldsymbol{\alpha}.\label{eq:Glauber-Sudarshan P Representation}
 \end{equation}

If $P\left(\boldsymbol{\alpha}\right)$ is  non-negative and no more singular than a delta function, the state is termed classical~\cite{Titulaer1965Correlation, Reid1986}, and $P(\boldsymbol{\alpha})$ can be interpreted as a probability distribution over coherent state projectors. Thus, the photon-count distribution of $\hat{\rho}$ is just a $P(\boldsymbol{\alpha})$ mixture over coherent state number distributions $|\braket{n|\alpha}|^2=e^{-|\alpha|^2}|\alpha|^{2n}/n!$, and is efficiently simulable whenever $P$ can be sampled efficiently. A Gaussian state has a classical $P$ representation if and only if its normally ordered quadrature covariance is positive semidefinite. In the End Matter Theorem, we prove that linear loss cannot remove non-classicality if the transmission matrix has full column rank, although thermal noise can. Hence the output of a lossy GBS network with squeezed inputs remains non-classical.

In order to get a non-singular and positive distribution for quantum states we can use the $M$-mode positive-$P$ representation~\cite{drummond1980generalised}, which expands an arbitrary density matrix as a probability distribution over a non-classical phase space with $4M$ real dimensions, as:
\begin{equation}
\hat{\rho}=\int\int P\left(\boldsymbol{\alpha},\boldsymbol{\beta}\right)\frac{\left|\boldsymbol{\alpha}\right\rangle \left\langle \boldsymbol{\beta}^{*}\right|}{\left\langle \boldsymbol{\beta}^{*}\mid\boldsymbol{\alpha}\right\rangle }d^{2M}\boldsymbol{\alpha}d^{2M}\boldsymbol{\beta}\label{eq:positive P Representation}.
\end{equation}

For a multi-mode pure input state with squeezing parameters $r_{i}$, the coherences and photon numbers are~\cite{walls2008quantum}:
\begin{align}
\left\langle \hat{a}_{i}^{2}\right\rangle  & =\bar{m}_{i}=\sinh\left(r_{i}\right)\cosh\left(r_{i}\right)\nonumber \\
\left\langle \hat{a}_{i}^{\dagger}\hat{a}_{i}\right\rangle  & =\bar{n}_{i}=\sinh^{2}\left(r_{i}\right),
\end{align}
with $\left\langle \cdot\right\rangle $ denoting operator expectation values. Of the many phase-space distributions that represent the same state, we want the one closest to the $\alpha=\beta^{*}$ subspace, to minimize the estimator variance and the projection error described later. We use a known compact distribution for squeezed states~\cite{drummondSimulatingComplexNetworks2022} with
\begin{align}
\alpha_{j} & =\delta_{+j}w_{j}+i\,\delta_{-j}w_{j+M},\nonumber \\
\beta_{j} & =\delta_{+j}w_{j}-i\,\delta_{-j}w_{j+M},\label{eq:compact_dist}
\end{align}
where $\delta_{\pm j}=\sqrt{(\bar{n}_{j}\pm\bar{m}_{j})/2}$ ($\bar{m}_{j}$ is replaced by $\tilde{m}_{j}$ of Eq.~\eqref{eq:thermal_coherence} for the corrected ground truths), and $w_{j}$ are independent real standard normal noises. For squeezed inputs $\bar{n}_{j}<\bar{m}_{j}$, so $\alpha_{j}$ and $\beta_{j}$ are real, and $\beta_{j}\neq\alpha_{j}^{*}$. Each sample is then propagated through the linear network,
\begin{equation}
\boldsymbol{\alpha}\rightarrow\boldsymbol{T}\boldsymbol{\alpha},\qquad\boldsymbol{\beta}\rightarrow\boldsymbol{T}^{*}\boldsymbol{\beta},\label{eq:propagation}
\end{equation}
where $\boldsymbol{T}$ is the transmission matrix of the network (a Haar-random unitary in the lossless case). From here on, $\alpha_j$ and $\beta_j$ denote these output amplitudes. Averages over the positive-$P$ samples, written $\langle\cdot\rangle_P$, equal normally ordered quantum expectation values in the large sample number limit~\cite{drummondSimulatingComplexNetworks2022}.

\paragraph*{GBS sampling from phase space---}

Why is the positive-$P$ method not a GBS sampler? Its samples have the moments of the GBS output distribution, so why can they not be mapped to photon counts?
The difference is apparent from Eqs.~\eqref{eq:Glauber-Sudarshan P Representation} and~\eqref{eq:positive P Representation}.  Only in the former case does a sample correspond to a state and thus have a photon counting distribution. The restriction $\alpha=\beta^{*}$ defines a $2M$-dimensional subspace in which the positive-$P$ representation collapses to the Glauber--Sudarshan representation and the phase-space samples define an equally weighted mixture distribution over coherent states, which is easy to sample from~\cite{RahimiKeshari2016}.

Approximate sampling from a GBS distribution simply requires projecting our positive-$P$ phase-space samples onto this set $\{\alpha,\beta\}\rightarrow\{\alpha^{(cl)},\beta^{(cl)}\}$ where $\alpha^{(cl)}=(\beta^{(cl)})^*=f(\alpha,\beta)$. There are countless possible projections, and the best to use varies by detector type and experiment. The \textit{Jiuzhang} experiments use threshold detectors, which are sensitive to even a single photon but cannot distinguish the number of photons. For such detectors, the probability of any observation can be determined from the phase-space vacuum probability $\pi_j=e^{-\alpha_j\beta_j}$, whose positive-$P$ mean is the $j$th mode no-click probability $p_j(0)=\langle\pi_j\rangle_P$. The minimum-distance projection in $\pi_j$ is:
\begin{equation}
\pi_{j}^{(cl)}\equiv\mathcal{T}_{p}(\pi_{j})\equiv\min\left(1,\max\left(\operatorname{Re}\pi_j,0\right)\right).\label{eq:threshold_projection}
\end{equation}
For full PNR detectors, such as those used by the Borealis experiment, there are two choices. Projecting the phase-space photon-number variables $n_{j}=\alpha_{j}\beta_{j}$ yields only mixtures of Poisson distributions. Projecting instead the per-mode number category probabilities $\pi_{j,c} = e^{-n_j} n_j^c/c!$ can give non-Poissonian statistics. 

Since none of these projections preserves the low-order moments, we improve them by applying whitening--coloring ($\mathcal{W}$) transformations~\cite{kessy2018optimal}.
Given the stochastic vector $\boldsymbol{x}$, whose components are either $\pi_i$, $n_i$ or $\pi_{i,c}$, and the corresponding projection $\mathcal{T}$, the transformation is:
\begin{align}
\boldsymbol{x}^{(\ell+1)}
&=\mathcal{T}\!\left[\mathcal{W}(\boldsymbol{x}^{(\ell)},\boldsymbol{x})\right],
\qquad \boldsymbol{x}^{(0)}=\mathcal{T}(\boldsymbol{x}),\label{eq:wc_iter}\\
\mathcal{W}(\boldsymbol{x}^{(\ell)},\boldsymbol{x})
&=\left(\boldsymbol{x}^{(\ell)}-\langle\boldsymbol{x}^{(\ell)}\rangle_P\right)
\left[\Sigma^{(\ell)}\right]^{-1/2} \Sigma^{1/2}
+\langle\boldsymbol{x}\rangle_P,\label{eq:wc_W}
\end{align}
where $\Sigma^{(\ell)}_{ij}=\langle\Delta x_i^{(\ell)}\Delta x_j^{(\ell)}\rangle_P$ and $\Sigma_{ij}=\langle\Delta x_i\Delta x_j\rangle_P$ are the covariance matrices of the current projected samples and the target variables, respectively, with $\Delta x=x-\langle x\rangle_P$. If $\Sigma$ has negative eigenvalues, $\Sigma^{1/2}$ is complex, and the next projection keeps the real part of the iterate (Supplemental Sec.~S5\,A\@). 

 This transformation replaces the means, variances, and covariances of the projected samples with those of the target variables. Doing this can move samples outside the physical range $[0,1]$ for $\pi_j$ or $\pi_{j,c}$, or $[0,\infty)$ for $n_j$, requiring the transformed samples to be projected again. Iterating the procedure reduces the proportion outside the physical range over the iterations used here (Supplemental Sec.~S5\,A\@).

After the final iteration, we derive one detector sample per phase-space sample. For threshold detectors this is
\begin{align}
    c_j = \boldsymbol{1}\{u_j>x_j^{(\ell)}\}, \qquad u_j \sim U(0,1),
    \label{eq:draw_threshold}
\end{align}
which is used for all our threshold-detector data. For PNR detectors, each count is drawn either from a Poisson distribution or from each mode's whitened probability vector (End Matter). In practice, the Poisson output step worked best for the smallest Borealis dataset, B16, and the categorical output step worked best for all other Borealis datasets (Supplemental Sec.~S5\,B\@).

\begin{figure*}
\centering
\includegraphics[width=0.995\textwidth]{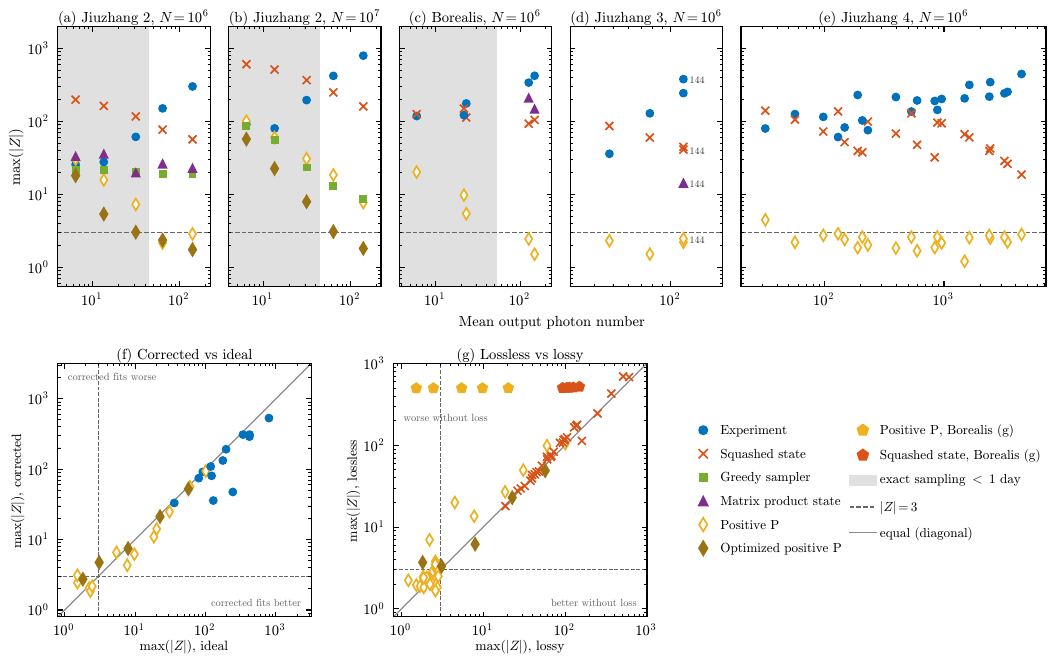}
\caption{Summary of all comparisons; smaller values are better in every panel. (a)--(e) Maximum absolute $Z$-score, $\max|Z|$, over all observables against the ideal ground truth ($t=1$, $\epsilon=0$), plotted against the mean number of photons reaching the detectors: (a) \textit{Jiuzhang~2} with $N=10^{6}$ samples, for comparison with the MPS sampler, (b) \textit{Jiuzhang~2} with $N=10^{7}$, (c) Borealis, (d) \textit{Jiuzhang~3}, with its 144-mode version labeled 144 (Supplemental Table~S6), and (e) \textit{Jiuzhang~4}. Below the dashed line, $\max|Z|=3$, there is no statistically significant disagreement. Gray shading marks datasets whose $N$ samples could be generated exactly in under one day (Supplemental Sec.~S6); the unshaded datasets define the quantum-advantage regime. (f) $\max|Z|$ against the corrected ground truth ($t\neq1$ or $\epsilon>0$) versus against the ideal one, for every experiment but \textit{Jiuzhang~4}; the correction is fitted to the experimental $\mathcal{G}^{M}_{M}$ or its mean. (g) $\max|Z|$ of each dataset's lossless version versus the original, each against its own ideal ground truth (Supplemental Sec.~S8), with the lossless Borealis total-count distribution scored against its exact distribution. The Borealis positive-$P$ and squashed-state points, which use photon-number detection, are filled pentagons in each sampler's color. Filled diamonds are the optimized positive-$P$ sampler (Supplemental Sec.~S5\,A\@). For Borealis, the positive-$P$ sampler uses the categorical output step on every dataset except B16.\label{fig:Jall}}
\end{figure*}

\paragraph*{Experiments and ground truths---}
We consider all major GBS experiments excluding \textit{Jiuzhang~1}, as it is in the classically tractable regime (Supplemental Sec.~S6; datasets in Table~S1). Since the input and output states are Gaussian, we can characterize each experiment using the covariance matrix and mean vector given by its published transmission matrix and squeezing parameters~\cite{serafini_quantum_2017}. The experiments' published models include no error besides photon loss, except \textit{Jiuzhang~3}'s, which also includes partial distinguishability (\textit{Jiuzhang~4} reports it separately~\cite{liu2025robust}; Supplemental Sec.~S1). We take the lossy GBS task as the ground truth and treat partial distinguishability as an experimental error, not as part of the task: of our ground truths, only the corrected \textit{Jiuzhang~3} one includes it.

A genuine quantum system can dramatically deviate from the ground truth and still be classically hard to simulate, for example if the deviation is caused entirely by parameter errors in the transmission matrix. Therefore, we include physically realistic corrections to the ground truth. These include an overall transmission matrix correction, parameterized by $t$, and partial decoherence of the input states, parameterized by $\epsilon \in [0,1]$~\cite{drummondSimulatingComplexNetworks2022} (Supplemental Sec.~S3). 

\paragraph*{Means of comparison---}
The only measure that has complexity-theoretic guarantees of classical hardness is the full-system total variation distance~\cite{AaronsonArkhipov2013LV}. However, this is unusable in practice as it requires both an exponentially large number of samples and exponential computation time per sample to calculate~\cite{hangleiter2019sample,hangleiter2023computational}. Therefore, we focus on observables such as low-order moments, marginalized probability measures and grouped count distributions (GCDs). 

The GCDs come from (exact up to sampling errors) phase-space simulations~\cite{drummondSimulatingComplexNetworks2022}, while marginal distributions are obtained exactly from the ground-truth covariance matrix~\cite{serafini_quantum_2017,gupt2019walrus}. For the lossless Borealis datasets the phase-space simulation cannot resolve the photon-number parity of the total count, so there we use its exact distribution (Supplemental Sec.~S8). The phase-space simulation algorithm and positive-$P$ approximate sampler are available in the xQSIM software package~\cite{xqsim3}. 

To put all comparisons on one scale, we convert each to a {$Z$-score}. This is the standard-normal value with the same tail probability as a Pearson $\chi^{2}$ test of the sampled frequencies for a given observable against the ground truth (End Matter). We also report two measures used widely in GBS research~\cite{zhongPhaseProgrammableGaussianBoson2021,deng2023gaussian,madsenQuantumComputationalAdvantage2022,villalonga2021efficient,MartinezCifuentes2023classicalmodelsmay,oh2024classical}, the cross-entropy and the joint cumulants up to third order (Supplemental Secs.~S4\,B\@ and~S4\,C\@). Every sampler's mean cross-entropy is within about $0.15$~nats of the experiment's. At second order the positive-$P$ sampler's cumulants are closer to the exact values than the experiment's on most datasets. At third order the experiment is closer on low-power \textit{Jiuzhang~2} and on B16, and elsewhere it does not separate the samplers.

\paragraph*{Comparison with other samplers---}
We compare the positive-$P$ sampler against the experimental data for \textit{Jiuzhang~2--4} and Borealis, and against three other sampling algorithms: the {squashed-state sampler}~\cite{MartinezCifuentes2023classicalmodelsmay,dellios2023validationtestsgbsquantum}, which uses the classical state with the highest fidelity to a squeezed state as its input; a third-order greedy cumulant-matching sampler~\cite{villalonga2021efficient}, which builds samples deterministically to {maximize} agreement with up to third-order marginals and cumulants; and an MPS sampler~\cite{oh2024classical}, which uses tensor-network methods to construct a low-entanglement approximation to the GBS output state. 

For \textit{Jiuzhang~2}, we also implement an optimized version of the positive-$P$ sampler. The whitening--coloring target leaves the diagonal of its covariance matrix free, and different choices favor different observables: one favors the high-order marginals and another the total-count distribution. The optimized sampler draws a fraction of its samples with each choice, with the fraction chosen per dataset (Supplemental Sec.~S5\,A\@). 

\begin{table}[t]
\centering
\caption{Worst-performing observable, and its signed {$Z$-score}, for \textit{Jiuzhang~2} with $N=10^{7}$ against the ideal ground truth. A dash means no observable exceeds $|Z|=3$; the bracket then gives the largest $|Z|$. All datasets are tabulated in Supplemental Tables~S3--S8.\label{tab:worst_performance}}
\footnotesize
\setlength{\tabcolsep}{2.5pt}
\begin{tabular}{lccccc}
\hline
\textbf{Model} & \textbf{0.15~W} & \textbf{0.3~W} & \textbf{0.6~W} & \textbf{1.0~W} & \textbf{1.65~W} \\
\hline
Exp. & $\langle c_j\rangle$ & $\langle c_j\rangle$ & $\langle c_j\rangle$ & $\langle c_j\rangle$ & $\langle c_j\rangle$ \\
 & ($+91$) & ($+80$) & ($+196$) & ($+420$) & ($+796$) \\[3pt]
Squashed & $\mathcal{G}^{M}_{M/2,M/2}$ & $\mathcal{G}^{M}_{M/2,M/2}$ & $\mathcal{G}^{M}_{M}$ & $\mathcal{G}^{M}_{M}$ & $\langle c_j\rangle$ \\
 & ($+607$) & ($+513$) & ($+367$) & ($+249$) & ($+160$) \\[3pt]
Greedy & $\mathcal{G}^{M}_{M}$ & $\mathcal{G}^{M}_{M}$ & $\mathcal{G}^{M}_{M}$ & $\mathcal{G}^{M}_{M}$ & $\mathcal{M}^{4}$ \\
 & ($+86$) & ($+56$) & ($+23$) & ($+13$) & ($-8.7$) \\[3pt]
{Positive-$P$} & $\mathcal{G}^{M}_{M}$ & $\mathcal{G}^{M}_{M}$ & $\mathcal{G}^{M}_{M}$ & $\mathcal{G}^{M}_{M}$ & $\mathcal{G}^{M}_{M}$ \\
 & ($+101$) & ($+60$) & ($+31$) & ($+18$) & ($+7.7$) \\[3pt]
Optimized & $\mathcal{M}^{15}$ & $\mathcal{M}^{11}$ & $\mathcal{M}^{13}$ & $\mathcal{M}^{5}$ & --- \\
 & ($+57$) & ($+23$) & ($+8.0$) & ($+3.1$) & ($1.8$) \\[3pt]
\hline
\end{tabular}
\end{table}
Across all datasets (Fig.~\ref{fig:Jall}), our sampler, in its optimized form on \textit{Jiuzhang~2}, outperforms the experiments and the squashed-state, greedy and MPS samplers on $\max|Z|$ in all regimes. Table~\ref{tab:worst_performance} reports the observable with the largest $Z$-score for each \textit{Jiuzhang~2} dataset. Full results are in the Supplemental Material~\cite{SupplementalMaterial}. The experiment's worst observable is always the single-mode click probability $\langle c_j\rangle$, or for Borealis the photon-number distribution $P_j(n)$. Since few-parameter corrections do not remove this (Fig.~\ref{fig:Jall}(f)), this points to mode-specific errors in the transmission matrix and input states. Partial distinguishability, which the ideal ground truths omit, also contributes for \textit{Jiuzhang~3} and~\textit{4}; no correction was fitted for \textit{Jiuzhang~4}. By contrast, the classical samplers almost always fail first on many-mode statistics: GCDs or high-order marginals. The greedy sampler's negative $Z$-score on the $1.65~\mathrm{W}$ dataset means that it agrees with the ground truth better than an exact sampler would, i.e., underdispersion, a consequence of building samples to match ground-truth marginals rather than sampling from a distribution.

The positive-$P$ sampler requires only $O(LM^{2})$ time per sample and $O(BM+M^{2})$ memory, where $M$ is the number of output modes, $L$ the number of whitening--coloring iterations and $B$ the number of phase-space samples per batch (the set of samples whitened together), which must exceed $M$ for the batch covariance to be invertible. For the categorical output step, which whitens three categories per mode, $M$ is replaced by $3M$; the optimized sampler differs only by a constant factor and has the same scaling. 

Neither the MPS~\cite{liu2025robust,oh2024classical} nor the greedy sampler can scale accurately to current or larger experiments. If the transmission falls no faster than $S^{-b}$ for some $b<1/2$, where $S$ is the number of squeezed inputs, as it does at fixed transmission, the bond dimension of the MPS sampler grows superpolynomially~\cite{oh2024classical}, while the greedy sampler cannot produce marginal events rarer than its inverse sample number (Supplemental Sec.~S7).

\paragraph*{Precision---}

Three further results bear on future experiments.
First, the accuracy of the positive-$P$ sampler does not degrade with system size. Across all 21 \textit{Jiuzhang~4} datasets, from 496 to 8176 modes, its 483 $Z$-scores have mean $0.07$ and standard deviation $1.05$, close to the 0 and 1 of an exact sampler; the dataset with the fewest photons reaches $|Z|=4.5$. Its accuracy increases with output photon number (Supplemental Fig.~S1) and, surprisingly, also improves with dilution, i.e., when the same light is spread over more modes: for \textit{Jiuzhang~3} at high power with $N=10^{7}$, $\max|Z|=2.1$ at 1152 modes but 9.4 for its 144-mode coarse-grained version (Supplemental Tables~S5 and~S6).

Second, its accuracy is relatively insensitive to loss, at least for threshold detectors. To demonstrate this, we replaced the experimental transmission matrices with Haar-random unitaries of equal size. All \textit{Jiuzhang~3} and 19 of the 21 \textit{Jiuzhang~4} datasets remain within or close to sampling error (Fig.~\ref{fig:Jall}(g), Supplemental Sec.~S8).
The \textit{Jiuzhang~2} datasets, which are smaller, get worse, but moving to the optimized sampler substantially reduces this effect.
The Borealis datasets are the exception. All get markedly worse without loss, because each projected sample gives every mode both photon-number parities, so about half the samples have an odd total, which a lossless output cannot have. The matrix-$P$ representation~\cite{drummond2026matrix} avoids this, but adapting the sampler to it is left for future work.

Third, its accuracy is not fixed by the algorithm, although no single parameter, like an MPS bond dimension, trades cost for accuracy. The whitening--coloring target has a free diagonal, which the optimized sampler exploits to substantially improve performance on \textit{Jiuzhang~2} (Fig.~\ref{fig:Jall}, Supplemental Sec.~S5\,A\@).
Beyond that, there are other avenues for algorithmic improvement, such as extending the cumulant matching to higher order.

While we do not claim limitless accuracy improvements, we currently know of no upper limit to the accuracy of the positive-$P$ approximate sampler.

In summary, our positive-$P$ sampler, optimized for \textit{Jiuzhang~2}, is closer to the ideal ground truth than every current experiment, agrees with it within sampling error, at the sample sizes tested, on nearly all datasets in the quantum-advantage regime, and matches or outperforms every other classical sampler we considered. The positive-$P$ sampler's success does not mean that the output states are classical. Up to 8176 modes, neither system size, nor removing loss from threshold-detector simulations so they have pure states, nor few-parameter corrections to the ground truth take the experiments beyond its reach, so these results close the most direct routes to quantum advantage with these tests. The experiments' deviations instead indicate the need to address mode-specific errors in the transmission matrix and input states, and partial distinguishability.

How these errors can be removed depends on the goal. For a demonstration of quantum advantage the target matrix need not be chosen in advance: the optical elements and input states can be characterized to high accuracy, and the target matrix constructed from these measurements. Loss is then part of what is measured, and need not be removed. Applications of GBS are more demanding, because the matrix encodes the problem and is set in advance. This likely requires removing loss as well, since loss in real optical elements is rarely uniform across their ports, so it usually also changes their effective splitting ratios and phases~\cite{barnett1998quantum}. Errors in implementing an intended unitary are a recognized obstacle in quantum computing~\cite{kueng2016comparing}, and the tests used here give one way to measure them in photonic devices.

These results do not show that the experiments are classically easy. Their deviations may come from errors that leave exact sampling hard. An experiment that matched its ground truth would also pass these tests, and would outscore our sampler wherever the sampler's own error is resolved.

\paragraph*{Acknowledgments---}
This research was funded through a generous grant from the John Templeton Foundation, grant ID 62843. The simulations were performed on the Ngarrgu Tindebeek supercomputer at Swinburne University of Technology. We thank Benjamin Villalonga for his insightful discussions, feedback on the manuscript, and work running the greedy sampler algorithm. We thank the USTC and Xanadu groups for the public release of the raw data of \textit{Jiuzhang~2--4} and Borealis. We thank C.~Oh and coauthors for making the samples of their MPS sampler~\cite{oh2024classical} publicly available. The analysis, job-management and plotting code, and edits to the text of this manuscript, were drafted with the assistance of a large language model (Anthropic's Claude Opus 5.5) under the direction of the authors, who tested all code against known results, checked every result reported here, and take full responsibility for the content.

%
\onecolumngrid

\newpage{}

\section*{End Matter}
\twocolumngrid

\paragraph*{Non-classicality under linear loss---}
\label{sec:theorem}A thermalized squeezed input state~\cite{Fearn_JModOpt1988} is more realistic for experiments than a pure squeezed state, giving an improved fit to experimental data~\cite{drummondSimulatingComplexNetworks2022,dellios2023validationtestsgbsquantum,dellios2024validation}. We use a definition in which the thermalization, parameterized by $\epsilon_{i}\leq1$, affects only the coherence~\cite{drummondSimulatingComplexNetworks2022}:
\begin{equation}
\left\langle \hat{a}_{i}^{2}\right\rangle =\tilde{m}_{i}=\left(1-\epsilon_{i}\right)\sinh\left(r_{i}\right)\cosh\left(r_{i}\right).\label{eq:thermal_coherence}
\end{equation}
To analyze the effects of photon loss in multi-mode photonic experiments, we define the quadrature operators for a set of $M$ annihilation operators $\hat{a}_{i}$ as:
\begin{align}
\hat{X}_{i} & =\hat{a}_{i}+\hat{a}_{i}^{\dagger},\nonumber\\
\hat{X}_{i+M}=\hat{Y}_{i} & =-i\left(\hat{a}_{i}-\hat{a}_{i}^{\dagger}\right).
\end{align}
We use quadratures normalized such that $[\hat{X}_{i},\hat{Y}_{j}]=2i\delta_{ij}$, corresponding to $\hbar=2$. For the thermalized squeezed state of Eq.~\eqref{eq:thermal_coherence}, an input Gaussian mode is non-classical if and only if its normally ordered $\hat{Y}_{i}$-quadrature variance satisfies:
\begin{equation}
\left\langle :\hat{Y}_{i}^{2}:\right\rangle =2\sinh\left(r_{i}\right)\left[\sinh\left(r_{i}\right)-\left(1-\epsilon_{i}\right)\cosh\left(r_{i}\right)\right]<0.
\end{equation}
Hence, there is always a classical $P$ distribution with thermal noise if $\epsilon_{i}\geq1-\tanh\left(r_{i}\right)$~\cite{qi2020regimes}.
\\\textit{Theorem:} let $\sigma_{ij}=\langle{:}\Delta\hat{X}_i\Delta\hat{X}_j{:}\rangle$ be the normally ordered quadrature covariance, which equals $V-I$, with $V$ the symmetrically ordered covariance. Linear loss whose transmission matrix $\boldsymbol{T}$ has full column rank does not change the number of negative eigenvalues of $\sigma$, so a state whose $\sigma$ has a negative eigenvalue remains non-classical.
\\\textit{Proof:} for a classical state the Glauber--Sudarshan distribution is non-negative and its quadratures are real, so $\sigma$ is an ordinary covariance matrix: $\boldsymbol{u}^{T}\sigma\boldsymbol{u}=\langle(\boldsymbol{u}\cdot\Delta\boldsymbol{X})^{2}\rangle_{P}\geq0$ for every real vector $\boldsymbol{u}$. A negative eigenvalue therefore implies non-classicality. Loss maps $\hat{\boldsymbol{a}}^{(out)}=\boldsymbol{T}\hat{\boldsymbol{a}}^{(in)}+{}$vacuum terms, so the quadratures transform as $\Delta\boldsymbol{X}^{(out)}=\boldsymbol{\mathcal{T}}\Delta\boldsymbol{X}^{(in)}+\Delta\boldsymbol{X}^{(vac)}$, with the real matrix
\begin{equation}
\boldsymbol{\mathcal{T}}=\begin{pmatrix}\mathrm{Re}\,\boldsymbol{T} & -\mathrm{Im}\,\boldsymbol{T}\\
\mathrm{Im}\,\boldsymbol{T} & \mathrm{Re}\,\boldsymbol{T}
\end{pmatrix}.\label{eq:Tquad}
\end{equation}
The vacuum has zero normally ordered covariance and is uncorrelated with the input, so
\begin{equation}
\sigma^{(out)}=\boldsymbol{\mathcal{T}}\sigma^{(in)}\boldsymbol{\mathcal{T}}^{T}.\label{eq:sigma_out}
\end{equation}
$\boldsymbol{\mathcal{T}}$ has full column rank when $\boldsymbol{T}$ does. By Sylvester's law of inertia~\cite{Sylvester1852Demonstration}, this congruence leaves the numbers of positive and negative eigenvalues unchanged; a rectangular $\boldsymbol{\mathcal{T}}$ only adds zero eigenvalues.~$\blacksquare$ For Gaussian states the converse also holds, since $P$ is then a Gaussian with covariance $\sigma$.
\paragraph*{Photon-number-resolving projections and output steps---}
For photon-number-resolving (PNR) detectors we can use the minimum-distance truncation on $n_j$:
\begin{equation}
n_{j}^{(cl)}\equiv\mathcal{T}_{n}(n_{j})\equiv\max\left(\operatorname{Re}n_{j},0\right).\label{eq:pnr_projection}
\end{equation}
Alternatively, we can use the minimum-distance truncation $\pi_{j,c}^{(cl)}=[\mathcal{T}_{\Delta}(\boldsymbol{\pi}_j)]_c$ on $\pi_{j,c}$:
\begin{alignat}{1}
\mathcal{T}_{\Delta}(\boldsymbol{\pi}_j)&\equiv
\operatorname*{arg\,min}_{\boldsymbol{q}\in\Delta_{C}}
\bigl\|\boldsymbol{q}-\operatorname{Re}\boldsymbol{\pi}_j\bigr\|_{2},
\label{eq:simplex_projection}
\\\Delta_{C}&=\Bigl\{\boldsymbol{q}\in\mathbb{R}^{C+1}:\,q_c\geq0,\ \textstyle\sum_{c=0}^{C}q_c=1\Bigr\},\label{eq:simplex}
\end{alignat}
where $\boldsymbol{\pi}_j=(\pi_{j,0},\ldots,\pi_{j,C-1},\pi_{j,\geq C})$ with
$\pi_{j,\geq C}=1-\sum_{c<C}\pi_{j,c}$. Its solution is
$[\mathcal{T}_{\Delta}(\boldsymbol{\pi}_j)]_c=\max(\operatorname{Re}\pi_{j,c}-\tau_j,0)$,
with $\tau_j$ the unique value for which the entries sum to one~\cite{duchi2008efficient}.
For $C=1$, $\boldsymbol{\pi}_j=(\pi_j,1-\pi_j)$ and $\mathcal{T}_{\Delta}$ reduces to the clamp $\mathcal{T}_p$. In general, $\mathcal{T}_{\Delta}$ acts on count probabilities directly, so a projected sample need not be a coherent state. We take $C$ such that $P_j(n\geq C)\leq10^{-4}$ in every mode, where $P_j$ is the exact single-mode photon-number distribution, rather than the largest count a detector resolves.

After the whitening--coloring iteration of Eqs.~\eqref{eq:wc_iter}--\eqref{eq:wc_W}, for PNR detectors with the Poisson output step, the whitened variables are the photon numbers, $x_j^{(\ell)}=n_j^{(\ell)}\geq0$, and each count is drawn as
\begin{align}
    c_j \sim \mathrm{Poisson}\bigl(n_j^{(\ell)}\bigr),~ ~
    \Pr(c_j=c)=\frac{e^{-n_j^{(\ell)}}\bigl(n_j^{(\ell)}\bigr)^{c}}{c!} .
    \label{eq:draw_poisson}
\end{align}
With the categorical output step, the whitened variables are each mode's probability vector $\boldsymbol{q}^{(\ell)}_j=(q^{(\ell)}_{j,0},\ldots,q^{(\ell)}_{j,C-1},q^{(\ell)}_{j,\geq C})$, with cross-mode covariances targeted only for categories 0--2, and the count is drawn from its cumulative sum,
\begin{align}
    c_j &= \min\Bigl\{c\in\{0,\ldots,C\}:\ u_j\le \textstyle\sum_{c'=0}^{c} q^{(\ell)}_{j,c'}\Bigr\}\label{eq:draw_categorical}, 
    \\&u_j\sim U(0,1),\nonumber
\end{align}
where $c_j=C$ denotes the top category, ``$C$ or more''. A count in that category is replaced by one drawn from the exact conditional distribution $P_j(n\,|\,n\geq C)$, computed from the Gaussian covariance matrix. With $C=1$, $\boldsymbol{q}_j=(x_j,1-x_j)$ and Eq.~\eqref{eq:draw_categorical} reduces to the threshold rule, Eq.~\eqref{eq:draw_threshold}.

\paragraph*{Observables---}
Since the statistical tests used are not all independent, we report the maximum $Z$-score of our tests rather than combining them. Each dataset is scored on up to 23 observables. For threshold detectors these are the per-mode click probabilities $\langle c_{j}\rangle$; the distribution $\mathcal{G}^{M}_{M}$ of the total number of clicks; the joint distribution $\mathcal{G}^{M}_{M/2,M/2}$ of the click totals in the first and second halves of the modes; and, for $\mathcal{K}=1,\ldots,20$ (at most $M$), the outcome distribution $\mathcal{M}^{\mathcal{K}}$ on $\mathcal{K}$ randomly chosen modes. For Borealis, the click observables are replaced by the mean photon number per mode, $\langle\hat{n}_{j}\rangle$, the photon-number distribution $P_{j}(n)$ and the total photon-number distribution, with $\mathcal{M}^{\mathcal{K}}$ restricted to at most one photon per mode. Each mode set is drawn at random, with a fixed seed, so that no single group of neighboring modes drives the marginal results. When all the sets fit within the $M$ modes together (210 modes for $\mathcal{K}=1,\ldots,20$), they are disjoint; otherwise each set is drawn independently. Every sampler and both ground truths are tested on the same mode sets and outcomes. Up to tenth order we use the complete marginal distribution for threshold detectors; above tenth order we test 1024 distinct outcomes, drawn at random from an independent-mode model in which every input carries the mean input photon number (each mode's bit probability clamped to $[0.02,0.98]$). This favors higher-probability patterns and reduces the loss of statistical power caused by sparsity at high order. The Borealis restriction makes the marginals comparable with the threshold-detector experiments. These outcomes are exact photon numbers. A sample with two or more photons in a tested mode therefore falls in none of the selected outcomes, but still counts toward $N$.

\paragraph*{Pearson $\chi^{2}$ and {$Z$-score}---}

We use Pearson $\chi^{2}$ scores~\cite{nist_ehandbook} as our primary means of comparison. GCDs and complete marginals each give $k$ probabilities $\{\mathcal{G}_{i}\}$ with $\sum_{i}\mathcal{G}_{i}=1$, which we compare with the sampled frequencies $\{\mathcal{G}_{S,i}\}$ through
\begin{align}
\chi_{P}^{2} & =\sum_{i=1}^{k}\dfrac{(\mathcal{G}_{i}-\mathcal{G}_{S,i})^{2}}{\sigma_{i}^{2}+\sigma_{S,i}^{2}},
\end{align}
where $\sigma_{i}^{2}$ is the sampling variance of the phase-space simulation (zero for covariance matrix methods) and $\sigma_{S,i}^{2}$ is the variance an exact sampler of $N$ samples would give, computed from the ground truth, so it is the same for every sampler. For large $N$ the multinomial counts make $\chi_{P}^{2}$ approximately a $\chi^{2}$ variable. A complete marginal contains every outcome on its selected modes and is normalized; an incomplete marginal, used above tenth order and for Borealis, contains only a subset of those outcomes. For each bin of a distribution, $\sigma_{S,i}^{2}$ is $\mathcal{G}_{i}/N$, the standard Pearson form. For a per-mode click probability $\langle c_j\rangle$ it is $\langle c_j\rangle(1-\langle c_j\rangle)/N$, since each detector is a single yes/no trial. For a per-mode mean photon number $\langle\hat{n}_j\rangle$ it is $(\langle\hat{n}_j\rangle+\langle\hat{n}_j\rangle^{2}+|\langle\hat{a}_j^{2}\rangle|^{2})/N$, the photon-number variance of a Gaussian mode~\cite{vallone2019means}. Using the sample's own frequencies instead would give smaller error bars to bins that happen to be under-populated, which over-disperses $\chi_{P}^{2}$. Bins are kept when $N\mathcal{G}_{i}\geq10$. 

For an exact sampler, a kept bin contributes on average
\begin{equation}
\frac{\sigma_{i}^{2}+\mathcal{G}_{i}(1-\mathcal{G}_{i})/N}{\sigma_{i}^{2}+\sigma_{S,i}^{2}}\label{eq:null_bin}
\end{equation}
to $\chi_{P}^{2}$, from the multinomial variance of a sampled frequency~\cite{nist_ehandbook}. The sum of Eq.~\eqref{eq:null_bin} over kept bins is the expected value $\mu$ of $\chi_{P}^{2}$. With $\sigma_i=0$, $\mu$ equals the number of kept bins $k$ minus their total probability. This is $k-1$ for a complete distribution, somewhat more for an incomplete marginal, and about $k$ minus the number of modes for the per-mode photon-number distributions $P_j(n)$. For per-mode values ($\langle c_j\rangle$, $\langle\hat{n}_j\rangle$) each term has mean one, so $\mu=k$. We treat $\chi_{P}^{2}$ as a $\chi^{2}$ variable with $\mu$ degrees of freedom, and define $Z$ as the standard-normal value with the same tail probability. For upper-tail probability $Q$, $Z=\sqrt{2}\,\mathrm{erfc}^{-1}(2Q)$, and similarly in the lower tail, with $Q$ computed from the regularized incomplete gamma function~\cite{nist_handbook}. Where the tail probability falls below $10^{-300}$, which is $|Z|\gtrsim37$, we use the saddlepoint approximation~\cite{lugannani1980saddle} in the form~\cite{barndorffnielsen1991modified}
\begin{align}
Z&=\omega+\frac{1}{\omega}\ln\frac{v}{\omega},\nonumber\\
\omega&=\mathrm{sign}(\kappa-1)\sqrt{\mu\left(\kappa-1-\ln \kappa\right)},\quad v=(\kappa-1)\sqrt{\mu/2},
\end{align}
with $\kappa=\chi_{P}^{2}/\mu$. It agrees with the exact value to within $0.005$ at $|Z|=30$, even for one degree of freedom. A dataset's score, $\max|Z|$, is the largest of up to 23 $Z$-scores, so even an exact sampler exceeds $|Z|=3$ more often than in a single test; keeping the overall false-alarm rate of one $|Z|=3$ test would raise the threshold to about 3.85 (Bonferroni correction). We keep $|Z|=3$, which is stricter on every sampler.

We tested the whole procedure on samples drawn exactly from a 16-mode product state, whose statistics are known in closed form: 16 squeezed vacua, each sent with transmission 0.5 to its own threshold detector, with click probabilities spaced logarithmically from 0.005 to 0.5. Over 200 repetitions of $10^{6}$ samples, the $Z$-scores of $\langle c_j\rangle$, $\mathcal{G}^{M}_{M}$ and $\mathcal{M}^{10}$ had means $0.11$, $0.03$ and $-0.02$, and standard deviations $1.02$, $1.05$ and $1.01$, as expected for a standard normal variable. This test has independent modes. For correlated modes, treating $\chi_{P}^{2}$ as a $\chi^{2}$ variable is an approximation, because it ignores the correlations between detectors and between bins. The Mahalanobis distance, which uses the full covariance of the bins in place of their separate variances, would remove this approximation, at the cost of estimating and inverting that covariance.

For our phase-space simulated results, we estimate $\sigma_{i}$ by the batch means method. We divide the phase-space samples into blocks, compute the observable for each block, and estimate the standard error of the mean from the variance of the block means.


\begin{thebibliography}{57}%
\makeatletter
\providecommand \@ifxundefined [1]{%
 \@ifx{#1\undefined}
}%
\providecommand \@ifnum [1]{%
 \ifnum #1\expandafter \@firstoftwo
 \else \expandafter \@secondoftwo
 \fi
}%
\providecommand \@ifx [1]{%
 \ifx #1\expandafter \@firstoftwo
 \else \expandafter \@secondoftwo
 \fi
}%
\providecommand \natexlab [1]{#1}%
\providecommand \enquote  [1]{``#1''}%
\providecommand \bibnamefont  [1]{#1}%
\providecommand \bibfnamefont [1]{#1}%
\providecommand \citenamefont [1]{#1}%
\providecommand \href@noop [0]{\@secondoftwo}%
\providecommand \href [0]{\begingroup \@sanitize@url \@href}%
\providecommand \@href[1]{\@@startlink{#1}\@@href}%
\providecommand \@@href[1]{\endgroup#1\@@endlink}%
\providecommand \@sanitize@url [0]{\catcode `\\12\catcode `\$12\catcode
  `\&12\catcode `\#12\catcode `\^12\catcode `\_12\catcode `\%12\relax}%
\providecommand \@@startlink[1]{}%
\providecommand \@@endlink[0]{}%
\providecommand \url  [0]{\begingroup\@sanitize@url \@url }%
\providecommand \@url [1]{\endgroup\@href {#1}{\urlprefix }}%
\providecommand \urlprefix  [0]{URL }%
\providecommand \Eprint [0]{\href }%
\providecommand \doibase [0]{https://doi.org/}%
\providecommand \selectlanguage [0]{\@gobble}%
\providecommand \bibinfo  [0]{\@secondoftwo}%
\providecommand \bibfield  [0]{\@secondoftwo}%
\providecommand \translation [1]{[#1]}%
\providecommand \BibitemOpen [0]{}%
\providecommand \bibitemStop [0]{}%
\providecommand \bibitemNoStop [0]{.\EOS\space}%
\providecommand \EOS [0]{\spacefactor3000\relax}%
\providecommand \BibitemShut  [1]{\csname bibitem#1\endcsname}%
\let\auto@bib@innerbib\@empty
\bibitem [{\citenamefont {Zhong}\ \emph {et~al.}(2020)\citenamefont {Zhong},
  \citenamefont {Wang}, \citenamefont {Deng}, \citenamefont {Chen},
  \citenamefont {Peng}, \citenamefont {Luo}, \citenamefont {Qin}, \citenamefont
  {Wu}, \citenamefont {Ding}, \citenamefont {Hu}, \citenamefont {Hu},
  \citenamefont {Yang}, \citenamefont {Zhang}, \citenamefont {Li},
  \citenamefont {Li}, \citenamefont {Jiang}, \citenamefont {Gan}, \citenamefont
  {Yang}, \citenamefont {You}, \citenamefont {Wang}, \citenamefont {Li},
  \citenamefont {Liu}, \citenamefont {Lu},\ and\ \citenamefont
  {Pan}}]{zhong2020quantum}%
  \BibitemOpen
  \bibfield  {author} {\bibinfo {author} {\bibfnamefont {H.-S.}\ \bibnamefont
  {Zhong}}, \bibinfo {author} {\bibfnamefont {H.}~\bibnamefont {Wang}},
  \bibinfo {author} {\bibfnamefont {Y.-H.}\ \bibnamefont {Deng}}, \bibinfo
  {author} {\bibfnamefont {M.-C.}\ \bibnamefont {Chen}}, \bibinfo {author}
  {\bibfnamefont {L.-C.}\ \bibnamefont {Peng}}, \bibinfo {author}
  {\bibfnamefont {Y.-H.}\ \bibnamefont {Luo}}, \bibinfo {author} {\bibfnamefont
  {J.}~\bibnamefont {Qin}}, \bibinfo {author} {\bibfnamefont {D.}~\bibnamefont
  {Wu}}, \bibinfo {author} {\bibfnamefont {X.}~\bibnamefont {Ding}}, \bibinfo
  {author} {\bibfnamefont {Y.}~\bibnamefont {Hu}}, \bibinfo {author}
  {\bibfnamefont {P.}~\bibnamefont {Hu}}, \bibinfo {author} {\bibfnamefont
  {X.-Y.}\ \bibnamefont {Yang}}, \bibinfo {author} {\bibfnamefont {W.-J.}\
  \bibnamefont {Zhang}}, \bibinfo {author} {\bibfnamefont {H.}~\bibnamefont
  {Li}}, \bibinfo {author} {\bibfnamefont {Y.}~\bibnamefont {Li}}, \bibinfo
  {author} {\bibfnamefont {X.}~\bibnamefont {Jiang}}, \bibinfo {author}
  {\bibfnamefont {L.}~\bibnamefont {Gan}}, \bibinfo {author} {\bibfnamefont
  {G.}~\bibnamefont {Yang}}, \bibinfo {author} {\bibfnamefont {L.}~\bibnamefont
  {You}}, \bibinfo {author} {\bibfnamefont {Z.}~\bibnamefont {Wang}}, \bibinfo
  {author} {\bibfnamefont {L.}~\bibnamefont {Li}}, \bibinfo {author}
  {\bibfnamefont {N.-L.}\ \bibnamefont {Liu}}, \bibinfo {author} {\bibfnamefont
  {C.-Y.}\ \bibnamefont {Lu}},\ and\ \bibinfo {author} {\bibfnamefont {J.-W.}\
  \bibnamefont {Pan}},\ }\href {https://doi.org/10.1126/science.abe8770}
  {\bibfield  {journal} {\bibinfo  {journal} {Science}\ }\textbf {\bibinfo
  {volume} {370}},\ \bibinfo {pages} {1460} (\bibinfo {year}
  {2020})}\BibitemShut {NoStop}%
\bibitem [{\citenamefont {Zhong}\ \emph {et~al.}(2021)\citenamefont {Zhong},
  \citenamefont {Deng}, \citenamefont {Qin}, \citenamefont {Wang},
  \citenamefont {Chen}, \citenamefont {Peng}, \citenamefont {Luo},
  \citenamefont {Wu}, \citenamefont {Gong}, \citenamefont {Su}, \citenamefont
  {Hu}, \citenamefont {Hu}, \citenamefont {Yang}, \citenamefont {Zhang},
  \citenamefont {Li}, \citenamefont {Li}, \citenamefont {Jiang}, \citenamefont
  {Gan}, \citenamefont {Yang}, \citenamefont {You}, \citenamefont {Wang},
  \citenamefont {Li}, \citenamefont {Liu}, \citenamefont {Renema},
  \citenamefont {Lu},\ and\ \citenamefont
  {Pan}}]{zhongPhaseProgrammableGaussianBoson2021}%
  \BibitemOpen
  \bibfield  {author} {\bibinfo {author} {\bibfnamefont {H.-S.}\ \bibnamefont
  {Zhong}}, \bibinfo {author} {\bibfnamefont {Y.-H.}\ \bibnamefont {Deng}},
  \bibinfo {author} {\bibfnamefont {J.}~\bibnamefont {Qin}}, \bibinfo {author}
  {\bibfnamefont {H.}~\bibnamefont {Wang}}, \bibinfo {author} {\bibfnamefont
  {M.-C.}\ \bibnamefont {Chen}}, \bibinfo {author} {\bibfnamefont {L.-C.}\
  \bibnamefont {Peng}}, \bibinfo {author} {\bibfnamefont {Y.-H.}\ \bibnamefont
  {Luo}}, \bibinfo {author} {\bibfnamefont {D.}~\bibnamefont {Wu}}, \bibinfo
  {author} {\bibfnamefont {S.-Q.}\ \bibnamefont {Gong}}, \bibinfo {author}
  {\bibfnamefont {H.}~\bibnamefont {Su}}, \bibinfo {author} {\bibfnamefont
  {Y.}~\bibnamefont {Hu}}, \bibinfo {author} {\bibfnamefont {P.}~\bibnamefont
  {Hu}}, \bibinfo {author} {\bibfnamefont {X.-Y.}\ \bibnamefont {Yang}},
  \bibinfo {author} {\bibfnamefont {W.-J.}\ \bibnamefont {Zhang}}, \bibinfo
  {author} {\bibfnamefont {H.}~\bibnamefont {Li}}, \bibinfo {author}
  {\bibfnamefont {Y.}~\bibnamefont {Li}}, \bibinfo {author} {\bibfnamefont
  {X.}~\bibnamefont {Jiang}}, \bibinfo {author} {\bibfnamefont
  {L.}~\bibnamefont {Gan}}, \bibinfo {author} {\bibfnamefont {G.}~\bibnamefont
  {Yang}}, \bibinfo {author} {\bibfnamefont {L.}~\bibnamefont {You}}, \bibinfo
  {author} {\bibfnamefont {Z.}~\bibnamefont {Wang}}, \bibinfo {author}
  {\bibfnamefont {L.}~\bibnamefont {Li}}, \bibinfo {author} {\bibfnamefont
  {N.-L.}\ \bibnamefont {Liu}}, \bibinfo {author} {\bibfnamefont {J.~J.}\
  \bibnamefont {Renema}}, \bibinfo {author} {\bibfnamefont {C.-Y.}\
  \bibnamefont {Lu}},\ and\ \bibinfo {author} {\bibfnamefont {J.-W.}\
  \bibnamefont {Pan}},\ }\href {https://doi.org/10.1103/PhysRevLett.127.180502}
  {\bibfield  {journal} {\bibinfo  {journal} {Phys. Rev. Lett.}\ }\textbf
  {\bibinfo {volume} {127}},\ \bibinfo {pages} {180502} (\bibinfo {year}
  {2021})}\BibitemShut {NoStop}%
\bibitem [{\citenamefont {Deng}\ \emph {et~al.}(2023)\citenamefont {Deng},
  \citenamefont {Gu}, \citenamefont {Liu}, \citenamefont {Gong}, \citenamefont
  {Su}, \citenamefont {Zhang}, \citenamefont {Tang}, \citenamefont {Jia},
  \citenamefont {Xu}, \citenamefont {Chen}, \citenamefont {Qin}, \citenamefont
  {Peng}, \citenamefont {Yan}, \citenamefont {Hu}, \citenamefont {Huang},
  \citenamefont {Li}, \citenamefont {Li}, \citenamefont {Chen}, \citenamefont
  {Jiang}, \citenamefont {Gan}, \citenamefont {Yang}, \citenamefont {You},
  \citenamefont {Li}, \citenamefont {Zhong}, \citenamefont {Wang},
  \citenamefont {Liu}, \citenamefont {Renema}, \citenamefont {Lu},\ and\
  \citenamefont {Pan}}]{deng2023gaussian}%
  \BibitemOpen
  \bibfield  {author} {\bibinfo {author} {\bibfnamefont {Y.-H.}\ \bibnamefont
  {Deng}}, \bibinfo {author} {\bibfnamefont {Y.-C.}\ \bibnamefont {Gu}},
  \bibinfo {author} {\bibfnamefont {H.-L.}\ \bibnamefont {Liu}}, \bibinfo
  {author} {\bibfnamefont {S.-Q.}\ \bibnamefont {Gong}}, \bibinfo {author}
  {\bibfnamefont {H.}~\bibnamefont {Su}}, \bibinfo {author} {\bibfnamefont
  {Z.-J.}\ \bibnamefont {Zhang}}, \bibinfo {author} {\bibfnamefont {H.-Y.}\
  \bibnamefont {Tang}}, \bibinfo {author} {\bibfnamefont {M.-H.}\ \bibnamefont
  {Jia}}, \bibinfo {author} {\bibfnamefont {J.-M.}\ \bibnamefont {Xu}},
  \bibinfo {author} {\bibfnamefont {M.-C.}\ \bibnamefont {Chen}}, \bibinfo
  {author} {\bibfnamefont {J.}~\bibnamefont {Qin}}, \bibinfo {author}
  {\bibfnamefont {L.-C.}\ \bibnamefont {Peng}}, \bibinfo {author}
  {\bibfnamefont {J.}~\bibnamefont {Yan}}, \bibinfo {author} {\bibfnamefont
  {Y.}~\bibnamefont {Hu}}, \bibinfo {author} {\bibfnamefont {J.}~\bibnamefont
  {Huang}}, \bibinfo {author} {\bibfnamefont {H.}~\bibnamefont {Li}}, \bibinfo
  {author} {\bibfnamefont {Y.}~\bibnamefont {Li}}, \bibinfo {author}
  {\bibfnamefont {Y.}~\bibnamefont {Chen}}, \bibinfo {author} {\bibfnamefont
  {X.}~\bibnamefont {Jiang}}, \bibinfo {author} {\bibfnamefont
  {L.}~\bibnamefont {Gan}}, \bibinfo {author} {\bibfnamefont {G.}~\bibnamefont
  {Yang}}, \bibinfo {author} {\bibfnamefont {L.}~\bibnamefont {You}}, \bibinfo
  {author} {\bibfnamefont {L.}~\bibnamefont {Li}}, \bibinfo {author}
  {\bibfnamefont {H.-S.}\ \bibnamefont {Zhong}}, \bibinfo {author}
  {\bibfnamefont {H.}~\bibnamefont {Wang}}, \bibinfo {author} {\bibfnamefont
  {N.-L.}\ \bibnamefont {Liu}}, \bibinfo {author} {\bibfnamefont {J.~J.}\
  \bibnamefont {Renema}}, \bibinfo {author} {\bibfnamefont {C.-Y.}\
  \bibnamefont {Lu}},\ and\ \bibinfo {author} {\bibfnamefont {J.-W.}\
  \bibnamefont {Pan}},\ }\href {https://doi.org/10.1103/PhysRevLett.131.150601}
  {\bibfield  {journal} {\bibinfo  {journal} {Phys. Rev. Lett.}\ }\textbf
  {\bibinfo {volume} {131}},\ \bibinfo {pages} {150601} (\bibinfo {year}
  {2023})}\BibitemShut {NoStop}%
\bibitem [{\citenamefont {Liu}\ \emph {et~al.}(2026{\natexlab{a}})\citenamefont
  {Liu}, \citenamefont {Su}, \citenamefont {Deng}, \citenamefont {Gong},
  \citenamefont {Gu}, \citenamefont {Tang}, \citenamefont {Jia}, \citenamefont
  {Wei}, \citenamefont {Song}, \citenamefont {Wang}, \citenamefont {Zheng},
  \citenamefont {Chen}, \citenamefont {Li}, \citenamefont {Ren}, \citenamefont
  {Zhu}, \citenamefont {Wang}, \citenamefont {Chen}, \citenamefont {Liu},
  \citenamefont {Song}, \citenamefont {Yang}, \citenamefont {Chen},
  \citenamefont {An}, \citenamefont {Zhang}, \citenamefont {Gan}, \citenamefont
  {Yang}, \citenamefont {Xu}, \citenamefont {He}, \citenamefont {Wang},
  \citenamefont {Zhong}, \citenamefont {Chen}, \citenamefont {Jiang},
  \citenamefont {Li}, \citenamefont {Liu}, \citenamefont {Su}, \citenamefont
  {Zhang}, \citenamefont {Lu},\ and\ \citenamefont {Pan}}]{liu2025robust}%
  \BibitemOpen
  \bibfield  {author} {\bibinfo {author} {\bibfnamefont {H.-L.}\ \bibnamefont
  {Liu}}, \bibinfo {author} {\bibfnamefont {H.}~\bibnamefont {Su}}, \bibinfo
  {author} {\bibfnamefont {Y.-H.}\ \bibnamefont {Deng}}, \bibinfo {author}
  {\bibfnamefont {S.-Q.}\ \bibnamefont {Gong}}, \bibinfo {author}
  {\bibfnamefont {Y.-C.}\ \bibnamefont {Gu}}, \bibinfo {author} {\bibfnamefont
  {H.-Y.}\ \bibnamefont {Tang}}, \bibinfo {author} {\bibfnamefont {M.-H.}\
  \bibnamefont {Jia}}, \bibinfo {author} {\bibfnamefont {Q.}~\bibnamefont
  {Wei}}, \bibinfo {author} {\bibfnamefont {Y.-K.}\ \bibnamefont {Song}},
  \bibinfo {author} {\bibfnamefont {D.-Z.}\ \bibnamefont {Wang}}, \bibinfo
  {author} {\bibfnamefont {M.-Y.}\ \bibnamefont {Zheng}}, \bibinfo {author}
  {\bibfnamefont {F.-X.}\ \bibnamefont {Chen}}, \bibinfo {author}
  {\bibfnamefont {L.-B.}\ \bibnamefont {Li}}, \bibinfo {author} {\bibfnamefont
  {S.-Y.}\ \bibnamefont {Ren}}, \bibinfo {author} {\bibfnamefont {X.-Z.}\
  \bibnamefont {Zhu}}, \bibinfo {author} {\bibfnamefont {M.-H.}\ \bibnamefont
  {Wang}}, \bibinfo {author} {\bibfnamefont {Y.-J.}\ \bibnamefont {Chen}},
  \bibinfo {author} {\bibfnamefont {Y.-F.}\ \bibnamefont {Liu}}, \bibinfo
  {author} {\bibfnamefont {L.-S.}\ \bibnamefont {Song}}, \bibinfo {author}
  {\bibfnamefont {P.-Y.}\ \bibnamefont {Yang}}, \bibinfo {author}
  {\bibfnamefont {J.-S.}\ \bibnamefont {Chen}}, \bibinfo {author}
  {\bibfnamefont {H.}~\bibnamefont {An}}, \bibinfo {author} {\bibfnamefont
  {L.}~\bibnamefont {Zhang}}, \bibinfo {author} {\bibfnamefont
  {L.}~\bibnamefont {Gan}}, \bibinfo {author} {\bibfnamefont {G.-w.}\
  \bibnamefont {Yang}}, \bibinfo {author} {\bibfnamefont {J.-M.}\ \bibnamefont
  {Xu}}, \bibinfo {author} {\bibfnamefont {Y.-M.}\ \bibnamefont {He}}, \bibinfo
  {author} {\bibfnamefont {H.}~\bibnamefont {Wang}}, \bibinfo {author}
  {\bibfnamefont {H.-S.}\ \bibnamefont {Zhong}}, \bibinfo {author}
  {\bibfnamefont {M.-C.}\ \bibnamefont {Chen}}, \bibinfo {author}
  {\bibfnamefont {X.}~\bibnamefont {Jiang}}, \bibinfo {author} {\bibfnamefont
  {L.}~\bibnamefont {Li}}, \bibinfo {author} {\bibfnamefont {N.-L.}\
  \bibnamefont {Liu}}, \bibinfo {author} {\bibfnamefont {X.-L.}\ \bibnamefont
  {Su}}, \bibinfo {author} {\bibfnamefont {Q.}~\bibnamefont {Zhang}}, \bibinfo
  {author} {\bibfnamefont {C.-Y.}\ \bibnamefont {Lu}},\ and\ \bibinfo {author}
  {\bibfnamefont {J.-W.}\ \bibnamefont {Pan}},\ }\href
  {https://doi.org/10.1038/s41586-026-10523-6} {\bibfield  {journal} {\bibinfo
  {journal} {Nature}\ }\textbf {\bibinfo {volume} {653}},\ \bibinfo {pages}
  {687} (\bibinfo {year} {2026}{\natexlab{a}})}\BibitemShut {NoStop}%
\bibitem [{\citenamefont {Madsen}\ \emph {et~al.}(2022)\citenamefont {Madsen},
  \citenamefont {Laudenbach}, \citenamefont {Askarani}, \citenamefont
  {Rortais}, \citenamefont {Vincent}, \citenamefont {Bulmer}, \citenamefont
  {Miatto}, \citenamefont {Neuhaus}, \citenamefont {Helt}, \citenamefont
  {Collins}, \citenamefont {Lita}, \citenamefont {Gerrits}, \citenamefont
  {Nam}, \citenamefont {Vaidya}, \citenamefont {Menotti}, \citenamefont
  {Dhand}, \citenamefont {Vernon}, \citenamefont {Quesada},\ and\ \citenamefont
  {Lavoie}}]{madsenQuantumComputationalAdvantage2022}%
  \BibitemOpen
  \bibfield  {author} {\bibinfo {author} {\bibfnamefont {L.~S.}\ \bibnamefont
  {Madsen}}, \bibinfo {author} {\bibfnamefont {F.}~\bibnamefont {Laudenbach}},
  \bibinfo {author} {\bibfnamefont {M.~F.}\ \bibnamefont {Askarani}}, \bibinfo
  {author} {\bibfnamefont {F.}~\bibnamefont {Rortais}}, \bibinfo {author}
  {\bibfnamefont {T.}~\bibnamefont {Vincent}}, \bibinfo {author} {\bibfnamefont
  {J.~F.~F.}\ \bibnamefont {Bulmer}}, \bibinfo {author} {\bibfnamefont {F.~M.}\
  \bibnamefont {Miatto}}, \bibinfo {author} {\bibfnamefont {L.}~\bibnamefont
  {Neuhaus}}, \bibinfo {author} {\bibfnamefont {L.~G.}\ \bibnamefont {Helt}},
  \bibinfo {author} {\bibfnamefont {M.~J.}\ \bibnamefont {Collins}}, \bibinfo
  {author} {\bibfnamefont {A.~E.}\ \bibnamefont {Lita}}, \bibinfo {author}
  {\bibfnamefont {T.}~\bibnamefont {Gerrits}}, \bibinfo {author} {\bibfnamefont
  {S.~W.}\ \bibnamefont {Nam}}, \bibinfo {author} {\bibfnamefont {V.~D.}\
  \bibnamefont {Vaidya}}, \bibinfo {author} {\bibfnamefont {M.}~\bibnamefont
  {Menotti}}, \bibinfo {author} {\bibfnamefont {I.}~\bibnamefont {Dhand}},
  \bibinfo {author} {\bibfnamefont {Z.}~\bibnamefont {Vernon}}, \bibinfo
  {author} {\bibfnamefont {N.}~\bibnamefont {Quesada}},\ and\ \bibinfo {author}
  {\bibfnamefont {J.}~\bibnamefont {Lavoie}},\ }\href
  {https://doi.org/10.1038/s41586-022-04725-x} {\bibfield  {journal} {\bibinfo
  {journal} {Nature}\ }\textbf {\bibinfo {volume} {606}},\ \bibinfo {pages}
  {75} (\bibinfo {year} {2022})}\BibitemShut {NoStop}%
\bibitem [{\citenamefont {Hamilton}\ \emph {et~al.}(2017)\citenamefont
  {Hamilton}, \citenamefont {Kruse}, \citenamefont {Sansoni}, \citenamefont
  {Barkhofen}, \citenamefont {Silberhorn},\ and\ \citenamefont
  {Jex}}]{hamiltonGaussian2017}%
  \BibitemOpen
  \bibfield  {author} {\bibinfo {author} {\bibfnamefont {C.~S.}\ \bibnamefont
  {Hamilton}}, \bibinfo {author} {\bibfnamefont {R.}~\bibnamefont {Kruse}},
  \bibinfo {author} {\bibfnamefont {L.}~\bibnamefont {Sansoni}}, \bibinfo
  {author} {\bibfnamefont {S.}~\bibnamefont {Barkhofen}}, \bibinfo {author}
  {\bibfnamefont {C.}~\bibnamefont {Silberhorn}},\ and\ \bibinfo {author}
  {\bibfnamefont {I.}~\bibnamefont {Jex}},\ }\href
  {https://doi.org/10.1103/PhysRevLett.119.170501} {\bibfield  {journal}
  {\bibinfo  {journal} {Phys. Rev. Lett.}\ }\textbf {\bibinfo {volume} {119}},\
  \bibinfo {pages} {170501} (\bibinfo {year} {2017})}\BibitemShut {NoStop}%
\bibitem [{\citenamefont {Quesada}\ \emph {et~al.}(2018)\citenamefont
  {Quesada}, \citenamefont {Arrazola},\ and\ \citenamefont
  {Killoran}}]{quesadaGaussian2018}%
  \BibitemOpen
  \bibfield  {author} {\bibinfo {author} {\bibfnamefont {N.}~\bibnamefont
  {Quesada}}, \bibinfo {author} {\bibfnamefont {J.~M.}\ \bibnamefont
  {Arrazola}},\ and\ \bibinfo {author} {\bibfnamefont {N.}~\bibnamefont
  {Killoran}},\ }\href {https://doi.org/10.1103/PhysRevA.98.062322} {\bibfield
  {journal} {\bibinfo  {journal} {Phys. Rev. A}\ }\textbf {\bibinfo {volume}
  {98}},\ \bibinfo {pages} {062322} (\bibinfo {year} {2018})}\BibitemShut
  {NoStop}%
\bibitem [{\citenamefont {Kruse}\ \emph {et~al.}(2019)\citenamefont {Kruse},
  \citenamefont {Hamilton}, \citenamefont {Sansoni}, \citenamefont {Barkhofen},
  \citenamefont {Silberhorn},\ and\ \citenamefont {Jex}}]{kruseDetailed2019}%
  \BibitemOpen
  \bibfield  {author} {\bibinfo {author} {\bibfnamefont {R.}~\bibnamefont
  {Kruse}}, \bibinfo {author} {\bibfnamefont {C.~S.}\ \bibnamefont {Hamilton}},
  \bibinfo {author} {\bibfnamefont {L.}~\bibnamefont {Sansoni}}, \bibinfo
  {author} {\bibfnamefont {S.}~\bibnamefont {Barkhofen}}, \bibinfo {author}
  {\bibfnamefont {C.}~\bibnamefont {Silberhorn}},\ and\ \bibinfo {author}
  {\bibfnamefont {I.}~\bibnamefont {Jex}},\ }\href
  {https://doi.org/10.1103/PhysRevA.100.032326} {\bibfield  {journal} {\bibinfo
   {journal} {Phys. Rev. A}\ }\textbf {\bibinfo {volume} {100}},\ \bibinfo
  {pages} {032326} (\bibinfo {year} {2019})}\BibitemShut {NoStop}%
\bibitem [{\citenamefont {Aaronson}\ and\ \citenamefont
  {Arkhipov}(2013)}]{AaronsonArkhipov2013LV}%
  \BibitemOpen
  \bibfield  {author} {\bibinfo {author} {\bibfnamefont {S.}~\bibnamefont
  {Aaronson}}\ and\ \bibinfo {author} {\bibfnamefont {A.}~\bibnamefont
  {Arkhipov}},\ }\href {https://doi.org/10.4086/toc.2013.v009a004} {\bibfield
  {journal} {\bibinfo  {journal} {Theory Comput.}\ }\textbf {\bibinfo {volume}
  {9}},\ \bibinfo {pages} {143} (\bibinfo {year} {2013})}\BibitemShut {NoStop}%
\bibitem [{\citenamefont {Grier}\ \emph {et~al.}(2022)\citenamefont {Grier},
  \citenamefont {Brod}, \citenamefont {Arrazola}, \citenamefont
  {de~Andrade~Alonso},\ and\ \citenamefont {Quesada}}]{grier2022complexity}%
  \BibitemOpen
  \bibfield  {author} {\bibinfo {author} {\bibfnamefont {D.}~\bibnamefont
  {Grier}}, \bibinfo {author} {\bibfnamefont {D.~J.}\ \bibnamefont {Brod}},
  \bibinfo {author} {\bibfnamefont {J.~M.}\ \bibnamefont {Arrazola}}, \bibinfo
  {author} {\bibfnamefont {M.~B.}\ \bibnamefont {de~Andrade~Alonso}},\ and\
  \bibinfo {author} {\bibfnamefont {N.}~\bibnamefont {Quesada}},\ }\href
  {https://doi.org/10.22331/q-2022-11-28-863} {\bibfield  {journal} {\bibinfo
  {journal} {Quantum}\ }\textbf {\bibinfo {volume} {6}},\ \bibinfo {pages}
  {863} (\bibinfo {year} {2022})}\BibitemShut {NoStop}%
\bibitem [{\citenamefont {Deshpande}\ \emph {et~al.}(2022)\citenamefont
  {Deshpande}, \citenamefont {Mehta}, \citenamefont {Vincent}, \citenamefont
  {Quesada}, \citenamefont {Hinsche}, \citenamefont {Ioannou}, \citenamefont
  {Madsen}, \citenamefont {Lavoie}, \citenamefont {Qi}, \citenamefont {Eisert},
  \citenamefont {Hangleiter}, \citenamefont {Fefferman},\ and\ \citenamefont
  {Dhand}}]{deshpandeQuantumComputationalAdvantage2022a}%
  \BibitemOpen
  \bibfield  {author} {\bibinfo {author} {\bibfnamefont {A.}~\bibnamefont
  {Deshpande}}, \bibinfo {author} {\bibfnamefont {A.}~\bibnamefont {Mehta}},
  \bibinfo {author} {\bibfnamefont {T.}~\bibnamefont {Vincent}}, \bibinfo
  {author} {\bibfnamefont {N.}~\bibnamefont {Quesada}}, \bibinfo {author}
  {\bibfnamefont {M.}~\bibnamefont {Hinsche}}, \bibinfo {author} {\bibfnamefont
  {M.}~\bibnamefont {Ioannou}}, \bibinfo {author} {\bibfnamefont
  {L.}~\bibnamefont {Madsen}}, \bibinfo {author} {\bibfnamefont
  {J.}~\bibnamefont {Lavoie}}, \bibinfo {author} {\bibfnamefont
  {H.}~\bibnamefont {Qi}}, \bibinfo {author} {\bibfnamefont {J.}~\bibnamefont
  {Eisert}}, \bibinfo {author} {\bibfnamefont {D.}~\bibnamefont {Hangleiter}},
  \bibinfo {author} {\bibfnamefont {B.}~\bibnamefont {Fefferman}},\ and\
  \bibinfo {author} {\bibfnamefont {I.}~\bibnamefont {Dhand}},\ }\href
  {https://doi.org/10.1126/sciadv.abi7894} {\bibfield  {journal} {\bibinfo
  {journal} {Sci. Adv.}\ }\textbf {\bibinfo {volume} {8}},\ \bibinfo {pages}
  {eabi7894} (\bibinfo {year} {2022})}\BibitemShut {NoStop}%
\bibitem [{\citenamefont {Qi}\ \emph {et~al.}(2020)\citenamefont {Qi},
  \citenamefont {Brod}, \citenamefont {Quesada},\ and\ \citenamefont
  {Garc{\'\i}a-Patr{\'o}n}}]{qi2020regimes}%
  \BibitemOpen
  \bibfield  {author} {\bibinfo {author} {\bibfnamefont {H.}~\bibnamefont
  {Qi}}, \bibinfo {author} {\bibfnamefont {D.~J.}\ \bibnamefont {Brod}},
  \bibinfo {author} {\bibfnamefont {N.}~\bibnamefont {Quesada}},\ and\ \bibinfo
  {author} {\bibfnamefont {R.}~\bibnamefont {Garc{\'\i}a-Patr{\'o}n}},\ }\href
  {https://doi.org/10.1103/PhysRevLett.124.100502} {\bibfield  {journal}
  {\bibinfo  {journal} {Phys. Rev. Lett.}\ }\textbf {\bibinfo {volume} {124}},\
  \bibinfo {pages} {100502} (\bibinfo {year} {2020})}\BibitemShut {NoStop}%
\bibitem [{\citenamefont {Arrazola}\ \emph {et~al.}(2021)\citenamefont
  {Arrazola}, \citenamefont {Bergholm}, \citenamefont {Br{\'a}dler},
  \citenamefont {Bromley}, \citenamefont {Collins}, \citenamefont {Dhand},
  \citenamefont {Fumagalli}, \citenamefont {Gerrits}, \citenamefont {Goussev},
  \citenamefont {Helt}, \citenamefont {Hundal}, \citenamefont {Isacsson},
  \citenamefont {Israel}, \citenamefont {Izaac}, \citenamefont {Jahangiri},
  \citenamefont {Janik}, \citenamefont {Killoran}, \citenamefont {Kumar},
  \citenamefont {Lavoie}, \citenamefont {Lita}, \citenamefont {Mahler},
  \citenamefont {Menotti}, \citenamefont {Morrison}, \citenamefont {Nam},
  \citenamefont {Neuhaus}, \citenamefont {Qi}, \citenamefont {Quesada},
  \citenamefont {Repingon}, \citenamefont {Sabapathy}, \citenamefont {Schuld},
  \citenamefont {Su}, \citenamefont {Swinarton}, \citenamefont {Sz{\'a}va},
  \citenamefont {Tan}, \citenamefont {Tan}, \citenamefont {Vaidya},
  \citenamefont {Vernon}, \citenamefont {Zabaneh},\ and\ \citenamefont
  {Zhang}}]{arrazola2021quantum}%
  \BibitemOpen
  \bibfield  {author} {\bibinfo {author} {\bibfnamefont {J.~M.}\ \bibnamefont
  {Arrazola}}, \bibinfo {author} {\bibfnamefont {V.}~\bibnamefont {Bergholm}},
  \bibinfo {author} {\bibfnamefont {K.}~\bibnamefont {Br{\'a}dler}}, \bibinfo
  {author} {\bibfnamefont {T.~R.}\ \bibnamefont {Bromley}}, \bibinfo {author}
  {\bibfnamefont {M.~J.}\ \bibnamefont {Collins}}, \bibinfo {author}
  {\bibfnamefont {I.}~\bibnamefont {Dhand}}, \bibinfo {author} {\bibfnamefont
  {A.}~\bibnamefont {Fumagalli}}, \bibinfo {author} {\bibfnamefont
  {T.}~\bibnamefont {Gerrits}}, \bibinfo {author} {\bibfnamefont
  {A.}~\bibnamefont {Goussev}}, \bibinfo {author} {\bibfnamefont {L.~G.}\
  \bibnamefont {Helt}}, \bibinfo {author} {\bibfnamefont {J.}~\bibnamefont
  {Hundal}}, \bibinfo {author} {\bibfnamefont {T.}~\bibnamefont {Isacsson}},
  \bibinfo {author} {\bibfnamefont {R.~B.}\ \bibnamefont {Israel}}, \bibinfo
  {author} {\bibfnamefont {J.}~\bibnamefont {Izaac}}, \bibinfo {author}
  {\bibfnamefont {S.}~\bibnamefont {Jahangiri}}, \bibinfo {author}
  {\bibfnamefont {R.}~\bibnamefont {Janik}}, \bibinfo {author} {\bibfnamefont
  {N.}~\bibnamefont {Killoran}}, \bibinfo {author} {\bibfnamefont {S.~P.}\
  \bibnamefont {Kumar}}, \bibinfo {author} {\bibfnamefont {J.}~\bibnamefont
  {Lavoie}}, \bibinfo {author} {\bibfnamefont {A.~E.}\ \bibnamefont {Lita}},
  \bibinfo {author} {\bibfnamefont {D.~H.}\ \bibnamefont {Mahler}}, \bibinfo
  {author} {\bibfnamefont {M.}~\bibnamefont {Menotti}}, \bibinfo {author}
  {\bibfnamefont {B.}~\bibnamefont {Morrison}}, \bibinfo {author}
  {\bibfnamefont {S.~W.}\ \bibnamefont {Nam}}, \bibinfo {author} {\bibfnamefont
  {L.}~\bibnamefont {Neuhaus}}, \bibinfo {author} {\bibfnamefont {H.~Y.}\
  \bibnamefont {Qi}}, \bibinfo {author} {\bibfnamefont {N.}~\bibnamefont
  {Quesada}}, \bibinfo {author} {\bibfnamefont {A.}~\bibnamefont {Repingon}},
  \bibinfo {author} {\bibfnamefont {K.~K.}\ \bibnamefont {Sabapathy}}, \bibinfo
  {author} {\bibfnamefont {M.}~\bibnamefont {Schuld}}, \bibinfo {author}
  {\bibfnamefont {D.}~\bibnamefont {Su}}, \bibinfo {author} {\bibfnamefont
  {J.}~\bibnamefont {Swinarton}}, \bibinfo {author} {\bibfnamefont
  {A.}~\bibnamefont {Sz{\'a}va}}, \bibinfo {author} {\bibfnamefont
  {K.}~\bibnamefont {Tan}}, \bibinfo {author} {\bibfnamefont {P.}~\bibnamefont
  {Tan}}, \bibinfo {author} {\bibfnamefont {V.~D.}\ \bibnamefont {Vaidya}},
  \bibinfo {author} {\bibfnamefont {Z.}~\bibnamefont {Vernon}}, \bibinfo
  {author} {\bibfnamefont {Z.}~\bibnamefont {Zabaneh}},\ and\ \bibinfo {author}
  {\bibfnamefont {Y.}~\bibnamefont {Zhang}},\ }\href
  {https://doi.org/10.1038/s41586-021-03202-1} {\bibfield  {journal} {\bibinfo
  {journal} {Nature}\ }\textbf {\bibinfo {volume} {591}},\ \bibinfo {pages}
  {54} (\bibinfo {year} {2021})}\BibitemShut {NoStop}%
\bibitem [{\citenamefont {Stefszky}\ \emph {et~al.}(2026)\citenamefont
  {Stefszky}, \citenamefont {Luo}, \citenamefont {Eickmann}, \citenamefont
  {Atzeni}, \citenamefont {L{\"u}tkewitte}, \citenamefont {Pandey},
  \citenamefont {Schlue}, \citenamefont {Lammers}, \citenamefont {Roiz},
  \citenamefont {Schapeler}, \citenamefont {Ares}, \citenamefont {Yahyapour},
  \citenamefont {Kastner}, \citenamefont {Martinek}, \citenamefont
  {Mittermair}, \citenamefont {Sevilla-Guti{\'e}rrez}, \citenamefont
  {Leyendecker}, \citenamefont {Kohout}, \citenamefont {Mitin}, \citenamefont
  {Holzwarth}, \citenamefont {Sperling}, \citenamefont {Bartley}, \citenamefont
  {Steinlechner}, \citenamefont {Brecht},\ and\ \citenamefont
  {Silberhorn}}]{stefszky2025benchmarking}%
  \BibitemOpen
  \bibfield  {author} {\bibinfo {author} {\bibfnamefont {M.}~\bibnamefont
  {Stefszky}}, \bibinfo {author} {\bibfnamefont {K.-H.}\ \bibnamefont {Luo}},
  \bibinfo {author} {\bibfnamefont {J.-L.}\ \bibnamefont {Eickmann}}, \bibinfo
  {author} {\bibfnamefont {S.}~\bibnamefont {Atzeni}}, \bibinfo {author}
  {\bibfnamefont {F.}~\bibnamefont {L{\"u}tkewitte}}, \bibinfo {author}
  {\bibfnamefont {C.}~\bibnamefont {Pandey}}, \bibinfo {author} {\bibfnamefont
  {F.}~\bibnamefont {Schlue}}, \bibinfo {author} {\bibfnamefont
  {J.}~\bibnamefont {Lammers}}, \bibinfo {author} {\bibfnamefont
  {M.}~\bibnamefont {Roiz}}, \bibinfo {author} {\bibfnamefont {T.}~\bibnamefont
  {Schapeler}}, \bibinfo {author} {\bibfnamefont {L.}~\bibnamefont {Ares}},
  \bibinfo {author} {\bibfnamefont {M.}~\bibnamefont {Yahyapour}}, \bibinfo
  {author} {\bibfnamefont {A.}~\bibnamefont {Kastner}}, \bibinfo {author}
  {\bibfnamefont {J.}~\bibnamefont {Martinek}}, \bibinfo {author}
  {\bibfnamefont {M.}~\bibnamefont {Mittermair}}, \bibinfo {author}
  {\bibfnamefont {C.}~\bibnamefont {Sevilla-Guti{\'e}rrez}}, \bibinfo {author}
  {\bibfnamefont {M.}~\bibnamefont {Leyendecker}}, \bibinfo {author}
  {\bibfnamefont {O.}~\bibnamefont {Kohout}}, \bibinfo {author} {\bibfnamefont
  {D.}~\bibnamefont {Mitin}}, \bibinfo {author} {\bibfnamefont
  {R.}~\bibnamefont {Holzwarth}}, \bibinfo {author} {\bibfnamefont
  {J.}~\bibnamefont {Sperling}}, \bibinfo {author} {\bibfnamefont
  {T.}~\bibnamefont {Bartley}}, \bibinfo {author} {\bibfnamefont
  {F.}~\bibnamefont {Steinlechner}}, \bibinfo {author} {\bibfnamefont
  {B.}~\bibnamefont {Brecht}},\ and\ \bibinfo {author} {\bibfnamefont
  {C.}~\bibnamefont {Silberhorn}},\ }\href {https://doi.org/10.1103/kqyv-lvvy}
  {\bibfield  {journal} {\bibinfo  {journal} {PRX Quantum}\ }\textbf {\bibinfo
  {volume} {7}},\ \bibinfo {pages} {043033} (\bibinfo {year}
  {2026})}\BibitemShut {NoStop}%
\bibitem [{\citenamefont {Drummond}\ \emph {et~al.}(2022)\citenamefont
  {Drummond}, \citenamefont {Opanchuk}, \citenamefont {Dellios},\ and\
  \citenamefont {Reid}}]{drummondSimulatingComplexNetworks2022}%
  \BibitemOpen
  \bibfield  {author} {\bibinfo {author} {\bibfnamefont {P.~D.}\ \bibnamefont
  {Drummond}}, \bibinfo {author} {\bibfnamefont {B.}~\bibnamefont {Opanchuk}},
  \bibinfo {author} {\bibfnamefont {A.}~\bibnamefont {Dellios}},\ and\ \bibinfo
  {author} {\bibfnamefont {M.~D.}\ \bibnamefont {Reid}},\ }\href
  {https://doi.org/10.1103/PhysRevA.105.012427} {\bibfield  {journal} {\bibinfo
   {journal} {Phys. Rev. A}\ }\textbf {\bibinfo {volume} {105}},\ \bibinfo
  {pages} {012427} (\bibinfo {year} {2022})}\BibitemShut {NoStop}%
\bibitem [{\citenamefont {Mart{\'{i}}nez-Cifuentes}\ \emph
  {et~al.}(2023)\citenamefont {Mart{\'{i}}nez-Cifuentes}, \citenamefont
  {Fonseca-Romero},\ and\ \citenamefont
  {Quesada}}]{MartinezCifuentes2023classicalmodelsmay}%
  \BibitemOpen
  \bibfield  {author} {\bibinfo {author} {\bibfnamefont {J.}~\bibnamefont
  {Mart{\'{i}}nez-Cifuentes}}, \bibinfo {author} {\bibfnamefont {K.~M.}\
  \bibnamefont {Fonseca-Romero}},\ and\ \bibinfo {author} {\bibfnamefont
  {N.}~\bibnamefont {Quesada}},\ }\href
  {https://doi.org/10.22331/q-2023-08-08-1076} {\bibfield  {journal} {\bibinfo
  {journal} {{Quantum}}\ }\textbf {\bibinfo {volume} {7}},\ \bibinfo {pages}
  {1076} (\bibinfo {year} {2023})}\BibitemShut {NoStop}%
\bibitem [{\citenamefont {Dellios}\ \emph
  {et~al.}(2025{\natexlab{a}})\citenamefont {Dellios}, \citenamefont
  {Opanchuk}, \citenamefont {Goodman}, \citenamefont {Reid},\ and\
  \citenamefont {Drummond}}]{dellios2023validationtestsgbsquantum}%
  \BibitemOpen
  \bibfield  {author} {\bibinfo {author} {\bibfnamefont {A.~S.}\ \bibnamefont
  {Dellios}}, \bibinfo {author} {\bibfnamefont {B.}~\bibnamefont {Opanchuk}},
  \bibinfo {author} {\bibfnamefont {N.}~\bibnamefont {Goodman}}, \bibinfo
  {author} {\bibfnamefont {M.~D.}\ \bibnamefont {Reid}},\ and\ \bibinfo
  {author} {\bibfnamefont {P.~D.}\ \bibnamefont {Drummond}},\ }\href
  {https://doi.org/10.1016/j.physleta.2025.130529} {\bibfield  {journal}
  {\bibinfo  {journal} {Phys. Lett. A}\ }\textbf {\bibinfo {volume} {549}},\
  \bibinfo {pages} {130529} (\bibinfo {year} {2025}{\natexlab{a}})}\BibitemShut
  {NoStop}%
\bibitem [{\citenamefont {Dellios}\ \emph
  {et~al.}(2025{\natexlab{b}})\citenamefont {Dellios}, \citenamefont {Reid},\
  and\ \citenamefont {Drummond}}]{dellios2024validation}%
  \BibitemOpen
  \bibfield  {author} {\bibinfo {author} {\bibfnamefont {A.~S.}\ \bibnamefont
  {Dellios}}, \bibinfo {author} {\bibfnamefont {M.~D.}\ \bibnamefont {Reid}},\
  and\ \bibinfo {author} {\bibfnamefont {P.~D.}\ \bibnamefont {Drummond}},\
  }\href {https://doi.org/10.1088/2058-9565/adfe16} {\bibfield  {journal}
  {\bibinfo  {journal} {Quantum Sci. Technol.}\ }\textbf {\bibinfo {volume}
  {10}},\ \bibinfo {pages} {045030} (\bibinfo {year}
  {2025}{\natexlab{b}})}\BibitemShut {NoStop}%
\bibitem [{Sup()}]{SupplementalMaterial}%
  \BibitemOpen
  \href@noop {} {}\bibinfo {note} {See Supplemental Material appended to this paper for full results for every dataset and sampler,
  additional tests and derivations, which includes
  Refs.~\cite{vonzurgathen2013modern,walker2011anderson,liu2026efficient,aaronson2020classical,aharonov2023polynomial,ehrenberg2025transition,gao2024limitations,martinezcifuentes2024linear,ohSpoofingCrossEntropy2022a,martinezcifuentes2022torontonianjulia}}\BibitemShut
  {NoStop}%
\bibitem [{\citenamefont {Bulmer}\ \emph {et~al.}(2022)\citenamefont {Bulmer},
  \citenamefont {Bell}, \citenamefont {Chadwick}, \citenamefont {Jones},
  \citenamefont {Moise}, \citenamefont {Rigazzi}, \citenamefont {Thorbecke},
  \citenamefont {Haus}, \citenamefont {Van~Vaerenbergh}, \citenamefont {Patel},
  \citenamefont {Walmsley},\ and\ \citenamefont
  {Laing}}]{bulmerBoundaryQuantumAdvantage2022a}%
  \BibitemOpen
  \bibfield  {author} {\bibinfo {author} {\bibfnamefont {J.~F.~F.}\
  \bibnamefont {Bulmer}}, \bibinfo {author} {\bibfnamefont {B.~A.}\
  \bibnamefont {Bell}}, \bibinfo {author} {\bibfnamefont {R.~S.}\ \bibnamefont
  {Chadwick}}, \bibinfo {author} {\bibfnamefont {A.~E.}\ \bibnamefont {Jones}},
  \bibinfo {author} {\bibfnamefont {D.}~\bibnamefont {Moise}}, \bibinfo
  {author} {\bibfnamefont {A.}~\bibnamefont {Rigazzi}}, \bibinfo {author}
  {\bibfnamefont {J.}~\bibnamefont {Thorbecke}}, \bibinfo {author}
  {\bibfnamefont {U.-U.}\ \bibnamefont {Haus}}, \bibinfo {author}
  {\bibfnamefont {T.}~\bibnamefont {Van~Vaerenbergh}}, \bibinfo {author}
  {\bibfnamefont {R.~B.}\ \bibnamefont {Patel}}, \bibinfo {author}
  {\bibfnamefont {I.~A.}\ \bibnamefont {Walmsley}},\ and\ \bibinfo {author}
  {\bibfnamefont {A.}~\bibnamefont {Laing}},\ }\href
  {https://doi.org/10.1126/sciadv.abl9236} {\bibfield  {journal} {\bibinfo
  {journal} {Sci. Adv.}\ }\textbf {\bibinfo {volume} {8}},\ \bibinfo {pages}
  {eabl9236} (\bibinfo {year} {2022})}\BibitemShut {NoStop}%
\bibitem [{\citenamefont {Villalonga}\ \emph {et~al.}(2021)\citenamefont
  {Villalonga}, \citenamefont {Niu}, \citenamefont {Li}, \citenamefont {Neven},
  \citenamefont {Platt}, \citenamefont {Smelyanskiy},\ and\ \citenamefont
  {Boixo}}]{villalonga2021efficient}%
  \BibitemOpen
  \bibfield  {author} {\bibinfo {author} {\bibfnamefont {B.}~\bibnamefont
  {Villalonga}}, \bibinfo {author} {\bibfnamefont {M.~Y.}\ \bibnamefont {Niu}},
  \bibinfo {author} {\bibfnamefont {L.}~\bibnamefont {Li}}, \bibinfo {author}
  {\bibfnamefont {H.}~\bibnamefont {Neven}}, \bibinfo {author} {\bibfnamefont
  {J.~C.}\ \bibnamefont {Platt}}, \bibinfo {author} {\bibfnamefont {V.~N.}\
  \bibnamefont {Smelyanskiy}},\ and\ \bibinfo {author} {\bibfnamefont
  {S.}~\bibnamefont {Boixo}},\ }\href@noop {} {\bibfield  {journal} {\bibinfo
  {journal} {arXiv:2109.11525}\ } (\bibinfo {year} {2021})}\BibitemShut
  {NoStop}%
\bibitem [{\citenamefont {Oh}\ \emph {et~al.}(2024)\citenamefont {Oh},
  \citenamefont {Liu}, \citenamefont {Alexeev}, \citenamefont {Fefferman},\
  and\ \citenamefont {Jiang}}]{oh2024classical}%
  \BibitemOpen
  \bibfield  {author} {\bibinfo {author} {\bibfnamefont {C.}~\bibnamefont
  {Oh}}, \bibinfo {author} {\bibfnamefont {M.}~\bibnamefont {Liu}}, \bibinfo
  {author} {\bibfnamefont {Y.}~\bibnamefont {Alexeev}}, \bibinfo {author}
  {\bibfnamefont {B.}~\bibnamefont {Fefferman}},\ and\ \bibinfo {author}
  {\bibfnamefont {L.}~\bibnamefont {Jiang}},\ }\href
  {https://doi.org/10.1038/s41567-024-02535-8} {\bibfield  {journal} {\bibinfo
  {journal} {Nat. Phys.}\ }\textbf {\bibinfo {volume} {20}},\ \bibinfo {pages}
  {1461} (\bibinfo {year} {2024})}\BibitemShut {NoStop}%
\bibitem [{\citenamefont {Dodd}\ \emph {et~al.}(2026)\citenamefont {Dodd},
  \citenamefont {Mart{\'\i}nez-Cifuentes}, \citenamefont {Brown}, \citenamefont
  {Quesada},\ and\ \citenamefont {Garc{\'\i}a-Patr{\'o}n}}]{dodd2025fast}%
  \BibitemOpen
  \bibfield  {author} {\bibinfo {author} {\bibfnamefont {T.}~\bibnamefont
  {Dodd}}, \bibinfo {author} {\bibfnamefont {J.}~\bibnamefont
  {Mart{\'\i}nez-Cifuentes}}, \bibinfo {author} {\bibfnamefont {O.~T.}\
  \bibnamefont {Brown}}, \bibinfo {author} {\bibfnamefont {N.}~\bibnamefont
  {Quesada}},\ and\ \bibinfo {author} {\bibfnamefont {R.}~\bibnamefont
  {Garc{\'\i}a-Patr{\'o}n}},\ }\href {https://doi.org/10.1126/sciadv.aef8431}
  {\bibfield  {journal} {\bibinfo  {journal} {Sci. Adv.}\ }\textbf {\bibinfo
  {volume} {12}},\ \bibinfo {pages} {eaef8431} (\bibinfo {year} {2026})},\
  \bibinfo {note} {arXiv:2511.14923}\BibitemShut {NoStop}%
\bibitem [{\citenamefont {Glauber}(1963)}]{Glauber_1963_P-Rep}%
  \BibitemOpen
  \bibfield  {author} {\bibinfo {author} {\bibfnamefont {R.~J.}\ \bibnamefont
  {Glauber}},\ }\href {https://doi.org/10.1103/PhysRev.131.2766} {\bibfield
  {journal} {\bibinfo  {journal} {Phys. Rev.}\ }\textbf {\bibinfo {volume}
  {131}},\ \bibinfo {pages} {2766} (\bibinfo {year} {1963})}\BibitemShut
  {NoStop}%
\bibitem [{\citenamefont {Sudarshan}(1963)}]{Sudarshan_1963_P-Rep}%
  \BibitemOpen
  \bibfield  {author} {\bibinfo {author} {\bibfnamefont {E.~C.~G.}\
  \bibnamefont {Sudarshan}},\ }\href
  {https://doi.org/10.1103/PhysRevLett.10.277} {\bibfield  {journal} {\bibinfo
  {journal} {Phys. Rev. Lett.}\ }\textbf {\bibinfo {volume} {10}},\ \bibinfo
  {pages} {277} (\bibinfo {year} {1963})}\BibitemShut {NoStop}%
\bibitem [{\citenamefont {Walls}\ and\ \citenamefont
  {Milburn}(2008)}]{walls2008quantum}%
  \BibitemOpen
  \bibfield  {author} {\bibinfo {author} {\bibfnamefont {D.}~\bibnamefont
  {Walls}}\ and\ \bibinfo {author} {\bibfnamefont {G.}~\bibnamefont
  {Milburn}},\ }\href@noop {} {\emph {\bibinfo {title} {Quantum Optics}}}\
  (\bibinfo  {publisher} {Springer},\ \bibinfo {year} {2008})\BibitemShut
  {NoStop}%
\bibitem [{\citenamefont {Titulaer}\ and\ \citenamefont
  {Glauber}(1965)}]{Titulaer1965Correlation}%
  \BibitemOpen
  \bibfield  {author} {\bibinfo {author} {\bibfnamefont {U.~M.}\ \bibnamefont
  {Titulaer}}\ and\ \bibinfo {author} {\bibfnamefont {R.~J.}\ \bibnamefont
  {Glauber}},\ }\href {https://doi.org/10.1103/PhysRev.140.B676} {\bibfield
  {journal} {\bibinfo  {journal} {Phys. Rev.}\ }\textbf {\bibinfo {volume}
  {140}},\ \bibinfo {pages} {B676} (\bibinfo {year} {1965})}\BibitemShut
  {NoStop}%
\bibitem [{\citenamefont {Reid}\ and\ \citenamefont {Walls}(1986)}]{Reid1986}%
  \BibitemOpen
  \bibfield  {author} {\bibinfo {author} {\bibfnamefont {M.~D.}\ \bibnamefont
  {Reid}}\ and\ \bibinfo {author} {\bibfnamefont {D.~F.}\ \bibnamefont
  {Walls}},\ }\href {https://doi.org/10.1103/PhysRevA.34.1260} {\bibfield
  {journal} {\bibinfo  {journal} {Phys. Rev. A}\ }\textbf {\bibinfo {volume}
  {34}},\ \bibinfo {pages} {1260} (\bibinfo {year} {1986})}\BibitemShut
  {NoStop}%
\bibitem [{\citenamefont {Drummond}\ and\ \citenamefont
  {Gardiner}(1980)}]{drummond1980generalised}%
  \BibitemOpen
  \bibfield  {author} {\bibinfo {author} {\bibfnamefont {P.~D.}\ \bibnamefont
  {Drummond}}\ and\ \bibinfo {author} {\bibfnamefont {C.~W.}\ \bibnamefont
  {Gardiner}},\ }\href {https://doi.org/10.1088/0305-4470/13/7/018} {\bibfield
  {journal} {\bibinfo  {journal} {J. Phys. A: Math. Gen.}\ }\textbf {\bibinfo
  {volume} {13}},\ \bibinfo {pages} {2353} (\bibinfo {year}
  {1980})}\BibitemShut {NoStop}%
\bibitem [{\citenamefont {Rahimi-Keshari}\ \emph {et~al.}(2016)\citenamefont
  {Rahimi-Keshari}, \citenamefont {Ralph},\ and\ \citenamefont
  {Caves}}]{RahimiKeshari2016}%
  \BibitemOpen
  \bibfield  {author} {\bibinfo {author} {\bibfnamefont {S.}~\bibnamefont
  {Rahimi-Keshari}}, \bibinfo {author} {\bibfnamefont {T.~C.}\ \bibnamefont
  {Ralph}},\ and\ \bibinfo {author} {\bibfnamefont {C.~M.}\ \bibnamefont
  {Caves}},\ }\href {https://doi.org/10.1103/PhysRevX.6.021039} {\bibfield
  {journal} {\bibinfo  {journal} {Phys. Rev. X}\ }\textbf {\bibinfo {volume}
  {6}},\ \bibinfo {pages} {021039} (\bibinfo {year} {2016})}\BibitemShut
  {NoStop}%
\bibitem [{\citenamefont {Kessy}\ \emph {et~al.}(2018)\citenamefont {Kessy},
  \citenamefont {Lewin},\ and\ \citenamefont {Strimmer}}]{kessy2018optimal}%
  \BibitemOpen
  \bibfield  {author} {\bibinfo {author} {\bibfnamefont {A.}~\bibnamefont
  {Kessy}}, \bibinfo {author} {\bibfnamefont {A.}~\bibnamefont {Lewin}},\ and\
  \bibinfo {author} {\bibfnamefont {K.}~\bibnamefont {Strimmer}},\ }\href
  {https://doi.org/10.1080/00031305.2016.1277159} {\bibfield  {journal}
  {\bibinfo  {journal} {Am. Stat.}\ }\textbf {\bibinfo {volume} {72}},\
  \bibinfo {pages} {309} (\bibinfo {year} {2018})}\BibitemShut {NoStop}%
\bibitem [{\citenamefont {Serafini}(2017)}]{serafini_quantum_2017}%
  \BibitemOpen
  \bibfield  {author} {\bibinfo {author} {\bibfnamefont {A.}~\bibnamefont
  {Serafini}},\ }\href
  {https://www.taylorfrancis.com/books/mono/10.1201/9781315118727/quantum-continuous-variables-alessio-serafini}
  {\emph {\bibinfo {title} {Quantum continuous variables: a primer of
  theoretical methods}}}\ (\bibinfo  {publisher} {CRC Press},\ \bibinfo
  {address} {Boca Raton},\ \bibinfo {year} {2017})\BibitemShut {NoStop}%
\bibitem [{\citenamefont {Hangleiter}\ \emph {et~al.}(2019)\citenamefont
  {Hangleiter}, \citenamefont {Kliesch}, \citenamefont {Eisert},\ and\
  \citenamefont {Gogolin}}]{hangleiter2019sample}%
  \BibitemOpen
  \bibfield  {author} {\bibinfo {author} {\bibfnamefont {D.}~\bibnamefont
  {Hangleiter}}, \bibinfo {author} {\bibfnamefont {M.}~\bibnamefont {Kliesch}},
  \bibinfo {author} {\bibfnamefont {J.}~\bibnamefont {Eisert}},\ and\ \bibinfo
  {author} {\bibfnamefont {C.}~\bibnamefont {Gogolin}},\ }\href
  {https://doi.org/10.1103/PhysRevLett.122.210502} {\bibfield  {journal}
  {\bibinfo  {journal} {Phys. Rev. Lett.}\ }\textbf {\bibinfo {volume} {122}},\
  \bibinfo {pages} {210502} (\bibinfo {year} {2019})}\BibitemShut {NoStop}%
\bibitem [{\citenamefont {Hangleiter}\ and\ \citenamefont
  {Eisert}(2023)}]{hangleiter2023computational}%
  \BibitemOpen
  \bibfield  {author} {\bibinfo {author} {\bibfnamefont {D.}~\bibnamefont
  {Hangleiter}}\ and\ \bibinfo {author} {\bibfnamefont {J.}~\bibnamefont
  {Eisert}},\ }\href {https://doi.org/10.1103/RevModPhys.95.035001} {\bibfield
  {journal} {\bibinfo  {journal} {Rev. Mod. Phys.}\ }\textbf {\bibinfo {volume}
  {95}},\ \bibinfo {pages} {035001} (\bibinfo {year} {2023})}\BibitemShut
  {NoStop}%
\bibitem [{\citenamefont {Gupt}\ \emph {et~al.}(2019)\citenamefont {Gupt},
  \citenamefont {Izaac},\ and\ \citenamefont {Quesada}}]{gupt2019walrus}%
  \BibitemOpen
  \bibfield  {author} {\bibinfo {author} {\bibfnamefont {B.}~\bibnamefont
  {Gupt}}, \bibinfo {author} {\bibfnamefont {J.}~\bibnamefont {Izaac}},\ and\
  \bibinfo {author} {\bibfnamefont {N.}~\bibnamefont {Quesada}},\ }\href
  {https://doi.org/10.21105/joss.01705} {\bibfield  {journal} {\bibinfo
  {journal} {J. Open Source Softw.}\ }\textbf {\bibinfo {volume} {4}},\
  \bibinfo {pages} {1705} (\bibinfo {year} {2019})}\BibitemShut {NoStop}%
\bibitem [{\citenamefont {Drummond}\ \emph {et~al.}(2025)\citenamefont
  {Drummond}, \citenamefont {Dellios},\ and\ \citenamefont {Goodman}}]{xqsim3}%
  \BibitemOpen
  \bibfield  {author} {\bibinfo {author} {\bibfnamefont {P.~D.}\ \bibnamefont
  {Drummond}}, \bibinfo {author} {\bibfnamefont {A.~S.}\ \bibnamefont
  {Dellios}},\ and\ \bibinfo {author} {\bibfnamefont {N.}~\bibnamefont
  {Goodman}},\ }\href {https://github.com/peterddrummond/xqsim} {\bibinfo
  {title} {{xQSIM3}: phase-space simulation software, written in {MATLAB}}}
  (\bibinfo {year} {2025}),\ \bibinfo {note}
  {\url{https://github.com/peterddrummond/xqsim}}\BibitemShut {NoStop}%
\bibitem [{\citenamefont {Drummond}\ \emph {et~al.}(2026)\citenamefont
  {Drummond}, \citenamefont {Dellios},\ and\ \citenamefont
  {Reid}}]{drummond2026matrix}%
  \BibitemOpen
  \bibfield  {author} {\bibinfo {author} {\bibfnamefont {P.~D.}\ \bibnamefont
  {Drummond}}, \bibinfo {author} {\bibfnamefont {A.~S.}\ \bibnamefont
  {Dellios}},\ and\ \bibinfo {author} {\bibfnamefont {M.~D.}\ \bibnamefont
  {Reid}},\ }\href {https://doi.org/10.1103/htk1-rn8v} {\bibfield  {journal}
  {\bibinfo  {journal} {Phys. Rev. Lett.}\ }\textbf {\bibinfo {volume} {136}},\
  \bibinfo {pages} {013601} (\bibinfo {year} {2026})}\BibitemShut {NoStop}%
\bibitem [{\citenamefont {Barnett}\ \emph {et~al.}(1998)\citenamefont
  {Barnett}, \citenamefont {Jeffers}, \citenamefont {Gatti},\ and\
  \citenamefont {Loudon}}]{barnett1998quantum}%
  \BibitemOpen
  \bibfield  {author} {\bibinfo {author} {\bibfnamefont {S.~M.}\ \bibnamefont
  {Barnett}}, \bibinfo {author} {\bibfnamefont {J.}~\bibnamefont {Jeffers}},
  \bibinfo {author} {\bibfnamefont {A.}~\bibnamefont {Gatti}},\ and\ \bibinfo
  {author} {\bibfnamefont {R.}~\bibnamefont {Loudon}},\ }\href
  {https://doi.org/10.1103/PhysRevA.57.2134} {\bibfield  {journal} {\bibinfo
  {journal} {Phys. Rev. A}\ }\textbf {\bibinfo {volume} {57}},\ \bibinfo
  {pages} {2134} (\bibinfo {year} {1998})}\BibitemShut {NoStop}%
\bibitem [{\citenamefont {Kueng}\ \emph {et~al.}(2016)\citenamefont {Kueng},
  \citenamefont {Long}, \citenamefont {Doherty},\ and\ \citenamefont
  {Flammia}}]{kueng2016comparing}%
  \BibitemOpen
  \bibfield  {author} {\bibinfo {author} {\bibfnamefont {R.}~\bibnamefont
  {Kueng}}, \bibinfo {author} {\bibfnamefont {D.~M.}\ \bibnamefont {Long}},
  \bibinfo {author} {\bibfnamefont {A.~C.}\ \bibnamefont {Doherty}},\ and\
  \bibinfo {author} {\bibfnamefont {S.~T.}\ \bibnamefont {Flammia}},\ }\href
  {https://doi.org/10.1103/PhysRevLett.117.170502} {\bibfield  {journal}
  {\bibinfo  {journal} {Phys. Rev. Lett.}\ }\textbf {\bibinfo {volume} {117}},\
  \bibinfo {pages} {170502} (\bibinfo {year} {2016})}\BibitemShut {NoStop}%
\bibitem [{\citenamefont {von~zur Gathen}\ and\ \citenamefont
  {Gerhard}(2013)}]{vonzurgathen2013modern}%
  \BibitemOpen
  \bibfield  {author} {\bibinfo {author} {\bibfnamefont {J.}~\bibnamefont
  {von~zur Gathen}}\ and\ \bibinfo {author} {\bibfnamefont {J.}~\bibnamefont
  {Gerhard}},\ }\href@noop {} {\emph {\bibinfo {title} {Modern Computer
  Algebra}}},\ \bibinfo {edition} {3rd}\ ed.\ (\bibinfo  {publisher} {Cambridge
  University Press},\ \bibinfo {address} {Cambridge},\ \bibinfo {year}
  {2013})\BibitemShut {NoStop}%
\bibitem [{\citenamefont {Walker}\ and\ \citenamefont
  {Ni}(2011)}]{walker2011anderson}%
  \BibitemOpen
  \bibfield  {author} {\bibinfo {author} {\bibfnamefont {H.~F.}\ \bibnamefont
  {Walker}}\ and\ \bibinfo {author} {\bibfnamefont {P.}~\bibnamefont {Ni}},\
  }\href {https://doi.org/10.1137/10078356X} {\bibfield  {journal} {\bibinfo
  {journal} {SIAM J. Numer. Anal.}\ }\textbf {\bibinfo {volume} {49}},\
  \bibinfo {pages} {1715} (\bibinfo {year} {2011})}\BibitemShut {NoStop}%
\bibitem [{\citenamefont {Liu}\ \emph {et~al.}(2026{\natexlab{b}})\citenamefont
  {Liu}, \citenamefont {Jin}, \citenamefont {Xiang},\ and\ \citenamefont
  {Tu}}]{liu2026efficient}%
  \BibitemOpen
  \bibfield  {author} {\bibinfo {author} {\bibfnamefont {T.}~\bibnamefont
  {Liu}}, \bibinfo {author} {\bibfnamefont {H.-K.}\ \bibnamefont {Jin}},
  \bibinfo {author} {\bibfnamefont {T.}~\bibnamefont {Xiang}},\ and\ \bibinfo
  {author} {\bibfnamefont {H.-H.}\ \bibnamefont {Tu}},\ }\href
  {https://doi.org/10.1038/s41534-026-01254-7} {\bibfield  {journal} {\bibinfo
  {journal} {npj Quantum Inf.}\ }\textbf {\bibinfo {volume} {12}},\ \bibinfo
  {pages} {110} (\bibinfo {year} {2026}{\natexlab{b}})},\ \bibinfo {note}
  {arXiv:2512.10643}\BibitemShut {NoStop}%
\bibitem [{\citenamefont {Aaronson}\ and\ \citenamefont
  {Gunn}(2020)}]{aaronson2020classical}%
  \BibitemOpen
  \bibfield  {author} {\bibinfo {author} {\bibfnamefont {S.}~\bibnamefont
  {Aaronson}}\ and\ \bibinfo {author} {\bibfnamefont {S.}~\bibnamefont
  {Gunn}},\ }\href {https://doi.org/10.4086/toc.2020.v016a011} {\bibfield
  {journal} {\bibinfo  {journal} {Theory Comput.}\ }\textbf {\bibinfo {volume}
  {16}},\ \bibinfo {pages} {1} (\bibinfo {year} {2020})}\BibitemShut {NoStop}%
\bibitem [{\citenamefont {Aharonov}\ \emph {et~al.}(2023)\citenamefont
  {Aharonov}, \citenamefont {Gao}, \citenamefont {Landau}, \citenamefont
  {Liu},\ and\ \citenamefont {Vazirani}}]{aharonov2023polynomial}%
  \BibitemOpen
  \bibfield  {author} {\bibinfo {author} {\bibfnamefont {D.}~\bibnamefont
  {Aharonov}}, \bibinfo {author} {\bibfnamefont {X.}~\bibnamefont {Gao}},
  \bibinfo {author} {\bibfnamefont {Z.}~\bibnamefont {Landau}}, \bibinfo
  {author} {\bibfnamefont {Y.}~\bibnamefont {Liu}},\ and\ \bibinfo {author}
  {\bibfnamefont {U.}~\bibnamefont {Vazirani}},\ }in\ \href
  {https://doi.org/10.1145/3564246.3585234} {\emph {\bibinfo {booktitle}
  {Proceedings of the 55th Annual ACM Symposium on Theory of Computing (STOC
  2023)}}}\ (\bibinfo {year} {2023})\ pp.\ \bibinfo {pages}
  {945--957}\BibitemShut {NoStop}%
\bibitem [{\citenamefont {Ehrenberg}\ \emph {et~al.}(2025)\citenamefont
  {Ehrenberg}, \citenamefont {Iosue}, \citenamefont {Deshpande}, \citenamefont
  {Hangleiter},\ and\ \citenamefont {Gorshkov}}]{ehrenberg2025transition}%
  \BibitemOpen
  \bibfield  {author} {\bibinfo {author} {\bibfnamefont {A.}~\bibnamefont
  {Ehrenberg}}, \bibinfo {author} {\bibfnamefont {J.~T.}\ \bibnamefont
  {Iosue}}, \bibinfo {author} {\bibfnamefont {A.}~\bibnamefont {Deshpande}},
  \bibinfo {author} {\bibfnamefont {D.}~\bibnamefont {Hangleiter}},\ and\
  \bibinfo {author} {\bibfnamefont {A.~V.}\ \bibnamefont {Gorshkov}},\ }\href
  {https://doi.org/10.1103/PhysRevLett.134.140601} {\bibfield  {journal}
  {\bibinfo  {journal} {Phys. Rev. Lett.}\ }\textbf {\bibinfo {volume} {134}},\
  \bibinfo {pages} {140601} (\bibinfo {year} {2025})}\BibitemShut {NoStop}%
\bibitem [{\citenamefont {Gao}\ \emph {et~al.}(2024)\citenamefont {Gao},
  \citenamefont {Kalinowski}, \citenamefont {Chou}, \citenamefont {Lukin},
  \citenamefont {Barak},\ and\ \citenamefont {Choi}}]{gao2024limitations}%
  \BibitemOpen
  \bibfield  {author} {\bibinfo {author} {\bibfnamefont {X.}~\bibnamefont
  {Gao}}, \bibinfo {author} {\bibfnamefont {M.}~\bibnamefont {Kalinowski}},
  \bibinfo {author} {\bibfnamefont {C.-N.}\ \bibnamefont {Chou}}, \bibinfo
  {author} {\bibfnamefont {M.~D.}\ \bibnamefont {Lukin}}, \bibinfo {author}
  {\bibfnamefont {B.}~\bibnamefont {Barak}},\ and\ \bibinfo {author}
  {\bibfnamefont {S.}~\bibnamefont {Choi}},\ }\href
  {https://doi.org/10.1103/PRXQuantum.5.010334} {\bibfield  {journal} {\bibinfo
   {journal} {PRX Quantum}\ }\textbf {\bibinfo {volume} {5}},\ \bibinfo {pages}
  {010334} (\bibinfo {year} {2024})}\BibitemShut {NoStop}%
\bibitem [{\citenamefont {Mart{\'\i}nez-Cifuentes}\ \emph
  {et~al.}(2024)\citenamefont {Mart{\'\i}nez-Cifuentes}, \citenamefont
  {de~Guise},\ and\ \citenamefont {Quesada}}]{martinezcifuentes2024linear}%
  \BibitemOpen
  \bibfield  {author} {\bibinfo {author} {\bibfnamefont {J.}~\bibnamefont
  {Mart{\'\i}nez-Cifuentes}}, \bibinfo {author} {\bibfnamefont
  {H.}~\bibnamefont {de~Guise}},\ and\ \bibinfo {author} {\bibfnamefont
  {N.}~\bibnamefont {Quesada}},\ }\href
  {https://doi.org/10.1103/PRXQuantum.5.040312} {\bibfield  {journal} {\bibinfo
   {journal} {PRX Quantum}\ }\textbf {\bibinfo {volume} {5}},\ \bibinfo {pages}
  {040312} (\bibinfo {year} {2024})}\BibitemShut {NoStop}%
\bibitem [{\citenamefont {Oh}\ \emph {et~al.}(2023)\citenamefont {Oh},
  \citenamefont {Jiang},\ and\ \citenamefont
  {Fefferman}}]{ohSpoofingCrossEntropy2022a}%
  \BibitemOpen
  \bibfield  {author} {\bibinfo {author} {\bibfnamefont {C.}~\bibnamefont
  {Oh}}, \bibinfo {author} {\bibfnamefont {L.}~\bibnamefont {Jiang}},\ and\
  \bibinfo {author} {\bibfnamefont {B.}~\bibnamefont {Fefferman}},\ }\href
  {https://doi.org/10.1103/PhysRevLett.131.010401} {\bibfield  {journal}
  {\bibinfo  {journal} {Phys. Rev. Lett.}\ }\textbf {\bibinfo {volume} {131}},\
  \bibinfo {pages} {010401} (\bibinfo {year} {2023})}\BibitemShut {NoStop}%
\bibitem [{\citenamefont {Mart{\'i}nez-Cifuentes}\ and\ \citenamefont
  {Quesada}(2022)}]{martinezcifuentes2022torontonianjulia}%
  \BibitemOpen
  \bibfield  {author} {\bibinfo {author} {\bibfnamefont {J.}~\bibnamefont
  {Mart{\'i}nez-Cifuentes}}\ and\ \bibinfo {author} {\bibfnamefont
  {N.}~\bibnamefont {Quesada}},\ }\href
  {https://github.com/polyquantique/torontonian-julia} {\bibinfo {title}
  {torontonian-julia}} (\bibinfo {year} {2022}),\ \bibinfo {note}
  {\url{https://github.com/polyquantique/torontonian-julia}}\BibitemShut
  {NoStop}%
\bibitem [{\citenamefont {Fearn}\ and\ \citenamefont
  {Collett}(1988)}]{Fearn_JModOpt1988}%
  \BibitemOpen
  \bibfield  {author} {\bibinfo {author} {\bibfnamefont {H.}~\bibnamefont
  {Fearn}}\ and\ \bibinfo {author} {\bibfnamefont {M.}~\bibnamefont
  {Collett}},\ }\href {https://doi.org/10.1080/09500348814550571} {\bibfield
  {journal} {\bibinfo  {journal} {J. Mod. Opt.}\ }\textbf {\bibinfo {volume}
  {35}},\ \bibinfo {pages} {553} (\bibinfo {year} {1988})}\BibitemShut
  {NoStop}%
\bibitem [{\citenamefont {Sylvester}(1852)}]{Sylvester1852Demonstration}%
  \BibitemOpen
  \bibfield  {author} {\bibinfo {author} {\bibfnamefont {J.}~\bibnamefont
  {Sylvester}},\ }\href {https://doi.org/10.1080/14786445208647087} {\bibfield
  {journal} {\bibinfo  {journal} {Philos. Mag.}\ }\textbf {\bibinfo {volume}
  {4}},\ \bibinfo {pages} {138} (\bibinfo {year} {1852})}\BibitemShut {NoStop}%
\bibitem [{\citenamefont {Duchi}\ \emph {et~al.}(2008)\citenamefont {Duchi},
  \citenamefont {Shalev-Shwartz}, \citenamefont {Singer},\ and\ \citenamefont
  {Chandra}}]{duchi2008efficient}%
  \BibitemOpen
  \bibfield  {author} {\bibinfo {author} {\bibfnamefont {J.}~\bibnamefont
  {Duchi}}, \bibinfo {author} {\bibfnamefont {S.}~\bibnamefont
  {Shalev-Shwartz}}, \bibinfo {author} {\bibfnamefont {Y.}~\bibnamefont
  {Singer}},\ and\ \bibinfo {author} {\bibfnamefont {T.}~\bibnamefont
  {Chandra}},\ }in\ \href {https://doi.org/10.1145/1390156.1390191} {\emph
  {\bibinfo {booktitle} {Proceedings of the 25th International Conference on
  Machine Learning}}}\ (\bibinfo {year} {2008})\ pp.\ \bibinfo {pages}
  {272--279}\BibitemShut {NoStop}%
\bibitem [{\citenamefont {Guthrie}(2020)}]{nist_ehandbook}%
  \BibitemOpen
  \bibfield  {author} {\bibinfo {author} {\bibfnamefont {W.~F.}\ \bibnamefont
  {Guthrie}},\ }\href {https://doi.org/10.18434/M32189} {\bibinfo {title}
  {{NIST/SEMATECH e-Handbook of Statistical Methods (NIST Handbook 151)}}},\
  \bibinfo {howpublished} {National Institute of Standards and Technology}
  (\bibinfo {year} {2020})\BibitemShut {NoStop}%
\bibitem [{\citenamefont {Vallone}\ \emph {et~al.}(2019)\citenamefont
  {Vallone}, \citenamefont {Cariolaro},\ and\ \citenamefont
  {Pierobon}}]{vallone2019means}%
  \BibitemOpen
  \bibfield  {author} {\bibinfo {author} {\bibfnamefont {G.}~\bibnamefont
  {Vallone}}, \bibinfo {author} {\bibfnamefont {G.}~\bibnamefont {Cariolaro}},\
  and\ \bibinfo {author} {\bibfnamefont {G.}~\bibnamefont {Pierobon}},\ }\href
  {https://doi.org/10.1103/PhysRevA.99.023817} {\bibfield  {journal} {\bibinfo
  {journal} {Phys. Rev. A}\ }\textbf {\bibinfo {volume} {99}},\ \bibinfo
  {pages} {023817} (\bibinfo {year} {2019})}\BibitemShut {NoStop}%
\bibitem [{\citenamefont {Olver}\ \emph {et~al.}(2010)\citenamefont {Olver},
  \citenamefont {Lozier}, \citenamefont {Boisvert},\ and\ \citenamefont
  {Clark}}]{nist_handbook}%
  \BibitemOpen
  \bibinfo {editor} {\bibfnamefont {F.~W.~J.}\ \bibnamefont {Olver}}, \bibinfo
  {editor} {\bibfnamefont {D.~W.}\ \bibnamefont {Lozier}}, \bibinfo {editor}
  {\bibfnamefont {R.~F.}\ \bibnamefont {Boisvert}},\ and\ \bibinfo {editor}
  {\bibfnamefont {C.~W.}\ \bibnamefont {Clark}},\ eds.,\ \href@noop {} {\emph
  {\bibinfo {title} {{NIST} Handbook of Mathematical Functions}}}\ (\bibinfo
  {publisher} {Cambridge University Press},\ \bibinfo {address} {Cambridge},\
  \bibinfo {year} {2010})\BibitemShut {NoStop}%
\bibitem [{\citenamefont {Lugannani}\ and\ \citenamefont
  {Rice}(1980)}]{lugannani1980saddle}%
  \BibitemOpen
  \bibfield  {author} {\bibinfo {author} {\bibfnamefont {R.}~\bibnamefont
  {Lugannani}}\ and\ \bibinfo {author} {\bibfnamefont {S.}~\bibnamefont
  {Rice}},\ }\href {https://doi.org/10.2307/1426607} {\bibfield  {journal}
  {\bibinfo  {journal} {Adv. Appl. Probab.}\ }\textbf {\bibinfo {volume}
  {12}},\ \bibinfo {pages} {475} (\bibinfo {year} {1980})}\BibitemShut
  {NoStop}%
\bibitem [{\citenamefont
  {Barndorff-Nielsen}(1991)}]{barndorffnielsen1991modified}%
  \BibitemOpen
  \bibfield  {author} {\bibinfo {author} {\bibfnamefont {O.~E.}\ \bibnamefont
  {Barndorff-Nielsen}},\ }\href {https://doi.org/10.1093/biomet/78.3.557}
  {\bibfield  {journal} {\bibinfo  {journal} {Biometrika}\ }\textbf {\bibinfo
  {volume} {78}},\ \bibinfo {pages} {557} (\bibinfo {year} {1991})}\BibitemShut
  {NoStop}%
\end{thebibliography}
\end{document}


\title{Supplemental Material for ``Gaussian boson sampling: Benchmarking quantum advantage''}
\author{Ned Goodman}
\author{Alexander S. Dellios}
\author{Margaret D. Reid}
\author{Peter D. Drummond}
\email[Contact author: ]{peterddrummond@protonmail.com}
\affiliation{Centre for Quantum Science and Technology Theory, Swinburne University of Technology, Melbourne 3122, Australia}
\maketitle
\renewcommand{\thesection}{S\arabic{section}}
\renewcommand{\theequation}{S\arabic{equation}}
\renewcommand{\thefigure}{S\arabic{figure}}
\renewcommand{\thetable}{S\arabic{table}}
\renewcommand{\thepage}{S\arabic{page}}
\setcounter{page}{1}
This Supplemental Material gives the complete results behind the Letter. The signed $Z$-score of every observable, for every dataset and sampler, is in Sec.~\ref{sec:full} and Figs.~\ref{fig:obs_J2_1e7_ID}--\ref{fig:obs_J4}, with each sampler's worst observable in Tables~\ref{tab:worst_J2_1e7}--\ref{tab:worst_J4}. The \textit{Jiuzhang~4} results are in Fig.~\ref{fig:obs_J4} and Table~\ref{tab:worst_J4}, and without loss in Fig.~\ref{fig:obs_LL_J4}. Whether the sampler's accuracy can be improved is examined in Sec.~\ref{sec:improve}, and its accuracy without loss in Sec.~\ref{sec:lossless}.

\section{Datasets and simulation parameters}
\label{sec:datasets_params}
Table~\ref{tab:datasets} lists every dataset analyzed and Table~\ref{tab:simparams} the parameters of the simulations.
{
\begin{table*}[t]
\centering
\caption{The datasets analyzed. ``Modes'' is the number of output modes $M$ that are resolved, time bins included; it is not the number of physical detectors. In the 144-mode version of \textit{Jiuzhang~3}, each mode merges eight of the 1152 modes and clicks if any of them clicks. ``Click'' marks threshold detectors and ``PNR'' photon-number-resolving detectors. The mean output photon number, from the ideal ground truth, is the horizontal axis of Fig.~1(a)--(e) and of Fig.~\ref{fig:photons_modes}(a); where a row covers several datasets, the values are listed in the same order, or as a range. $N$ is the number of samples per dataset. \textit{Jiuzhang~2} also has $10^{6}$ for the comparison that includes the MPS sampler. The high-power \textit{Jiuzhang~3} dataset and its 144-mode version also have $10^{7}$ (Tables~\ref{tab:worst_J3} and~\ref{tab:worst_J3_144}).\label{tab:datasets}}
\renewcommand{\arraystretch}{1.2}
\begin{tabular}{llcccc}
\hline
\textbf{Experiment} & \textbf{Datasets} & \textbf{Modes} & \textbf{Detector} & \textbf{Mean output photons} & $\boldsymbol{N}$ \\
\hline
\textit{Jiuzhang~2} & 150, 300, 600, 1000, 1650~mW & 144 & click & 6.39, 13.5, 31.3, 63.6, 140 & $10^{7}$, $10^{6}$ \\
\textit{Jiuzhang~3} & low, mid, high & 1152 & click & 47.9, 77.8, 116 & $10^{6}$ (high: also $10^{7}$) \\
\textit{Jiuzhang~3} & high, 144-mode version & 144 (merged) & click & 116 & $10^{6}$, $10^{7}$ \\
Borealis & B16 & 16 & PNR & 5.98 & $10^{6}$ \\
Borealis & B72 & 72 & PNR & 23.1 & $10^{6}$ \\
Borealis & B216L, B216H & 216 & PNR & 21.6, 126 & $10^{6}$ \\
Borealis & B288 & 288 & PNR & 148 & $10^{6}$ \\
\textit{Jiuzhang~4} & Miniature 6~nJ & 496 & click & 189 & $10^{6}$ \\
\textit{Jiuzhang~4} & A S64 2--6~nJ & 4336 & click & 31.9--207 & $10^{6}$ \\
\textit{Jiuzhang~4} & A M256 2--6~nJ & 5104 & click & 130--836 & $10^{6}$ \\
\textit{Jiuzhang~4} & A L1024 2--6~nJ & 8176 & click & 533--3410 & $10^{6}$ \\
\textit{Jiuzhang~4} & B L1024 3--7~nJ & 8176 & click & 879--4461 & $10^{6}$ \\
\hline
\end{tabular}
\end{table*}
}
{
\begin{table*}[t]
\centering
\caption{Simulation parameters. $B$ is the number of phase-space samples whitened together by the positive-$P$ sampler and $L$ the number of whitening--coloring iterations. Each phase-space ground truth uses as many positive-$P$ trajectories as the samples it is compared with ($N$ in Table~\ref{tab:datasets}), except in the two comparisons with $N=10^{6}$ that include the MPS sampler, whose ground truths use $10^{7}$; and ``Blocks'' is the number of independent blocks whose means give its standard errors. The corrected ground truths use the fitted amplitude correction $t$ and thermal fraction $\epsilon$ listed (Sec.~\ref{sec:corrected_gt}), in the order of the datasets in Table~\ref{tab:datasets}; the corrected \textit{Jiuzhang~3} ground truth also includes partial distinguishability. $^{*}$The optimized sampler uses 30 ($\gamma=0$) and 80 ($\gamma=0.9$) iterations (Sec.~\ref{sec:diagonal}). $^{\dagger}$10 for B16 (Poisson output step); 150, followed by mean-correcting iterations, for the categorical output step on B72, B216L, B216H and B288 (Sec.~\ref{sec:categorical}).\label{tab:simparams}}
\renewcommand{\arraystretch}{1.25}\small
\begin{tabular}{lcccc}
\hline
\textbf{Experiment} & $\boldsymbol{B}$ & $\boldsymbol{L}$ & \textbf{Blocks} & \textbf{Fitted $t$, $\epsilon$} \\
\hline
\textit{Jiuzhang~2} & $10^{6}$ & 10$^{*}$ & 1000 & \begin{tabular}[t]{@{}c@{}}$t=1.0011$, $1.0008$, $1.0015$, $1.0094$, $1.0184$\\ $\epsilon=0.0128$, $0.0173$, $0.026$, $0.032$, $0.0368$\end{tabular} \\
\textit{Jiuzhang~3} & $10^{4}$ & 10 & 1000 (high, $N=10^{7}$: $10^{4}$) & $t=1.0009$, $1.0076$, $1.0120$; $\epsilon=0$ \\
\textit{Jiuzhang~3}, 144-mode & $10^{6}$ & 10 & 1000 & --- \\
Borealis & 2000 & 10, 150$^{\dagger}$ & 500 & \begin{tabular}[t]{@{}c@{}}$t=0.9841$, $0.9861$, $0.9882$, $0.9941$, $0.9848$\\ $\epsilon=0.0648$, $0.0507$, $0.0105$, $0.0510$, $0.0547$\end{tabular} \\
\textit{Jiuzhang~4} & $10^{5}$ & 10 & 1000 & --- (ideal only) \\
\hline
\end{tabular}
\end{table*}
}
\paragraph{Experimental samples---}
The 144-mode \textit{Jiuzhang~3} samples are built from the 1152-mode samples by merging each group of eight detectors into one mode that clicks if any of them clicks.

\paragraph{Random seeds---}
Every random-number stream is seeded from a hash of strings that identify the dataset, ground truth, sampler and stage, so a rerun reproduces the same samples. The mode sets of the marginals and the samples used for the cross-entropy are seeded from the dataset's name, so every sampler of a dataset is tested on the same ones.

\paragraph{Squeezing parameters---}
The ground truths use the squeezing parameters released with each experiment. For \textit{Jiuzhang~4}, USTC released three values for each squeezer: the squeezing parameter $r$ of its ideal single-mode state, and those of a major and a minor spectral mode, $r_{1}=\lambda_{1}r$ and $r_{2}=\lambda_{2}r$. The fractions $\lambda_{1}$ and $\lambda_{2}$ are the same for every squeezer at a given pump energy. For B L1024 at 7~nJ only $r_{1}$ and $r_{2}$ are released. Its ratio $r_{2}/r_{1}=0.210$ equals that at 6~nJ, so we take the 6~nJ fractions, squeezer by squeezer: $r=(r_{1}+r_{2})\,r^{(6)}/(r^{(6)}_{1}+r^{(6)}_{2})$, where the superscript marks the 6~nJ values. The fraction $\lambda_{1}$ rises with pump energy, from $0.9857$ at 3~nJ to $0.9884$, $0.9903$ and $0.9919$ at 4, 5 and 6~nJ. Extrapolating it to 7~nJ with a quadratic or a linear fit in the energy instead lowers $r$ by $0.09\%$ or $0.23\%$. The experiment's worst $Z$-score at 7~nJ, on $\langle c_j\rangle$, then falls from $+446$ to $+387$ or $+364$ (Table~\ref{tab:worst_J4}), so the choice does not change any conclusion.

\section{Statistical tests}
\label{sec:stats}\label{sec:outprob}
\paragraph*{Calculating output probabilities, marginals and cumulants---}
Grouped count distributions are computed as in Refs.~\cite{drummondSimulatingComplexNetworks2022,dellios2024validation}; Sec.~\ref{sec:gcd_review} reviews the method and gives a faster evaluation for 8176 modes. Output probabilities, both full and {mode-marginalized}, are given as functions of the matrix hafnian~\cite{hamiltonGaussian2017} and Torontonian~\cite{quesadaGaussian2018} applied to the covariance matrix of the GBS output state.  Marginals and hafnians are computed with {The Walrus} code package~\cite{gupt2019walrus}; the Torontonians for the cross-entropy use our 256-bit Julia version of the code of Ref.~\cite{martinezcifuentes2022torontonianjulia}, extended to the product Torontonian for the corrected \textit{Jiuzhang~3} ground truth, since their alternating sums cancel heavily in double precision. These were cross-checked against the xQSIM~\cite{xqsim3} positive-$P$ {phase-space simulator}. Exact cumulants are also reported separately: orders one to three are computed from the same covariance matrix with The Walrus: through the one-, two- and three-mode click probabilities for threshold detectors, and from closed-form Gaussian moments, with no photon-number cutoff, for PNR detectors. Following Ref.~\cite{liu2025robust}, we fit the sampled joint cumulants against these exact values over many randomly chosen sets of modes, and report the slope error $\Delta K=|K-1|$, where $K$ is the fitted slope (Sec.~\ref{sec:cumulants}).

\paragraph*{Cross-entropy benchmarking---}
Cross-entropy is a common figure of merit for GBS experiments and samplers. However, calculating it requires the ground-truth probability of each output pattern, which is only tractable at small photon numbers. Hence, we restrict our calculations to a maximum of 15 total photon detections in our outputs. To achieve this, we truncate every sample $\mathbf{s}=(s_1,\ldots,s_M)$ to its first $M_{\max}$ output modes and assign it to the sector of its truncated total count:
\begin{equation}
\tilde{\mathbf{s}}=(s_1,\ldots,s_{M_{\max}}),\qquad n(\tilde{\mathbf{s}})=\sum_{j=1}^{M_{\max}}s_j.\label{eq:trunc}
\end{equation}
The ground-truth probability of a truncated pattern is the marginal over the discarded modes,
\begin{equation}
\Pr(\tilde{\mathbf{s}})=\sum_{s_{M_{\max}+1},\ldots,s_M}\Pr(\mathbf{s}),\label{eq:trunc_marginal}
\end{equation}
which we compute directly, as a Torontonian~\cite{quesadaGaussian2018} for threshold detectors or a loop-free repeated hafnian~\cite{gupt2019walrus} for Borealis, of the covariance matrix of the first $M_{\max}$ modes. The sum in Eq.~\eqref{eq:trunc_marginal} is therefore never carried out explicitly.

$M_{\max}$ is fixed once per dataset from the experimental samples $\mathbf{s}_i$, and every sampler is evaluated on the same modes. For threshold detectors,
\begin{equation}
M_{\max}=\operatorname*{median}_{i}\Bigl[\min\Bigl\{m:\sum_{j=1}^{m}s_{ij}=15\Bigr\}\Bigr],\label{eq:Mmax_click}
\end{equation}
the mode at which the median experimental sample registers its 15th click. The median is taken over the experimental samples that reach 15 clicks, so about half of them keep at least 15 clicks after truncation; if fewer than 1000 samples reach 15 clicks, we use all $M$ modes. For Borealis,
\begin{equation}
M_{\max}=\operatorname*{arg\,max}_{m}\;\#\Bigl\{i:\sum_{j=1}^{m}s_{ij}=15\Bigr\},\label{eq:Mmax_pnr}
\end{equation}
the number of modes that gives the most experimental samples with exactly 15 photons. This differs from the threshold rule because a photon count can jump past 15 when a single mode is added.

In each sector $n$ with at least 1000 truncated samples, we draw $N_{s}=1000$ of them at random, with a fixed seed, and compute
\begin{equation}
\mathrm{XE}_{n}=-\dfrac{1}{N_{s}}\sum_{i=1}^{N_{s}}\ln\Pr(\tilde{\mathbf{s}}_{i}),\qquad n(\tilde{\mathbf{s}}_{i})=n,\label{eq:XEB}
\end{equation}
for $n\leq15$ with threshold detectors and $n=10$--$15$ for Borealis. Its standard error is the sample standard deviation of $\ln\Pr(\tilde{\mathbf{s}}_{i})$ divided by $\sqrt{N_{s}}$. A sampler is compared with the experiment sector by sector through
\begin{equation}
\Delta\mathrm{XE}_n=\mathrm{XE}_n^{\rm sampler}-\mathrm{XE}_n^{\rm exp},\label{eq:dXE}
\end{equation}
whose standard error is the quadrature sum of the two standard errors. A positive $\Delta\mathrm{XE}_n$ means the experimental samples have a higher mean log-probability under the ground truth. For the corrected \textit{Jiuzhang~3} ground truth, the probabilities are product Torontonians (Sec.~\ref{sec:corrected_gt}). The Torontonian sums are computed in 256-bit floating-point arithmetic, with the matrix inverse in double precision (precision checks in Sec.~\ref{sec:xeb}); the Borealis hafnians are computed in double precision. Like other marginalized cross-entropy measures, $\mathrm{XE}_n$ can be spoofed and does not certify closeness to the ground-truth distribution~\cite{gao2024limitations,martinezcifuentes2024linear}.

\section{Corrected ground truths}
\label{sec:corrected_gt}
A comparison using a ground truth closer to the experimental GBS distribution is a more generous test for the experiment. Since we could {overparameterize} the ground-truth model and potentially fit any data, we restricted our investigation to few-parameter physically realistic corrections. As a first step in this direction, we use phase-space simulations to find more experimentally accurate ground-truth models to use as the basis of our comparison. These models are partially decoherent, as {parameterized} by $\epsilon$, and have the transmission efficiency corrected by a scalar factor: {$\boldsymbol{T}\rightarrow t\boldsymbol{T}$}, where $t$ is a field-amplitude correction factor. 
The decoherence and transmission correction factors are found by {minimizing} the Pearson $\chi^{2}_{P}$ of the experimental total count distribution against the corrected ground truth, so the experiment's total-count $Z$-score against a corrected ground truth is not an independent test. 
Fitting the total-count distribution only slightly reduces the discrepancies in the marginals. For \textit{Jiuzhang~3} the corrected ground truth also includes partial distinguishability, as in the experiment's own model~\cite{deng2023gaussian}: each squeezer emits into a major and a minor spectral mode, which pass through the same network but do not interfere, and the detectors cannot tell them apart. A detector therefore registers no click only when neither spectral mode delivers a photon. In phase space this gives $\pi_j=\exp[-\sum_s\alpha_{js}\beta_{js}]$, summed over the two spectral modes $s$. For exact pattern probabilities, each no-click probability inside the Torontonian's inclusion--exclusion sum becomes the product of the two spectral modes' no-click probabilities, which we call the product Torontonian. Only $t$ is corrected, chosen so that the mean number of clicks matches the experiment's. This gives $t=1.0009$, $1.0076$ and $1.0120$ for the low, mid and high datasets, within 0.001 of a fit to the whole total-count distribution. 
Since only its mean is fitted, $\mathcal{G}^{M}_{M}$ remains a partial test of the experiment on \textit{Jiuzhang~3}. We report \textit{Jiuzhang~4} scores against the ideal ground truth only: no correction was fitted, and its partial distinguishability (Sec.~\ref{sec:datasets_params}) is treated as an experimental error. There, the positive-$P$ sampler already agrees with the ideal ground truth up to sampling error on 20 of the 21 datasets, so there is no plausible way for the correction to alter the order of the results. On \textit{Jiuzhang~3}, where the corrected ground truth includes partial distinguishability, the experiment's largest $|Z|$ falls from 36, 129 and 244 to 33, 36 and 47, still far above the positive-$P$ sampler's 1.9 to 2.4 (Table~\ref{tab:worst_J3}).
On the other experiments, the correction lowered the experiment's largest $|Z|$ but never changed its ordering relative to the positive-$P$ sampler (Fig.~1(f)). 

\section{Full results}
\label{sec:fullresults}
Tables~\ref{tab:worst_J2_1e7}--\ref{tab:worst_J4} give each sampler's worst observable on each dataset of Table~\ref{tab:datasets}.

\subsection{$Z$-scores for every observable}
\label{sec:full}
Figures~\ref{fig:obs_J2_1e7_ID}--\ref{fig:obs_J4} give the signed $Z$-score of every observable behind Fig.~1. The worst-observable tables accompany the heat maps for each experiment and sample size. Figure~\ref{fig:obs_J3_144} and Table~\ref{tab:worst_J3_144} report the separate 144-mode \textit{Jiuzhang~3} comparison, and Fig.~\ref{fig:obs_J3_1e7} gives every observable for the high-power \textit{Jiuzhang~3} dataset and its 144-mode version with $N=10^{7}$. 
\paragraph{Sampler weaknesses---}

For the threshold-detector datasets (\textit{Jiuzhang~2--4}), the experiment's worst-performing observable is always the single-mode first moment $\langle c_j\rangle$. For Borealis it is the per-mode photon-number distribution $P_j(n)$, the photon-number-resolving analogue. The squashed-state sampler fails mainly on the total and two-dimensional count distributions. The positive-$P$ sampler's remaining failures are mainly on the total-count distribution $\mathcal{G}^{M}_{M}$: on \textit{Jiuzhang~2} at every power with $N=10^{7}$ (Fig.~\ref{fig:obs_J2_1e7_ID}) and at the three lowest powers with $N=10^{6}$ (Fig.~\ref{fig:obs_J2_1e6_ID}), on B72 and B216L (Fig.~\ref{fig:obs_B_ID}), on \textit{Jiuzhang~4} A S64 at 2~nJ (Fig.~\ref{fig:obs_J4}), and on the 144-mode \textit{Jiuzhang~3} dataset with $N=10^{7}$ (Fig.~\ref{fig:obs_J3_1e7}). With the Poisson output step it also fails on $P_j(n)$ and high-order marginals on Borealis. 
The categorical output step removes the $P_j(n)$ failures: with it, $P_j(n)$ has $|Z|<3$ on all four Borealis datasets where it is used, against both ground truths. On \textit{Jiuzhang~3} and~4 the positive-$P$ sampler is within sampling error at $N=10^6$ on every dataset but one. 
The greedy sampler is underdispersed: its worst observables have negative $Z$-scores at the highest \textit{Jiuzhang~2} power, and at every power with $N=10^{6}$ (Figs.~\ref{fig:obs_J2_1e7_ID} and~\ref{fig:obs_J2_1e6_ID}). The MPS sampler fails mainly on $\mathcal{G}^{M}_{M}$ on \textit{Jiuzhang~2} and on $P_j(n)$ on Borealis.

\paragraph{Low- and high-order observables---}
The experiments fail mainly on $\langle c_j\rangle$ (or $P_j(n)$), which is set largely by how much light reaches each detector. A single scalar transmission correction cannot remove such a mode-by-mode pattern of errors. The classical samplers are separated instead by observables that depend on correlations across many modes. The squashed-state sampler passes most marginals on the full-size \textit{Jiuzhang~3} and~4 datasets but fails $\mathcal{G}^{M}_{M}$ on every one of them (Figs.~\ref{fig:obs_J3_ID} and~\ref{fig:obs_J4}), so marginals alone do not expose a classical sampler; the grouped count distributions do. Of its 420 \textit{Jiuzhang~4} marginal $Z$-scores (mean $0.18$, standard deviation $1.14$), 9 exceed 3, 8 of them on Miniature 6~nJ (up to $5.9$), and it fails 12 of the 20 marginals on the 144-mode version of \textit{Jiuzhang~3} (Fig.~\ref{fig:obs_J3_144}). On \textit{Jiuzhang~4} each marginal order tests a single set of modes out of up to 8176, which limits its power: the experiment's marginal $Z$-scores there have mean $3.8$, while it fails $\langle c_j\rangle$ with $Z$ up to $+446$.

At high order the comparison is as follows. On \textit{Jiuzhang~4} (Fig.~\ref{fig:obs_J4}), none of the positive-$P$ sampler's 420 marginal $Z$-scores exceeds 3 in magnitude (the largest is 2.8), while more than 40\% of the experiment's do, up to 25. On \textit{Jiuzhang~2} with $N=10^{7}$ (Fig.~\ref{fig:obs_J2_1e7_ID}), the largest $|Z|$ among marginal orders 11 to 20 at 1000 and 1650~mW is 2.9 and 2.3 for the positive-$P$ sampler and 2.0 and 1.4 for the optimized one, where the experiment reaches 87 and 174. At the three lower powers, where the samples could be generated exactly in under one day (Sec.~\ref{sec:exact_cost}), the greedy sampler is the most accurate at these orders: 21, 5.4 and 4.3, against 66, 9.5 and 16 for the positive-$P$ sampler and 57, 23 and 8.0 for the optimized one, whose mixture gives up some high-order accuracy for the total-count distribution (Sec.~\ref{sec:diagonal}).

\subsection{Cross-entropy}
\label{sec:xeb}
Figures~\ref{fig:xe_ID} and~\ref{fig:xe_TD} summarize, for each dataset and sampler, the sector differences $\Delta\mathrm{XE}_n$ defined in Sec.~\ref{sec:stats} by their inverse-variance weighted mean,
\begin{equation}
\overline{\Delta\mathrm{XE}}=\frac{\sum_n w_n\,\Delta\mathrm{XE}_n}{\sum_n w_n},\qquad w_n=\sigma_n^{-2},\label{eq:XE_mean}
\end{equation}
with standard error $(\sum_n w_n)^{-1/2}$. Here $\sigma_n$ is the standard error of $\Delta\mathrm{XE}_n$, and the sums run over the sectors that are valid for both the sampler and the experiment. Zero is the experiment; a positive value means the sampler's samples have on average a lower log-probability under the ground truth than the experimental ones.

Against the ideal ground truth (Fig.~\ref{fig:xe_ID}), every sampler's mean cross-entropy is within about $0.15$~nats (natural-log units) of the experiment's, and most are within a few hundredths. The squashed-state sampler is the furthest above the experiment, on low-power \textit{Jiuzhang~2} and the smaller Borealis datasets. The positive-$P$ sampler is close to the experiment on \textit{Jiuzhang~2} and on the Borealis datasets that use the categorical step, and below it (its samples have on average a higher log-probability under the ground truth) on \textit{Jiuzhang~3} and almost every \textit{Jiuzhang~4} dataset. The corrected ground truth (Fig.~\ref{fig:xe_TD}) gives the same pattern.

These differences rank samplers by the mean ground-truth log-probability of their samples, which is not the same as closeness to the ground-truth distribution. An exact sampler has a definite expected $\mathrm{XE}_n$, but computing it needs a sum over all outcomes, so we cannot say which of two samplers is closer to it. A sampler that over-weights high-probability outcomes scores better than an exact one, and the cross-entropy can be spoofed at a cost comparable to computing it~\cite{ohSpoofingCrossEntropy2022a}. We therefore treat it as a secondary check. On its own it does not show which of the two is closer to the ground truth.

Cross-entropy benchmarking (XEB) is of interest primarily because of its use in random circuit sampling (RCS), a sampling task closely related to GBS, where XEB is the primary figure of merit and was argued to be classically hard to spoof~\cite{aaronson2020classical}, although the conjecture behind that argument fails for circuits of sublinear depth~\cite{aharonov2023polynomial}. However, the Porter--Thomas property that grounds the hardness of spoofing XEB in RCS has only a weak, regime-dependent analogue in GBS~\cite{ehrenberg2025transition}.

\paragraph*{Precision of the Torontonians---}
The threshold-detector probabilities are Torontonians whose sums are computed in 256-bit floating-point arithmetic (Sec.~\ref{sec:outprob}), because lower precision is not sufficient. With double-double arithmetic (about 32 significant digits), the \textit{Jiuzhang~4} A S64 2~nJ dataset gave positive but inaccurate probabilities at high click numbers, lowering $\mathrm{XE}_{15}$ by $1.7$~nats. Repeating 246 patterns in 384-bit arithmetic changed no result at double-precision accuracy.

\subsection{Cumulants}
\label{sec:cumulants}
For each order $\mathcal{K}=1,2,3$ we choose up to 10\,000 sets of $\mathcal{K}$ modes. At order one we use every mode. At order two we use every pair when there are at most 10\,000 of them (Borealis B16 and B72), and otherwise 10\,000 random pairs. At order three we use 10\,000 random triples (all 560 for B16). The sets are drawn with a random seed derived from the experiment's name, so every sampler and both ground truths use the same ones.

For each set we compute the joint cumulant of the {click variables for threshold detection or photon-number variables for PNR detection} from the samples, and exactly from the ground-truth covariance matrix (Sec.~\ref{sec:stats}). Following Ref.~\cite{liu2025robust}, we fit the sampled values against the exact ones by least squares, $\kappa_{\mathrm{sampled}}=a+K\kappa_{\mathrm{exact}}$. Figures~\ref{fig:dK_ID} and~\ref{fig:dK_TD} show the slope error $\Delta K=|K-1|$, the metric used in Ref.~\cite{liu2025robust}. Where $|K-1|$ is at least twice the standard error $\sigma_K$ of the fitted slope, the point is shown with $\pm1\sigma_K$; otherwise it is drawn as a downward arrow at the bottom of the panel.

The standard error $\sigma_K$ comes from each sampler's own residual scatter, not from the theoretical variance of the cumulants, so this is a heuristic comparison. It ignores correlations between mode sets that share modes, and it cannot detect underdispersion. Computing the theoretical variance would need the exact moments up to order $2\mathcal{K}$ (sixth order for the third-order cumulants) for every mode set, and the covariances between sets.

At first order the cumulants are the per-mode click probabilities (or mean photon numbers). The experiment's slope error is resolved on most datasets, which is the same failure as its $\langle c_j\rangle$ and $P_j(n)$ $Z$-scores, while the positive-$P$ slope error is unresolved on almost every dataset. The greedy sampler, which is built to match low-order statistics, has the smallest first-order slope errors on \textit{Jiuzhang~2}. At second order the positive-$P$ slope error is below those of the experiment and the squashed-state sampler on most datasets, and the optimized sampler's is comparable to or below the greedy sampler's. 
On \textit{Jiuzhang~4} the second-order slope errors of the experiment and the squashed-state sampler are each resolved on 19 of the 21 datasets, while the positive-$P$ sampler's are mostly unresolved. The experiment's values on the A datasets, about 0.03 to 0.08, are close to the 0.033 to 0.066 published in Ref.~\cite{liu2025robust}. At third order the slope is informative mainly for \textit{Jiuzhang~2}, which has $10^{7}$ samples, and for B16. There, at low pump power, the experiment is closer to the exact cumulants than the classical samplers, which matches the positive-$P$ sampler's weakness on $\mathcal{G}^{M}_{M}$ at low power; the classical samplers improve with pump power and overtake the experiment at the highest. 
Elsewhere the third-order slope errors of all samplers, the experiment included, are either unresolved or of order one and above, so the third-order slope does not discriminate between samplers. The corrected ground truth (Fig.~\ref{fig:dK_TD}) gives the same picture.

\section{Improvability}
\label{sec:improve}
Is the positive-$P$ sampler a fixed-accuracy approximation, or can its agreement with the ground truth be improved with additional computational resources? If its accuracy were fixed, it could eventually be overtaken as both experiments and competing classical methods improve. In this section we show that it can be improved with more whitening--coloring iterations, a better coloring target (Sec.~\ref{sec:diagonal}), and, for photon-number-resolving detectors, with a categorical output step (Sec.~\ref{sec:categorical}).

\subsection{Whitening--coloring iterations and the target diagonal}
\label{sec:diagonal}
For the unoptimized threshold positive-$P$ sampler of the main text we stop the iteration of Eqs.~\textup{(7)}--\textup{(8)} after $\ell=10$ iterations. 
For the ``optimized'' positive-$P$ sampler we introduce in this section and the categorical PNR positive-$P$ sampler of Sec.~\ref{sec:categorical}, we use 30--150 iterations depending on the target, and Anderson acceleration~\cite{walker2011anderson} to speed up fixed-point convergence in the longer runs.

The whitening--coloring transformation of Eqs.~\textup{(7)}--\textup{(8)} forces the projected samples to have a target mean vector and covariance matrix. Its accuracy depends on three choices: the number of phase-space samples subject to whitening--coloring at once (i.e., involved in deriving the whitening matrix~\footnote{A covariance matrix estimated from $B$ samples (batch sizes in Table~\ref{tab:simparams}) has rank at most $B-1$, so a batch must contain more samples than there are modes. The code stops with an error if a batch covariance matrix is singular, so every run reported here had a full-rank batch covariance.}), the number of iterations, and the coloring target itself. 
The last two are best considered together. Specifying that the whitening--coloring transformation acts on $\pi_{j}=\exp(-\alpha_{j}\beta_{j})$ is not sufficient to pick out a single transformation rule. First, the matrix square root used by a whitening transformation is non-unique. We use {zero-phase component analysis (ZCA) whitening}, which minimizes the total squared distance between the whitened and unwhitened data~\cite{kessy2018optimal}. 
Second, there is freedom in the diagonal elements of the target covariance matrix. For threshold detectors, the off-diagonal elements of the target are unbiased estimates of click moments: $\langle\pi_i\pi_j\rangle_P$ estimates the probability that neither mode $i$ nor mode $j$ clicks, which the analytic target below uses exactly. The phase-space estimate of each diagonal element, $\langle\pi_j^{2}\rangle_P-p_j(0)^{2}$ with $p_j(0)\equiv\langle\pi_j\rangle_P$ the mean no-click probability, is not a click moment: $\langle\pi_j^{2}\rangle_P$ estimates the photon-number parity $\langle(-1)^{\hat{n}_j}\rangle$. Since the final samples are drawn from a mixture of {Bernoulli} distributions, each mode's click variance is already fixed by its mean, so changing this diagonal does not alter the target first- or second-order click moments.

\paragraph*{The admissible range of diagonal elements---} Let $\Sigma^{\mathrm{off}}$ be the off-diagonal part of the target and $D$ the chosen diagonal. The target must be a covariance matrix and thus positive semidefinite ($\Sigma^{\mathrm{off}}+D\succeq0$). The smallest admissible diagonal is the minimum-trace completion, a small semidefinite program in $M$ variables,
\begin{alignat}{1}
  D^{\min} &= \arg\min \mathrm{tr}\, D\quad \text{subject to} \quad \Sigma^{\mathrm{off}} + D \succeq 0,\nonumber
  \\D_{jj}&\le p_j(0)[1-p_j(0)] .
  \label{eq:mintrace}
\end{alignat}
The natural upper choice is $D_{jj}=p_j(0)[1-p_j(0)]$, for which the target matrix is the covariance matrix of the true click indicators and thus positive semidefinite. The upper bound in Eq.~\eqref{eq:mintrace} is needed. No quantity confined to $[0,1]$ with mean $p_j(0)$ can have a variance above $p_j(0)[1-p_j(0)]$, and without the bound the minimum-trace diagonal can exceed it on some modes. Every diagonal between the two endpoints also gives a positive semidefinite matrix,
\begin{alignat}{1}
  D(\gamma) \;&=\; (1-\gamma)\, D^{\min} \;+\nonumber
  \\&\; \gamma \, \mathrm{diag}\big[p_j(0)[1-p_j(0)]\big],~ \gamma \in [0,1] .
  \label{eq:alpha}
\end{alignat}
A positive semidefinite matrix within these bounds is not always the covariance matrix of some set of values confined to $[0,1]$; the condition is necessary but not sufficient. We do not impose the exact restriction. A target that cannot be reached does not produce invalid samples, because every sample is projected back into $[0,1]$. The iteration simply does not converge to it, and the effect appears as a poorer $Z$-score, like any other defect of a sampler. The interpolation of Eq.~\eqref{eq:alpha} is used only for the optimized \textit{Jiuzhang~2} mixtures below. All other threshold-detector results use the phase-space estimate of the target. This estimate can have negative eigenvalues (see the Results below). A negative eigenvalue then gives a complex iterate, and the next projection keeps its real part. 

\paragraph*{Results---} Figure~\ref{fig:alpha_iter} applies this to \textit{Jiuzhang~2} at 150~mW, the dataset on which the baseline (unoptimized) positive-$P$ sampler performs worst. 
With the phase-space target, estimated from the phase-space samples and used elsewhere in this work, $\max|Z|$ does not improve with more iterations, and the per-mode click probabilities drift away from the ground truth (panel b).
 Beyond about 40 iterations the out-of-range amount for this target stops falling and repeatedly jumps back up (panel d). This happens because the target has negative eigenvalues, so the iteration cannot reach it. As it tries, the covariance of the projected samples approaches a singular matrix, and the inverse square root in Eq.~\textup{(8)} amplifies rounding errors along the directions with the smallest eigenvalues. The resulting large steps push many samples out of range at once. For this dataset the phase-space target has 33 negative eigenvalues. These are not only a finite-ensemble effect. Because the $\pi_j$ are complex, the real part of their covariance is $\mathrm{Cov}(\operatorname{Re}\boldsymbol{\pi})-\mathrm{Cov}(\operatorname{Im}\boldsymbol{\pi})$, which can be indefinite even for an infinite ensemble, and its diagonal is set by the photon-number parity. No set of values confined to $[0,1]$ has an indefinite covariance matrix, so the iteration cannot converge to this target. With the {analytic target} derived from the Gaussian covariance matrix, the target is positive semidefinite for every $\gamma$, and the per-mode click probabilities reach agreement with the ground truth within 300 iterations for $\gamma\le0.5$, and more slowly for $\gamma=0.75$ (panel b); a larger $\gamma$ lowers the total-count $Z$-score (panel c) at the cost of slower convergence (panel d). At $\gamma=0$ the out-of-range amount falls to the level of rounding error (panel d), so for this dataset even the smallest admissible diagonal is reached. 
With enough iterations the analytic target brings $\max|Z|$ below the experiment's, which the phase-space target never does (panel a).
 Accuracy is therefore not fixed by the algorithm: it improves with iterations and with the choice of $\gamma$. The improvement is limited by a trade-off: a larger $\gamma$ lowers the $Z$-score of $\mathcal{G}^{M}_{M}$ but raises those of the high-order marginals (for $\mathcal{M}^{15}$, about 11, 34 and 60 for $\gamma=0$, 0.25 and 0.5, and at least 75 for $\gamma=0.75$), so no single value improves every observable.

Whitening all $10^{7}$ trajectories together, rather than in ten batches of $10^{6}$, leaves the high-order discrepancy in place, so it is not caused by the finite batches. Because a small $\gamma$ favors the high-order marginals and a large $\gamma$ favors $\mathcal{G}^{M}_{M}$, mixing samples from the two can do better than either. We use this below.

\paragraph*{Optimized mixtures on \textit{Jiuzhang~2}---}
The optimized positive-$P$ sampler draws a fraction $w$ of its $N=10^{7}$ samples from a run with $\gamma=0$ and the rest from a run with $\gamma=0.9$, both with the analytic target and with independent phase-space trajectories. The $\gamma=0$ run uses 30 whitening iterations. The $\gamma=0.9$ run converges more slowly, so it uses 80 iterations with Anderson acceleration~\cite{walker2011anderson}, which extrapolates each iterate from the previous five. All observables are computed from the combined $N$ samples. The weight $w$ was chosen for each power from $w=0,0.05,\ldots,1$, as the value giving the smallest $\max|Z|$ against one phase-space estimate of the ground truth. The scores reported here are against a second, independently seeded estimate.

Table~\ref{tab:optimised_pp} gives the results. The optimized sampler is closer to the ground truth than the baseline, greedy and squashed-state samplers and the experiment at every power. Its worst observables are now high-order marginals at all but the highest power, where every observable has $|Z|<3$. 
A random subset of $10^{6}$ of its samples is also closer to the ground truth than every other sampler, including the MPS sampler, at every power but 1000~mW, where it and the unoptimized sampler are both within sampling error (Table~\ref{tab:worst_J2_1e6}).

Against the corrected ground truth (Table~\ref{tab:optimised_pp}, lower half) we chose the weights in the same way, from a separate pair of $\gamma=0$ and $\gamma=0.9$ runs whose phase-space samples were drawn from the corrected state. The optimized sampler is again closer to the ground truth than the baseline, greedy and squashed-state samplers and the experiment at every power. As against the ideal ground truth, its worst observables are high-order marginals ($\mathcal{M}^{6}$ to $\mathcal{M}^{15}$), 
and its scores are close to those against the ideal ground truth. In the quantum-advantage regime it is within sampling error at 1650~mW and close to it at 1000~mW. 

\subsection{A categorical output step for photon-number detection}
\label{sec:categorical}
With PNR detectors, the Poisson output step whitens the projected photon numbers $n_j^{(cl)}$ and draws each count from a Poisson distribution. This fixes the first two factorial moments of each mode's count, but not the rest of its distribution $P_j(n)$, the observable on which the Poisson step fails for Borealis. The categorical step instead acts on the probabilities of individual photon numbers, in the same way that the threshold-detector sampler acts on the no-click probability.

For each phase-space sample and mode we form the vector of category probabilities
\begin{align}
\pi_{j,c}&=\frac{e^{-n_j}n_j^{c}}{c!},\quad c=0,\ldots,C-1,\nonumber\\
\pi_{j,\geq C}&=1-\sum_{c<C}\pi_{j,c},
\end{align}
where $n_j=\alpha_j\beta_j$. Their positive-$P$ averages are the exact single-mode distribution, $\langle\pi_{j,c}\rangle_P=P_j(c)$, and the averages of products across two modes are the exact two-mode distributions, $\langle\pi_{i,c}\pi_{j,c'}\rangle_P=P_{ij}(c,c')$ for $i\neq j$. $C$ is the smallest value for which $P_j(n\geq C)\leq10^{-4}$ in every mode. The real part of each mode's vector is projected onto the nearest valid probability vector (non-negative entries that sum to one), End Matter Eq.~\textup{(16)}~\cite{duchi2008efficient}; with two categories this is the clamp $\mathcal{T}_p$. The projected vectors are then whitened--colored as in Eqs.~\textup{(7)}--\textup{(8)}. The target means are the exact $P_j(c)$, computed from the Gaussian covariance matrix. The target cross-mode covariances are the exact $P_{ij}(c,c')-P_i(c)P_j(c')$ for $c,c'<C_2$, where $C_2$ is the largest value, at most 3, for which $P_j(c)\geq10^{-3}$ in every mode for every $c<C_2$. Rarer categories, including all counts of three or more, are given only their mean, because their covariance blocks are too noisy, in a batch of 2000 phase-space samples, to whiten. As with the diagonal of Sec.~\ref{sec:diagonal}, the covariances within one mode are not fixed by the detector statistics. We set mode $j$'s within-mode block to $\nu_j\,(\mathrm{diag}\,\boldsymbol{P}_j-\boldsymbol{P}_j\boldsymbol{P}_j^{T})$, the covariance of a one-hot vector (one entry per category) scaled by $\nu_j$, where $\boldsymbol{P}_j$ holds the exact category probabilities. The $\nu_j$ are set together, as the analogue of $\gamma=0$ in Eq.~\eqref{eq:alpha}: they minimize the total within-mode variance subject to $\nu_j\leq1$ and a positive semidefinite target, a small semidefinite program that we solve by a log-barrier interior-point method. We use 150 iterations with Anderson acceleration, in batches of 2000 phase-space samples, and finish with one step that sets each category's sample mean to its exact value: in every vector, probability is moved from categories whose sample mean is too high to those whose mean is too low, in proportion to the shortfalls, so every vector stays a valid probability vector. Each output sample then takes a phase-space sample chosen at random from the batch, with replacement, and draws one count per mode from its categorical distribution, and a count in the top category is drawn from the exact conditional distribution $P_j(n\,|\,n\geq C)$.

Against the ideal ground truth, the categorical step lowers $\max|Z|$ on B72, B216L, B216H and B288 by factors of 2.8 to 25, bringing B216H and B288 within sampling error (Table~\ref{tab:worst_B}). 
 $P_j(n)$ is no longer the worst observable on any of these datasets; on B72 and B216L the worst is now $\mathcal{G}^{M}_{M}$. 
On B16 the categorical step is worse than the Poisson step, failing on $\mathcal{G}^{M}_{M}$, so B16 uses the Poisson step.
 These scores are against an independently seeded rerun of the phase-space ground truth; between that rerun and the estimate used to choose the output step, no $\max|Z|$ changed by more than 1.6, and the choice is the same under both. On B16 the categorical fit converges at every batch when the iteration limit is raised to 1000, yet the $\mathcal{G}^{M}_{M}$ and high-order marginal errors remain, so the B16 failure is not caused by too few iterations.

\section{Exact-sampling runtime}
\label{sec:exact_cost}
The shading in Fig.~1 marks the datasets that could be sampled exactly in under one day. We take the time to draw one exact sample with $n$ detected photons (clicks, for threshold detectors) from the Fugaku runtime model of Ref.~\cite{madsenQuantumComputationalAdvantage2022}, Eq.~(4),
\begin{equation}
t(n)=\tfrac{1}{2}\,c\,M\,n^{3}\,G^{n/2},\qquad c=4.41\times10^{-17}~{\rm s},
\end{equation}
with $M$ the number of output modes and $G=2$. The expected time for $N$ samples is $N\sum_n P(n)\,t(n)$, where $P(n)$ is the total-count distribution of the dataset's ideal ground truth. Because $t(n)$ grows exponentially, this sum is set by the upper tail of $P(n)$, and the mean photon number would underestimate it. $P(n)$ is a phase-space estimate, which is noisy where it is very small, so the sum runs over the counts around its peak where $P(n)$ is at least twice its standard error and $N\,P(n)\ge1$, that is, counts that $N$ samples would be expected to contain. A dataset is shaded if this time is under one day. 
The closest unshaded dataset is \textit{Jiuzhang~2} at 1000~mW with $N=10^{6}$, at about one day. For the larger datasets the sum is largest at the top of the counted range, so their times are lower bounds. For Borealis the model ignores the reduced cost of several photons arriving at one detector. For a single sample the model gives the scale of the largest experiment directly: the largest \textit{Jiuzhang~4} sample reported, with $n=3050$ clicks in $M=8176$ modes~\cite{liu2025robust}, would take $t(3050)\approx10^{457}$~s, about $10^{449}$ years. The same model places \textit{Jiuzhang~1} in the classically tractable regime: its largest reported sample, with $n=76$ clicks in $M=100$ modes~\cite{zhong2020quantum}, would take $t(76)\approx4$~minutes. For comparison, Ref.~\cite{liu2025robust} estimates that the approximate matrix-product-state method would need more than $10^{42}$ years on the El~Capitan supercomputer to build its tensor network.

\section{Computational scaling}
\label{sec:scaling}
Table~\ref{tab:scaling} gives the asymptotic cost of each sampler compared in the main text. $M$ is the number of output modes, $S$ the number of input squeezed states (called $K$ in Refs.~\cite{oh2024classical,liu2025robust}), $N$ the number of samples and $n$ the number of detected photons (clicks, for threshold detectors) in one sample.

For the positive-$P$ sampler, drawing one phase-space sample and propagating it through the $M\times S$ transmission matrix costs $O(MS)$. Each whitening--coloring iteration on a batch of $B$ samples forms the $M\times M$ batch covariance, at cost $O(BM^{2})$, diagonalizes it, at cost $O(M^{3})$, and applies the transform, at cost $O(BM^{2})$. With $L$ iterations, the cost per sample is therefore $O(MS+LM^{2}+LM^{3}/B)$. This is $O(MS+LM^{2})$, because the batch must contain more samples than there are modes for its covariance to be invertible, and since $S\le M$ for every dataset it is the $O(LM^{2})$ of the main text. Above that limit, we found no clear trend with batch size. For the 1152-mode high-power \textit{Jiuzhang~3} dataset with $N=10^{7}$, five batch sizes from 2500 to $10^{5}$ samples gave $\max|Z|$ between 2.1, for the batch of $10^{4}$ used in our other results, and 3.8, all within or close to sampling error. The phase-space target used for the unoptimized results is recomputed from each batch, at cost $O(BM^{2}+M^{3})$, which does not change the cost per sample. The analytic target of the optimized sampler is computed once per dataset, from the pairwise click probabilities at cost $O(M^{2})$ and the semidefinite program of Eq.~\eqref{eq:mintrace}. The memory needed is $O(BM+M^{2})$. The categorical output step whitens three categories per mode, which replaces $M$ by $3M$ in these expressions. In practice, each 8176-mode \textit{Jiuzhang~4} L1024 dataset of $10^{6}$ samples took about 130 core-hours (ten tasks on 18 cores, with a median of 44 minutes each).

The costs of the other samplers are taken from their papers, except where marked. The greedy sampler of order $\mathcal{K}$ fixes each of the $NM$ entries of its sample matrix by comparing $O(M^{\mathcal{K}-1})$ marginals of $2^{\mathcal{K}}$ outcomes, which gives its $O(NM^{\mathcal{K}}2^{\mathcal{K}})$ running time~\cite{villalonga2021efficient}. It first needs the $O(M^{\mathcal{K}})$ ground-truth marginals of order up to $\mathcal{K}$, which are computed from Torontonians of at most $\mathcal{K}$ clicks, so their cost grows exponentially only in $\mathcal{K}$~\cite{villalonga2021efficient}. The MPS sampler builds a matrix-product state of bond dimension $\chi$ and local photon-number cutoff $d$ in $O(Md\chi^{2}h)$ operations, where $h$ is the cost of one hafnian, which grows exponentially with the photon numbers kept in the truncated basis, and stores it in $O(Md\chi^{2})$ memory~\cite{oh2024classical}. For a fixed circuit the bond dimension needed for a total variation distance $\delta$ grows only as $\mathrm{polylog}(1/\delta)$, but over a family of experiments it is polynomial in $S$ if the transmission $\eta$ falls faster than about $1/\sqrt{S}$, and superpolynomial if it falls no faster than $S^{-b}$ for some $b<1/2$~\cite{oh2024classical}. Exact sampling costs $O(Mn^{3}2^{n/2})$ per sample, the cost of the hafnians of an $n$-photon pattern without collisions~\cite{madsenQuantumComputationalAdvantage2022,bulmerBoundaryQuantumAdvantage2022a}; this is the model behind the shading of Fig.~1 (Sec.~\ref{sec:exact_cost}).

\paragraph*{Limits of the MPS and greedy samplers---}
Because the bond dimension the MPS sampler needs grows faster than any polynomial once the transmission falls no faster than $S^{-b}$ for some $b<1/2$, experiments that scale up at fixed transmission, as \textit{Jiuzhang~4} does~\cite{liu2025robust}, leave the regime where it is efficient~\cite{oh2024classical}. A newer construction of the MPS~\cite{liu2026efficient} avoids the hafnian calculations that limit Ref.~\cite{oh2024classical}. For the smallest \textit{Jiuzhang~4} configuration (S64), its authors have built the MPS tensor of one central mode, with local photon-number cutoff $d=1$, and estimate about ten days to draw $10^{7}$ samples with the work spread over $4336d$ nodes. They judge the larger M256 and L1024 configurations to be beyond its current reach (Ref.~\cite{liu2026efficient}, Results, \textit{Jiuzhang~4.0}, and Fig.~4). It has not yet been used to generate samples, so we do not include it. We also do not include the emulator of Ref.~\cite{dodd2025fast}. It needs far fewer computational resources than the MPS sampler, but its authors' own comparison on \textit{Jiuzhang~2} and \textit{Jiuzhang~3} shows the MPS sampler slightly outperforming it (Figs.~S1 and~S2 of Ref.~\cite{dodd2025fast}). Since our sampler outperforms the MPS sampler on $\max|Z|$ wherever both were run, a separate comparison would add little.

The order-$\mathcal{K}$ greedy sampler needs only $O((n_{\max}+1)^{\mathcal{K}}M^{\mathcal{K}})$ time per sample, where $n_{\max}$ is the photon-number cutoff per mode ($O(2^{\mathcal{K}}M^{\mathcal{K}})$ for threshold detectors, Table~\ref{tab:scaling}). However, its deterministic algorithm only resolves probabilities to about $1/N_{gs}$, where $N_{gs}$ is the number of samples it generates in one run. Marginal events with smaller probabilities are never produced. Going from second to third order improved the sampler markedly for threshold detectors~\cite{villalonga2021efficient}, but made it slightly worse for Borealis, where a large part of the three-mode photon-number distribution lies below $1/N_{gs}$~\cite{madsenQuantumComputationalAdvantage2022}. Events above $1/N_{gs}$ are instead matched more closely than independent samples allow, which is the underdispersion in Table~I of the main text. A random subset of $N\ll N_{gs}$ samples restores the correct spread. The greedy samples with $N=10^{6}$ (Fig.~\ref{fig:obs_J2_1e6_ID} and Table~\ref{tab:worst_J2_1e6}) are not such a subset but a separate run with $N_{gs}=10^{6}$, so they show the same underdispersion. To generate $N$ samples that match the $\mathcal{K}$th-order marginals of the system, the greedy sampler therefore needs not a run of $N_{gs}=N$ samples but one larger than both $N$ and the inverse of the smallest marginal probabilities that matter. The second of these depends on the tails of the marginal distributions rather than on $N$, and grows quickly with $\mathcal{K}$ for PNR detectors. Memory is also a limit: the third-order histograms of the 8176-mode \textit{Jiuzhang~4} datasets would hold about $7\times10^{11}$ entries. Beyond $\mathcal{K}=2$ the greedy sampler is impractical for PNR experiments and for large threshold-detector experiments.

\section{Lossless experiments}
\label{sec:lossless}
To test whether removing loss would change our conclusions, we built a lossless version of every dataset. We replaced its transmission matrix with the first columns of a Haar-random unitary of the same size, so that no light is lost, and kept its squeezing parameters, mode numbers and detector type. Datasets that share an interferometer share one random matrix. We ran the positive-$P$ and squashed-state samplers on these lossless versions exactly as on the originals, and compared each with its own ideal ground truth. For Borealis, the positive-$P$ sampler uses the same output step on the lossy and lossless versions: the Poisson step for B16 and the categorical step for the other datasets. There are no experimental samples for these states. Every experiment aims to implement a Haar-random unitary, so this is an ideal implementation of each experiment, although its matrix differs from the experiment's own. Removing the loss also raises the mean photon number at the detectors, by a factor of 2.1 on \textit{Jiuzhang~2}, 2.4 on \textit{Jiuzhang~3}, 2.7 to 3.2 on Borealis and 1.7 to about 2 on \textit{Jiuzhang~4}; the high-power \textit{Jiuzhang~3} dataset, for example, rises from 116 to 274 photons. The sampler's accuracy improves with photon number (Fig.~\ref{fig:photons_modes}), so part of any change in accuracy may come from the higher photon number.

Figure~1(g) compares $\max|Z|$ for the lossy and lossless versions. The squashed-state sampler fails on both. The positive-$P$ sampler stays within or close to sampling error on \textit{Jiuzhang~3} and on 19 of the 21 \textit{Jiuzhang~4} datasets, whose 430 $Z$-scores (seven cells have too few bins to test) have mean $0.09$ and standard deviation $1.10$. Without loss it gets worse on \textit{Jiuzhang~2} and on the other two \textit{Jiuzhang~4} datasets, A S64 at 2 and 3~nJ, where $\max|Z|$ rises from 4.5 to 20 and from 2.2 to 7.0, both on $\mathcal{G}^{M}_{M}$. The largest deviations are on the lossless Borealis datasets, all of which get markedly worse. Without loss every output pattern has an even total photon number, but each projected phase-space sample gives every mode both parities, so the sampler produces odd totals; about half of its samples have one. The phase-space ground truth fails in the same way. Its $\mathcal{G}^{M}_{M}$ also puts about half of the probability on odd totals, with a standard error of about $10^{-4}$, where the exact value is zero. For lossless photon-number-resolving datasets the phase-space ground truth of the grouped count distributions is therefore unreliable. Here the exact distribution is easy to compute instead: a lossless network conserves the total photon number, so $\mathcal{G}^{M}_{M}$ is the distribution of the summed photon numbers of the independent squeezed inputs. For the lossless Borealis datasets we therefore score $\mathcal{G}^{M}_{M}$ against this exact distribution (Figs.~1(g) and~\ref{fig:obs_LL_B}). For Borealis, $\mathcal{G}^{M}_{M}$ is the only observable taken from the phase-space simulation; the others come from the covariance matrix, so every lossless Borealis score is against an exact ground truth. Both samplers then fail on $\mathcal{G}^{M}_{M}$ on every lossless Borealis dataset, with $Z$ between 499 and 520, where against the phase-space ground truth they scored 8.5 to 285. The parity mismatch alone sets a floor on these values. With about half of the samples in bins of zero probability, each even bin is about half empty, which contributes about $N/4$ to $\chi^2$, or $Z\approx\sqrt{N/4}=500$ for $N=10^{6}$, whatever else the sampler gets right. With loss, where the exact distribution follows from the covariance matrix, the two ground truths agree closely: on B72, B216L, B216H and B288 the sampler's $Z$-score on $\mathcal{G}^{M}_{M}$ changes by at most 0.4 when the exact distribution replaces the phase-space one. For the threshold-detector datasets no such failure appears: on every dataset, with and without loss, the phase-space $\mathcal{G}^{M}_{M}$ reproduces the exact mean and variance of the total click count, computed from the one- and two-detector no-click probabilities, to within 2.7 standard errors of an exact sampler at each dataset's $N$.

The optimized sampler of Sec.~\ref{sec:diagonal} removes most of the \textit{Jiuzhang~2} difference. We reran it on the lossless \textit{Jiuzhang~2} datasets, with new $\gamma=0$ and $\gamma=0.9$ runs and the weight chosen in the same way, giving $w=0.25$, $0.50$, $0.70$, $0.65$ and $0.45$. 
Its $\max|Z|$ is close to its lossy values at every power, rising at 1650~mW from 1.8 to 3.7, still close to sampling error, and far below the baseline sampler's without loss.
 On \textit{Jiuzhang~2}, most of the drop in the baseline sampler's accuracy when loss is removed is therefore a limit of the baseline settings, not of the method. Whether the same holds for the lossless Borealis datasets is not yet tested. For the large \textit{Jiuzhang~3} and~4 experiments, removing the loss does not change the sampler's agreement with the ground truth, except on two of the 21 \textit{Jiuzhang~4} datasets.

Figures~\ref{fig:obs_LL_J2}--\ref{fig:obs_LL_J4} give the signed $Z$-score of every observable for the lossless versions.

\section{Simulating grouped count distributions}
\label{sec:gcd_review}\label{sec:producttree}
Grouped count distributions (GCDs) are obtained by partitioning the GBS output modes into $d$ disjoint subsets $S_1,\ldots,S_d$ and determining the joint distribution of the total counts $\boldsymbol{m}=(m_1,\ldots,m_d)$ in those subsets~\cite{drummondSimulatingComplexNetworks2022,dellios2024validation}. The general GCD is defined as
\begin{equation}
\mathcal{G}_{\boldsymbol{S}}(\boldsymbol{m})=\left\langle \prod_{j=1}^{d}\left[\sum_{\sum_{k\in S_j} c_{k}=m_{j}}\hat{\mathcal{D}}_{S_{j}}(\boldsymbol{c})\right]\right\rangle,\label{eq:GCD_general}
\end{equation}
where $\boldsymbol{S}=(S_1,\ldots,S_d)$, each $S_j\subseteq\{1,\ldots,M\}$, and $M_j=|S_j|$. The operator $\hat{\mathcal{D}}_{S_{j}}(\boldsymbol{c})$ is a normally ordered pattern projector for the output modes in $S_j$. For threshold detection, $c_k\in\{0,1\}$ and $\hat{\mathcal{D}}_{S_{j}}(\boldsymbol{c})=\bigotimes_{k\in S_{j}}:e^{-\hat{n}_{k}}(e^{\hat{n}_{k}}-1)^{c_{k}}:$~\cite{drummondSimulatingComplexNetworks2022}. For PNR detection, $c_k\in\{0,1,\ldots,c_k^{(\mathrm{max})}\}$ and $\hat{\mathcal{D}}_{S_{j}}(\boldsymbol{c})=\bigotimes_{k\in S_{j}}(c_{k}!)^{-1}:(\hat{n}_{k})^{c_{k}}e^{-\hat{n}_{k}}:$~\cite{dellios2024validation}.

The total-count distributions and their lower-dimensional grouped marginals are special cases of this definition. In the shorthand used in the figures, $\mathcal{G}^{M}_{M}$ denotes one subset containing all $M$ modes, that is, $\mathcal{G}_{\boldsymbol{S}}$ with $d=1$ and $S_1=\{1,\ldots,M\}$, while $\mathcal{G}^{M}_{M/2,M/2}$ denotes two equal subsets that partition the $M$ modes.

Each GCD is estimated as an average over positive-$P$ trajectories, in which every output photon-number operator $\hat{n}_k$ is replaced by $n'_{k}=\alpha'_{k}\beta'_{k}$. Here $\boldsymbol{\alpha}'=\boldsymbol{T}\boldsymbol{\alpha}$ and $\boldsymbol{\beta}'=\boldsymbol{T}^{*}\boldsymbol{\beta}$ are the coherent amplitudes transformed by the transmission matrix $\boldsymbol{T}$. For PNR detectors, the per-trajectory distribution of the total photon number in $S_j$ then has the Poisson form with the complex parameter $n'_{S_j}=\sum_{k\in S_j}n'_k$~\cite{dellios2024validation}, which is evaluated in $O(M_j)$ operations per trajectory. The transition-edge sensors used in Borealis~\cite{madsenQuantumComputationalAdvantage2022} resolve up to 13 photons. We group counts of 11 or more into a single bin, $c_{k}^{(\textrm{max})}=11$, whose operator is the tail $1-\sum_{c=0}^{10}(c!)^{-1}:(\hat{n}_{k})^{c}e^{-\hat{n}_{k}}:$, in the same way for the ground truth and every sampler. This cap applies only to $P_j(n)$; the total-count distributions are computed from the uncapped counts.

\paragraph*{Threshold detectors---}
In one trajectory, mode $k$ contributes the factor $\pi_k(0)=\exp(-n_k')$ to the sum in Eq.~\eqref{eq:GCD_general} if it does not click and $\pi_k(1)=1-\exp(-n_k')$ if it does. The sum over all patterns in subset $S_j$ with exactly $m$ clicks is therefore the coefficient of $z^m$ in the polynomial
\begin{equation}
  F_j(z) = \prod_{k\in S_j}\bigl(\pi_k(0) + \pi_k(1) z\bigr) = \sum_{m=0}^{M_j} f_{j,m}\, z^{m}.
  \label{eq:S-genfn}
\end{equation}
The coefficients $f_{j,0},\dots,f_{j,M_j}$ give, for this trajectory, the total click distribution in $S_j$. The Fourier method of Ref.~\cite{drummondSimulatingComplexNetworks2022} obtains them by evaluating $F_j(z)$ at the $M_j+1$ roots of unity and applying an inverse discrete Fourier transform, which costs $O(M_j^2)$ per trajectory. This is difficult to run for the 8176 modes of \textit{Jiuzhang~4}, especially for the two-dimensional GCD.

\paragraph*{The subproduct tree---} We instead compute the coefficients directly, by multiplying the linear factors of Eq.~\eqref{eq:S-genfn} (i.e., each factor $\pi_k(0)+\pi_k(1)z$) in a balanced binary tree~\cite{vonzurgathen2013modern}. The $M_j$ factors are multiplied in pairs to give polynomials of degree two, these are multiplied in pairs to give polynomials of degree four, and so on, until a single polynomial of degree $M_j$ remains. To create a balanced tree the distribution is padded to $M_j'=2^{\lceil\log_2 M_j\rceil}$ modes, the next power of two, with trivial modes with $\pi_k(0)=1$ (for $k>M_j$). At each level, the products are computed as convolutions of coefficient vectors using the fast Fourier transform, so the whole tree takes $O(M_j\log_2^2{M_j})$ operations~\cite{vonzurgathen2013modern}.
\paragraph*{Several subsets---} A grouped count distribution with $d$ disjoint subsets $S_1,\dots,S_d$ requires the joint distribution of the click totals $(m_1,\dots,m_d)$. For one trajectory, the factors belonging to different subsets are separate, so the joint coefficient is the product of the single-subset coefficients, and the full $d$-dimensional array of joint coefficients, with entries $\prod_j f_{j,m_j}$, is the outer (tensor) product {$\boldsymbol f_1\otimes\cdots\otimes\boldsymbol f_d$}. This is not an independence assumption: correlations between subsets enter through the average over trajectories, so the product is taken before the average. The grouped count distribution is its average over trajectories,
\begin{equation}
  \mathcal{G}_{\boldsymbol{S}}(\boldsymbol{m}) =
  \mathrm{Re}\Bigl\langle \prod_{j=1}^{d} {f_{j,m_j}} \Bigr\rangle_{P}.
  \label{eq:S-outer}
\end{equation}

\paragraph*{Memory---} The outer product of each trajectory is added into a running sum as soon as it is formed, so only one array of size $N_G=\prod_j(M_j+1)$, the number of output probabilities, is held at a time, however many trajectories there are.

\paragraph*{Accuracy and speed---} The two methods compute the same coefficients by different routes, and their results agree to $10^{-15}$, relative to the largest entry, in every case we tested. Table~\ref{tab:S-timing} compares the time and memory of the two methods for a range of partitions. The advantage grows with the number of subsets and, for a fixed number of subsets, with the number of modes.

\onecolumngrid
\raggedbottom

\begin{herefigure}
\centering
\includegraphics[width=0.85\textwidth]{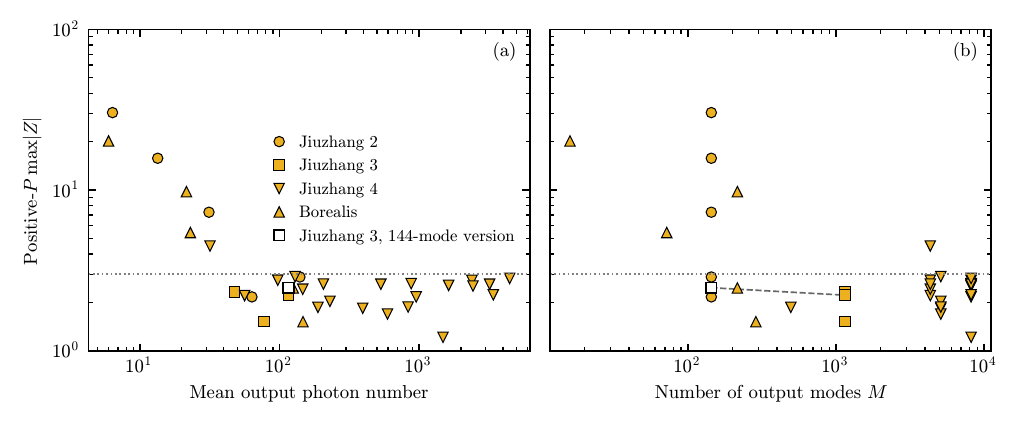}
\captionof{figure}{The positive-$P$ sampler's accuracy follows the photon number, not the system size. Positive-$P$ $\max|Z|$ against the ideal ground truth for every dataset, (a) against the mean output photon number and (b) against the number of output modes. In (a) the datasets of all four experiments fall on one trend; in (b) the same mode number spans the whole range of $\max|Z|$. Every dataset has $N=10^{6}$ samples, \textit{Jiuzhang~2} included, so the points are directly comparable. For the seven datasets with $\max|Z|>3$ (6 to 32 photons), a straight-line fit of $\log\max|Z|$ against $\log$ photon number gives $R^2=0.86$ and slope $-0.93$ ($R^2=0.92$ when extended to 70 photons). The open square is the 144-mode version of the high-power \textit{Jiuzhang~3} dataset (Table~\ref{tab:worst_J3_144}), which has the same photon number as the full dataset and one-eighth as many modes; the dashed line in (b) joins the two. Borealis uses the same output step as Fig.~1. The dotted line marks $|Z|=3$.\label{fig:photons_modes}}\par\kern0pt
\end{herefigure}

\bigskip

\begin{heretable}
\centering
\captionof{table}{Observable with the largest $|Z|$, and its signed $Z$-score, for each \textit{Jiuzhang~2} dataset with $N=10^{7}$. A dash means no observable exceeds $|Z|=3$; the bracket then gives the largest $|Z|$. Up to two other observables within 3 of the largest $|Z|$ (and above $|Z|=3$) are also listed. Observables are defined in Fig.~\ref{fig:obs_J2_1e7_ID}.\label{tab:worst_J2_1e7}}\par\kern0pt
\renewcommand{\arraystretch}{1.25}
\begin{tabular}{lcccc}
\hline
\textbf{Dataset} & \textbf{Experiment} & \textbf{Squashed} & \textbf{Greedy} & \textbf{Positive-$P$} \\
\hline
\multicolumn{5}{l}{\emph{Ideal ground truth}} \\
150~mW & \worstobs{$\langle c_j\rangle$}{$+91$} & \worstobs{$\mathcal{G}^{M}_{M/2,M/2}$}{$+607$} & \worstobs{$\mathcal{G}^{M}_{M}$}{$+86$} & \worstobs{$\mathcal{G}^{M}_{M}$}{$+101$} \\
300~mW & \worstobs{$\langle c_j\rangle$}{$+80$} & \worstobs{$\mathcal{G}^{M}_{M/2,M/2}$}{$+513$} & \worstobs{$\mathcal{G}^{M}_{M}$}{$+56$} & \worstobs{$\mathcal{G}^{M}_{M}$}{$+60$} \\
600~mW & \worstobs{$\langle c_j\rangle$}{$+196$} & \worstobs{$\mathcal{G}^{M}_{M}$}{$+367$}; \worstobs{$\mathcal{G}^{M}_{M/2,M/2}$}{$+367$} & \worstobs{$\mathcal{G}^{M}_{M}$}{$+23$} & \worstobs{$\mathcal{G}^{M}_{M}$}{$+31$} \\
1000~mW & \worstobs{$\langle c_j\rangle$}{$+420$} & \worstobs{$\mathcal{G}^{M}_{M}$}{$+249$} & \worstobs{$\mathcal{G}^{M}_{M}$}{$+13$} & \worstobs{$\mathcal{G}^{M}_{M}$}{$+18$} \\
1650~mW & \worstobs{$\langle c_j\rangle$}{$+796$} & \worstobs{$\langle c_j\rangle$}{$+160$} & \worstobs{$\mathcal{M}^{4}$}{$-8.7$}; \worstobs{$\langle c_j\rangle$}{$-7.9$}; \worstobs{$\mathcal{M}^{3}$}{$-7.5$} & \worstobs{$\mathcal{G}^{M}_{M}$}{$+7.7$} \\
\hline
\multicolumn{5}{l}{\emph{Corrected ground truth}} \\
150~mW & \worstobs{$\langle c_j\rangle$}{$+91$} & \worstobs{$\mathcal{G}^{M}_{M/2,M/2}$}{$+595$} & \worstobs{$\mathcal{G}^{M}_{M}$}{$+83$} & \worstobs{$\mathcal{G}^{M}_{M}$}{$+94$} \\
300~mW & \worstobs{$\langle c_j\rangle$}{$+74$} & \worstobs{$\mathcal{G}^{M}_{M/2,M/2}$}{$+491$} & \worstobs{$\mathcal{G}^{M}_{M}$}{$+55$} & \worstobs{$\mathcal{G}^{M}_{M}$}{$+56$} \\
600~mW & \worstobs{$\langle c_j\rangle$}{$+191$} & \worstobs{$\mathcal{G}^{M}_{M}$}{$+327$}; \worstobs{$\mathcal{G}^{M}_{M/2,M/2}$}{$+326$} & \worstobs{$\mathcal{G}^{M}_{M}$}{$+19$} & \worstobs{$\mathcal{G}^{M}_{M}$}{$+25$} \\
1000~mW & \worstobs{$\langle c_j\rangle$}{$+289$} & \worstobs{$\mathcal{G}^{M}_{M}$}{$+194$} & \worstobs{$\mathcal{G}^{M}_{M}$}{$+11$}; \worstobs{$\langle c_j\rangle$}{$-9.8$} & \worstobs{$\mathcal{G}^{M}_{M}$}{$+11$} \\
1650~mW & \worstobs{$\langle c_j\rangle$}{$+531$} & \worstobs{$\langle c_j\rangle$}{$+84$} & \worstobs{$\langle c_j\rangle$}{$-9.5$}; \worstobs{$\mathcal{M}^{4}$}{$-6.6$} & \worstobs{$\mathcal{G}^{M}_{M}$}{$+4.3$} \\
\hline
\end{tabular}
\end{heretable}

\medskip

\begin{herefigure}
\centering
\includegraphics[width=\textwidth]{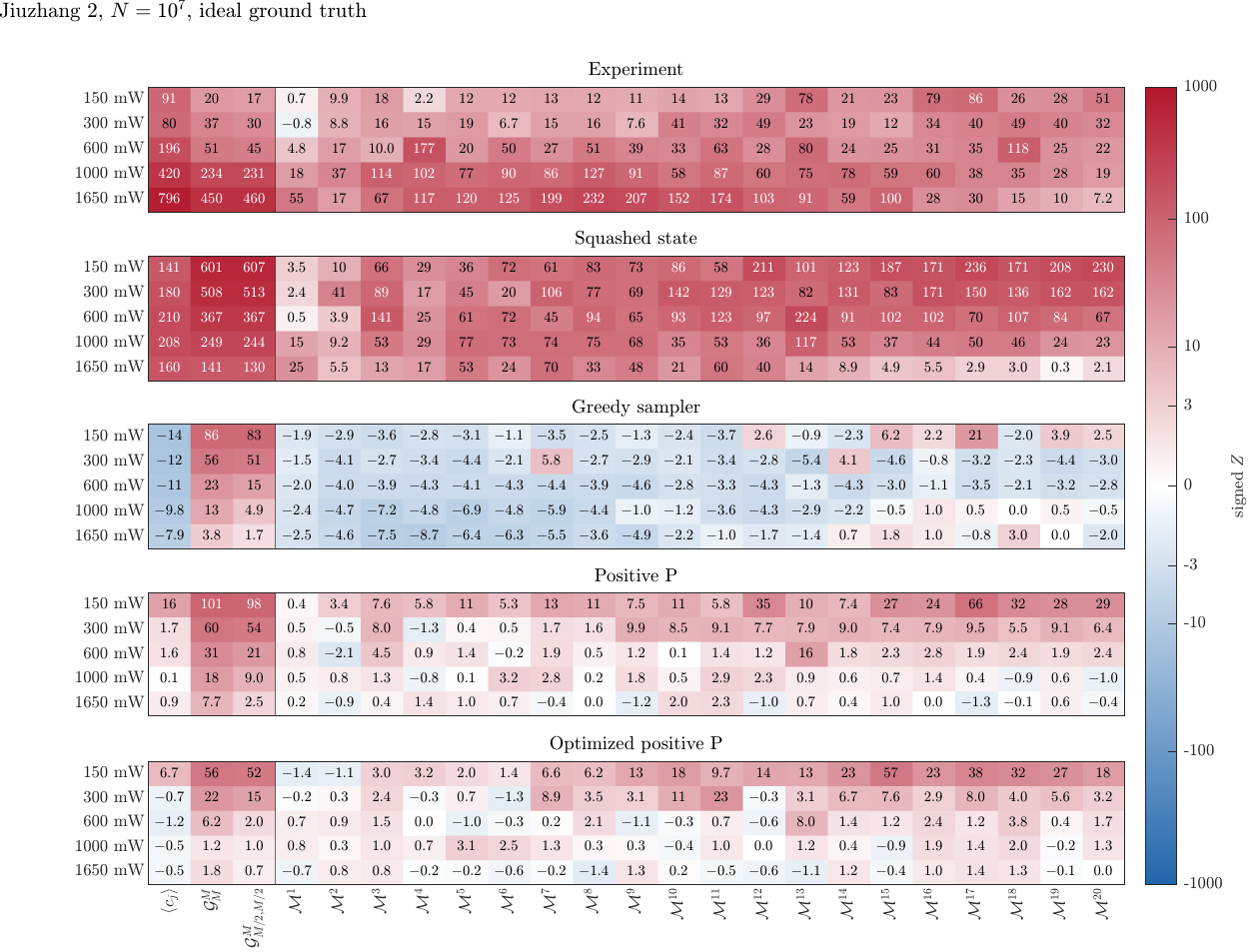}
\captionof{figure}{Signed $Z$-score of every observable for \textit{Jiuzhang~2} with $N=10^{7}$, against the ideal ground truth. Observables are the per-mode click probability $\langle c_j\rangle$, the total-count GCD $\mathcal{G}^{M}_{M}$, the two-dimensional GCD $\mathcal{G}^{M}_{M/2,M/2}$, and the $\mathcal{K}$th-order marginal distributions $\mathcal{M}^{\mathcal{K}}$ defined in the End Matter. Color is logarithmic in $|Z|$; {negative $Z$ indicates underdispersion relative to the expected sampling scatter}. Gray cells have too few bins to test. The bottom block is the optimized positive-$P$ sampler of Sec.~\ref{sec:diagonal}.\label{fig:obs_J2_1e7_ID}}\par\kern0pt
\end{herefigure}

\medskip

\begin{herefigure}
\centering
\includegraphics[width=\textwidth,height=0.59\textheight,keepaspectratio]{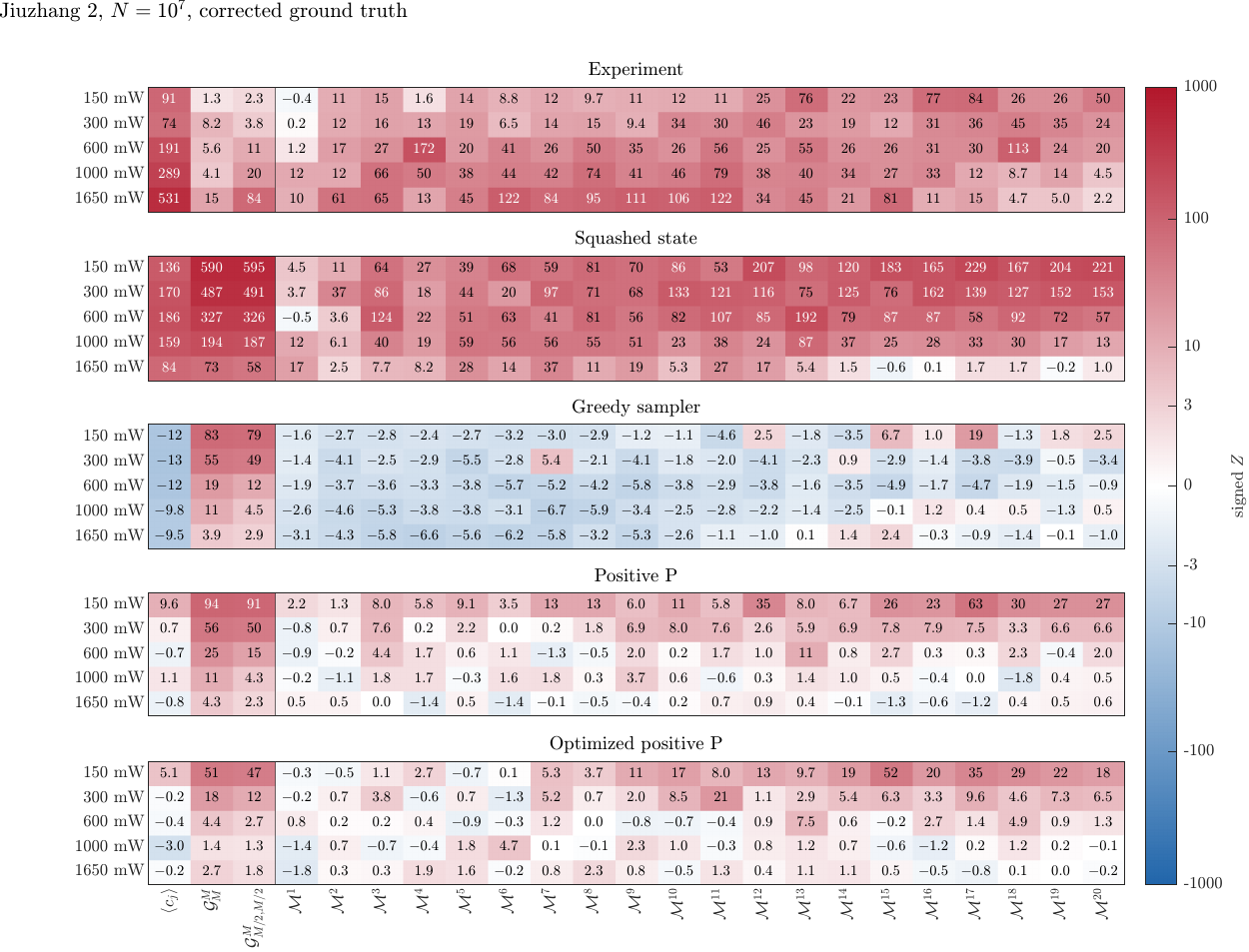}
\captionof{figure}{As Fig.~\ref{fig:obs_J2_1e7_ID}, against the corrected ground truth, including the optimized positive-$P$ sampler. The correction is fitted to the experimental $\mathcal{G}^{M}_{M}$, so for the experiment this is not a test.\label{fig:obs_J2_1e7_TD}}\par\kern0pt
\end{herefigure}

\clearpage

\begin{herefigure}
\centering
\includegraphics[width=\textwidth,height=0.71\textheight,keepaspectratio]{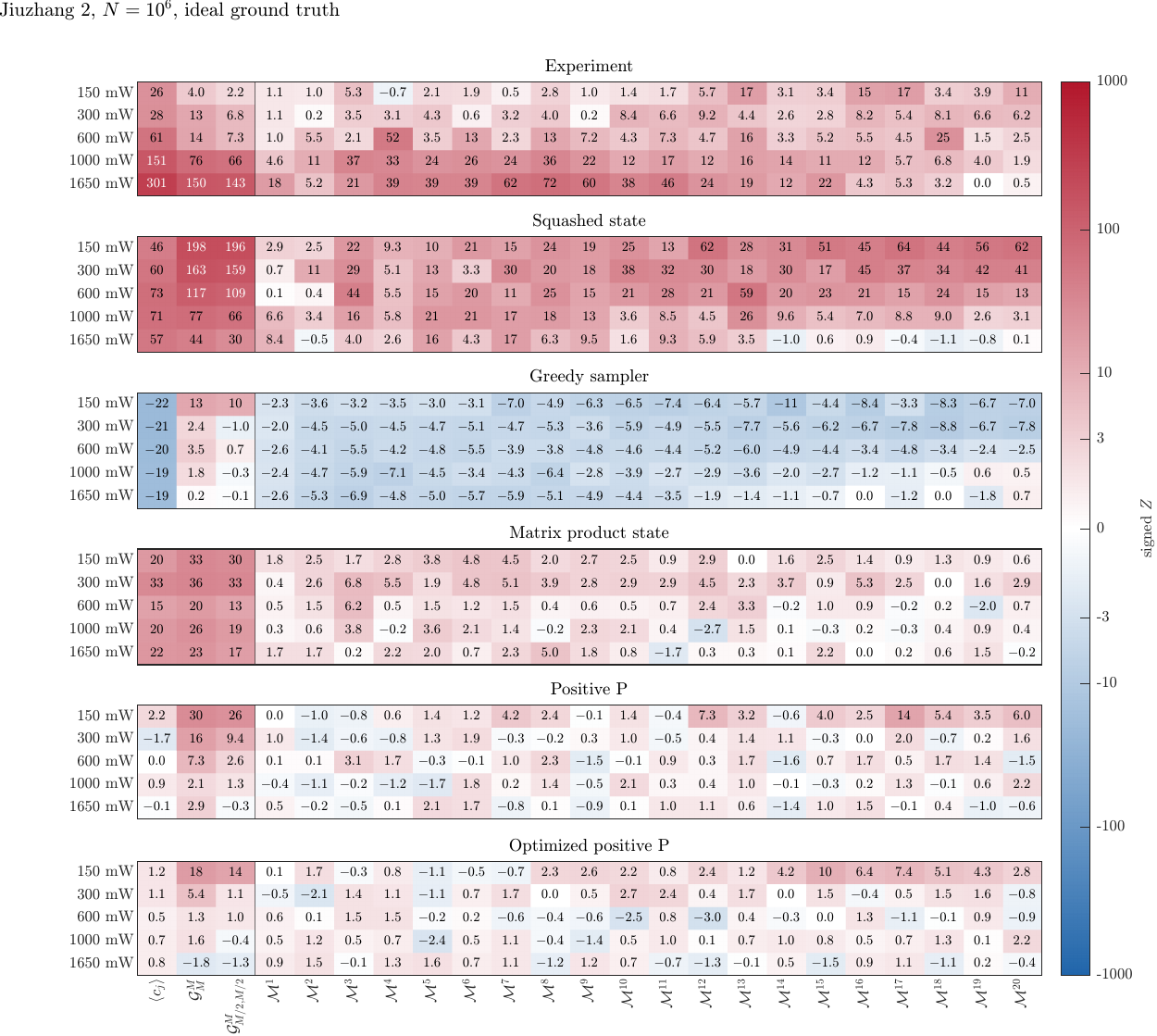}
\captionof{figure}{As Fig.~\ref{fig:obs_J2_1e7_ID}, with $N=10^{6}$, including the matrix-product-state sampler. The bottom block is a random subset of $10^{6}$ of the optimized positive-$P$ sampler's samples.\label{fig:obs_J2_1e6_ID}}\par\kern0pt
\end{herefigure}

\medskip

\begin{heretable}
\centering
\captionof{table}{As Table~\ref{tab:worst_J2_1e7}, for \textit{Jiuzhang~2} with $N=10^{6}$, against the ideal ground truth. ``Optimized'' is a random subset of $10^{6}$ of the optimized positive-$P$ sampler's $10^{7}$ samples (Sec.~\ref{sec:diagonal}). The greedy samples are a separate run of $10^{6}$.\label{tab:worst_J2_1e6}}\par\kern0pt
\setlength{\tabcolsep}{5.5pt}
\renewcommand{\arraystretch}{1.25}
\begin{tabular}{lcccccc}
\hline
\textbf{Dataset} & \textbf{Experiment} & \textbf{Squashed} & \textbf{Greedy} & \textbf{MPS} & \textbf{Positive-$P$} & \textbf{Optimized} \\
\hline
150~mW & \worstobs{$\langle c_j\rangle$}{$+26$} & \worstobs{$\mathcal{G}^{M}_{M}$}{$+198$}; \worstobs{$\mathcal{G}^{M}_{M/2,M/2}$}{$+196$} & \worstobs{$\langle c_j\rangle$}{$-22$} & \worstobs{$\mathcal{G}^{M}_{M}$}{$+33$} & \worstobs{$\mathcal{G}^{M}_{M}$}{$+30$} & \worstobs{$\mathcal{G}^{M}_{M}$}{$+18$} \\
300~mW & \worstobs{$\langle c_j\rangle$}{$+28$} & \worstobs{$\mathcal{G}^{M}_{M}$}{$+163$} & \worstobs{$\langle c_j\rangle$}{$-21$} & \worstobs{$\mathcal{G}^{M}_{M}$}{$+36$}; \worstobs{$\mathcal{G}^{M}_{M/2,M/2}$}{$+33$}; \worstobs{$\langle c_j\rangle$}{$+33$} & \worstobs{$\mathcal{G}^{M}_{M}$}{$+16$} & \worstobs{$\mathcal{G}^{M}_{M}$}{$+5.4$} \\
600~mW & \worstobs{$\langle c_j\rangle$}{$+62$} & \worstobs{$\mathcal{G}^{M}_{M}$}{$+117$} & \worstobs{$\langle c_j\rangle$}{$-20$} & \worstobs{$\mathcal{G}^{M}_{M}$}{$+20$} & \worstobs{$\mathcal{G}^{M}_{M}$}{$+7.3$} & \worstobs{$\mathcal{M}^{12}$}{$-3.0$} \\
1000~mW & \worstobs{$\langle c_j\rangle$}{$+151$} & \worstobs{$\mathcal{G}^{M}_{M}$}{$+77$} & \worstobs{$\langle c_j\rangle$}{$-19$} & \worstobs{$\mathcal{G}^{M}_{M}$}{$+26$} & \worstobs{---}{2.2} & \worstobs{---}{2.4} \\
1650~mW & \worstobs{$\langle c_j\rangle$}{$+301$} & \worstobs{$\langle c_j\rangle$}{$+57$} & \worstobs{$\langle c_j\rangle$}{$-19$} & \worstobs{$\mathcal{G}^{M}_{M}$}{$+23$}; \worstobs{$\langle c_j\rangle$}{$+22$} & \worstobs{---}{2.9} & \worstobs{---}{1.8} \\
\hline
\end{tabular}
\end{heretable}

\bigskip

\begin{herefigure}
\centering
\includegraphics[width=\textwidth]{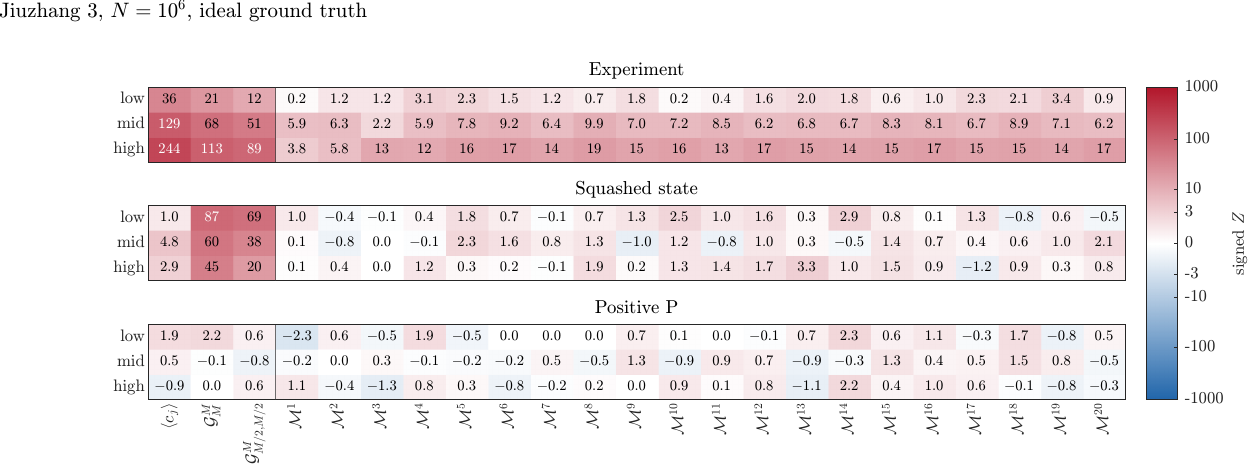}
\captionof{figure}{As Fig.~\ref{fig:obs_J2_1e7_ID}, for \textit{Jiuzhang~3}, with $N=10^{6}$.\label{fig:obs_J3_ID}}\par\kern0pt
\end{herefigure}

\medskip

\begin{herefigure}
\centering
\includegraphics[width=\textwidth]{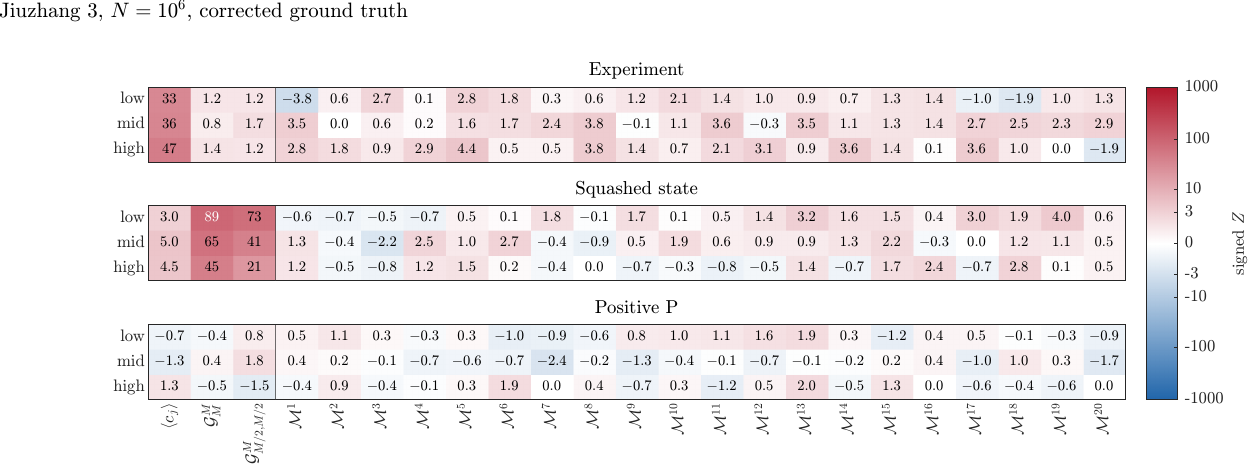}
\captionof{figure}{As Fig.~\ref{fig:obs_J3_ID}, against the corrected ground truth, which includes partial distinguishability. Its $t$ is fitted to the mean of the experimental $\mathcal{G}^{M}_{M}$, so for the experiment that observable is only a partial test.\label{fig:obs_J3_T}}\par\kern0pt
\end{herefigure}

\medskip

\begin{heretable}
\centering
\captionof{table}{As Table~\ref{tab:worst_J2_1e7}, for \textit{Jiuzhang~3}, with $N=10^{6}$, and for the high-power dataset also with $N=10^{7}$.\label{tab:worst_J3}}\par\kern0pt
\renewcommand{\arraystretch}{1.25}
\begin{tabular}{lccc}
\hline
\textbf{Dataset} & \textbf{Experiment} & \textbf{Squashed} & \textbf{Positive-$P$} \\
\hline
\multicolumn{4}{l}{\emph{Ideal ground truth}} \\
low & \worstobs{$\langle c_j\rangle$}{$+36$} & \worstobs{$\mathcal{G}^{M}_{M}$}{$+87$} & \worstobs{---}{2.3} \\
mid & \worstobs{$\langle c_j\rangle$}{$+129$} & \worstobs{$\mathcal{G}^{M}_{M}$}{$+60$} & \worstobs{---}{1.5} \\
high & \worstobs{$\langle c_j\rangle$}{$+244$} & \worstobs{$\mathcal{G}^{M}_{M}$}{$+45$} & \worstobs{---}{2.2} \\
high, $N=10^{7}$ & \worstobs{$\langle c_j\rangle$}{$+776$} & \worstobs{$\mathcal{G}^{M}_{M}$}{$+152$} & \worstobs{---}{2.1} \\
\hline
\multicolumn{4}{l}{\emph{Corrected ground truth}} \\
low & \worstobs{$\langle c_j\rangle$}{$+33$} & \worstobs{$\mathcal{G}^{M}_{M}$}{$+89$} & \worstobs{---}{1.9} \\
mid & \worstobs{$\langle c_j\rangle$}{$+36$} & \worstobs{$\mathcal{G}^{M}_{M}$}{$+65$} & \worstobs{---}{2.4} \\
high & \worstobs{$\langle c_j\rangle$}{$+47$} & \worstobs{$\mathcal{G}^{M}_{M}$}{$+45$} & \worstobs{---}{2.0} \\
\hline
\end{tabular}
\end{heretable}

\medskip

\begin{herefigure}
\centering
\includegraphics[width=\textwidth]{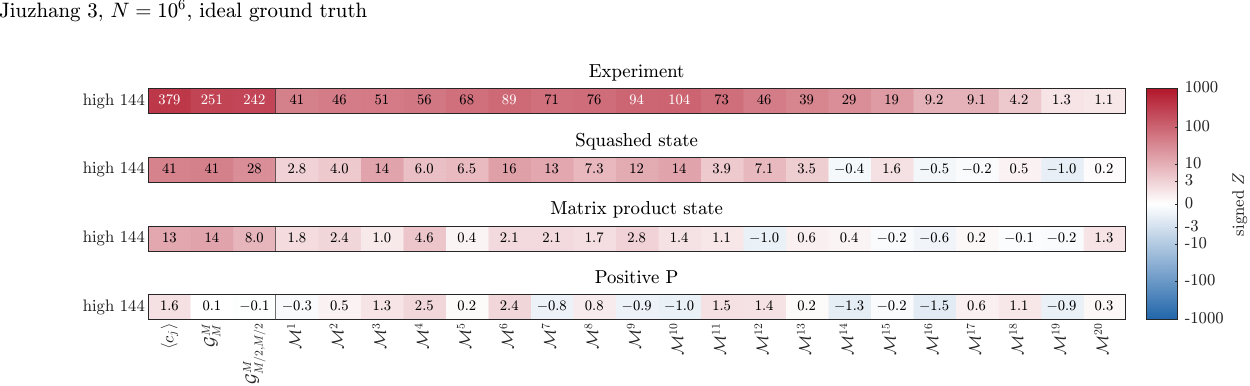}
\captionof{figure}{As Fig.~\ref{fig:obs_J2_1e7_ID}, with $N=10^{6}$, for the 144-mode version of the high-power \textit{Jiuzhang~3} dataset (Table~\ref{tab:worst_J3_144}), including the MPS sampler.\label{fig:obs_J3_144}}\par\kern0pt
\end{herefigure}

\medskip

\begin{heretable}
\centering
\captionof{table}{As Table~\ref{tab:worst_J2_1e7}, for the 144-mode version of the high-power \textit{Jiuzhang~3} dataset, in which each group of eight consecutive output modes is merged into one mode that clicks if any of its eight detectors clicks. Its ground truth replaces each group's eight rows of the transmission matrix by the single row that best reproduces the group's no-click probability: the leading right singular vector of the group's $8\times50$ block, scaled by its singular value. It is scored against the ideal ground truth, with $N=10^{6}$ and $N=10^{7}$. This is the only version of \textit{Jiuzhang~3} for which MPS samples are available, and only for $N=10^{6}$. They target the same 144-mode state as our ground truth (Sec.~S9 of the Supplementary Information of Ref.~\cite{oh2024classical}).\label{tab:worst_J3_144}}\par\kern0pt
\renewcommand{\arraystretch}{1.25}
\begin{tabular}{lcccc}
\hline
\textbf{Dataset} & \textbf{Experiment} & \textbf{Squashed} & \textbf{MPS} & \textbf{Positive-$P$} \\
\hline
$N=10^{6}$ & \worstobs{$\langle c_j\rangle$}{$+379$} & \worstobs{$\mathcal{G}^{M}_{M}$}{$+41$} & \worstobs{$\mathcal{G}^{M}_{M}$}{$+14$} & \worstobs{---}{2.5} \\
$N=10^{7}$ & \worstobs{$\langle c_j\rangle$}{$+1030$} & \worstobs{$\mathcal{G}^{M}_{M}$}{$+143$} & & \worstobs{$\mathcal{G}^{M}_{M}$}{$+9.4$} \\
\hline
\end{tabular}
\end{heretable}

\medskip

\begin{herefigure}
\centering
\includegraphics[width=\textwidth]{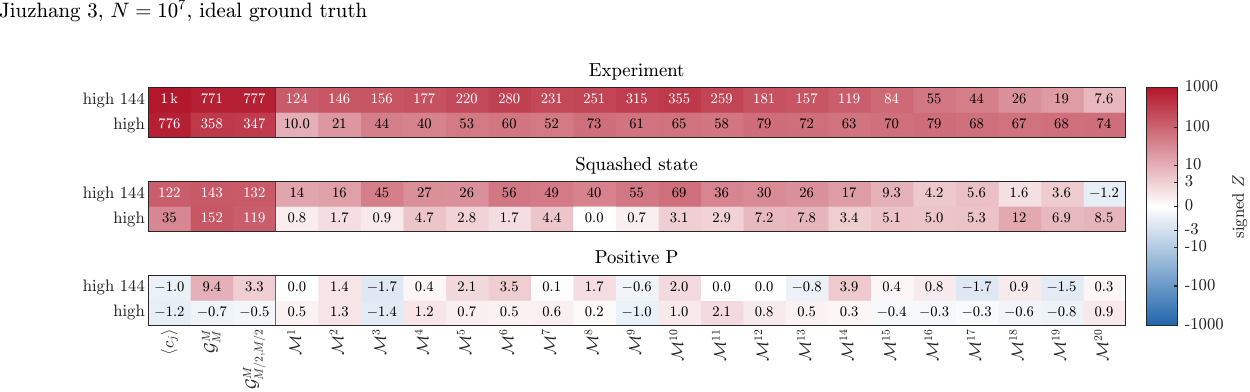}
\captionof{figure}{As Fig.~\ref{fig:obs_J2_1e7_ID}, for the high-power \textit{Jiuzhang~3} dataset with $N=10^{7}$, against the ideal ground truth: the full 1152-mode dataset (high) and its 144-mode version (high 144). The worst observables are in Tables~\ref{tab:worst_J3} and~\ref{tab:worst_J3_144}. There are no MPS samples at this $N$.\label{fig:obs_J3_1e7}}\par\kern0pt
\end{herefigure}

\bigskip

\begin{herefigure}
\centering
\includegraphics[width=\textwidth]{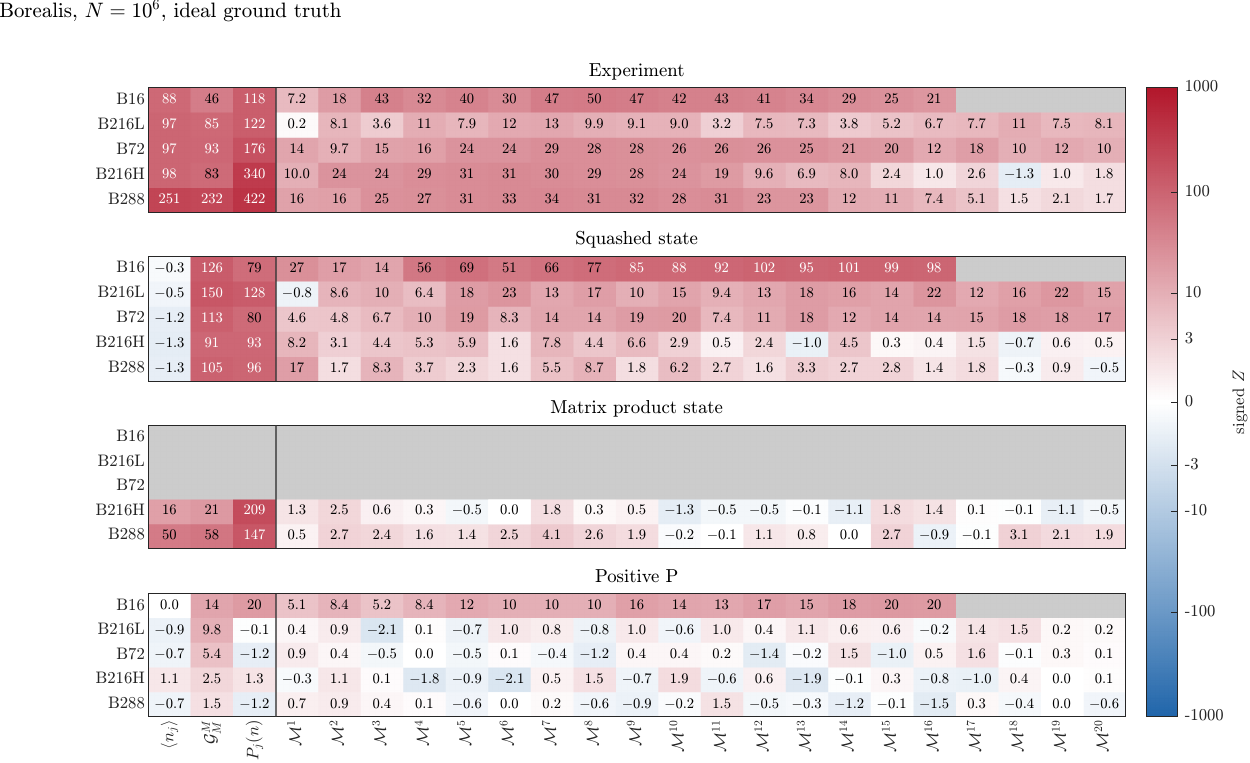}
\captionof{figure}{As Fig.~\ref{fig:obs_J2_1e7_ID}, for Borealis, with $N=10^{6}$. $\langle n_j\rangle$ is the per-mode mean photon number, $P_j(n)$ the per-mode photon-number distribution, and the marginals $\mathcal{M}^{\mathcal{K}}$ are restricted to at most one photon per mode. The positive-$P$ sampler uses the Poisson output step for B16 and the categorical output step (Sec.~\ref{sec:categorical}) for the other datasets. The MPS sampler is shown for B216H and B288, the only Borealis datasets for which its samples are available.\label{fig:obs_B_ID}}\par\kern0pt
\end{herefigure}

\medskip

\begin{herefigure}
\centering
\includegraphics[width=\textwidth]{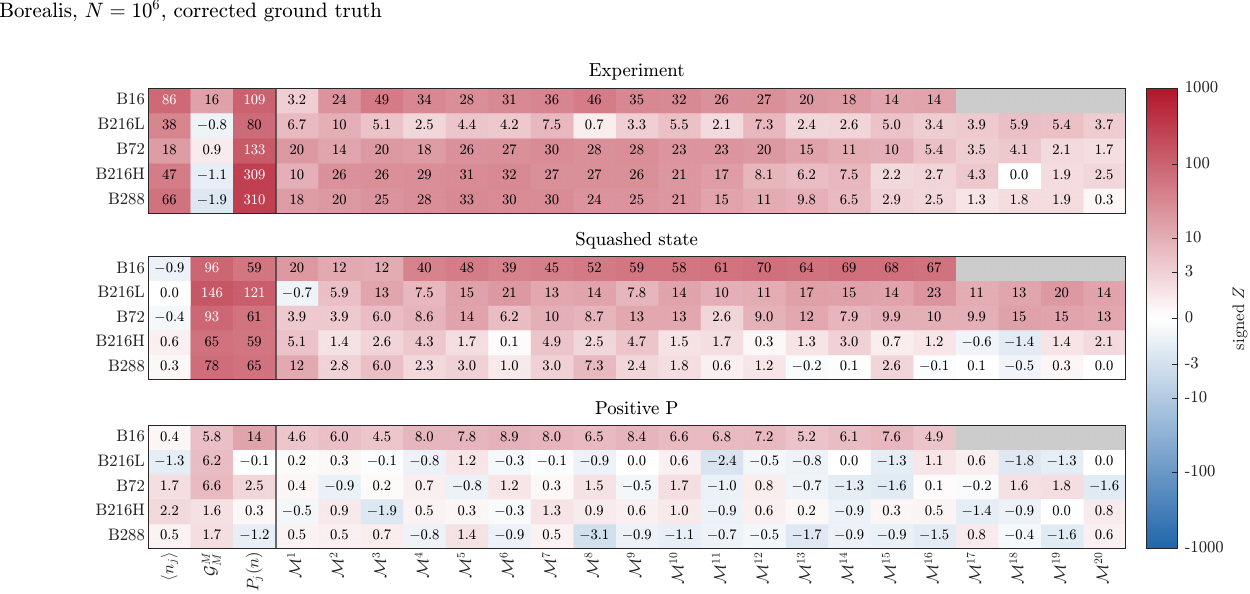}
\captionof{figure}{As Fig.~\ref{fig:obs_B_ID}, against the corrected ground truth, without the MPS sampler. As in Fig.~\ref{fig:obs_J2_1e7_TD}, the experimental $\mathcal{G}^{M}_{M}$ is fitted.\label{fig:obs_B_TD}}\par\kern0pt
\end{herefigure}

\medskip

\begin{heretable}
\centering
\captionof{table}{As Table~\ref{tab:worst_J2_1e7}, for Borealis, with $N=10^{6}$. ``Positive-$P$'' is the sampler shown in Fig.~1: the Poisson output step for B16 and the categorical output step (Sec.~\ref{sec:categorical}) for the other datasets. ``Poisson'' gives the Poisson output step on every dataset for comparison. The MPS sampler was run against the ideal ground truth only. All entries are scored against the independently seeded rerun of the phase-space ground truth. Rows are in order of mean output photon number (Table~\ref{tab:datasets}).\label{tab:worst_B}}\par\kern0pt
\renewcommand{\arraystretch}{1.25}
{
\begin{tabular}{lccccc}
\hline
\textbf{Dataset} & \textbf{Experiment} & \textbf{Squashed} & \textbf{MPS} & \textbf{Positive-$P$} & \textbf{Poisson} \\
\hline
\multicolumn{6}{l}{\emph{Ideal ground truth}} \\
B16 & \worstobs{$P_j(n)$}{$+118$} & \worstobs{$\mathcal{G}^{M}_{M}$}{$+126$} & & \worstobs{$\mathcal{M}^{15}$}{$+20$}; \worstobs{$P_j(n)$}{$+20$}; \worstobs{$\mathcal{M}^{16}$}{$+20$} & \worstobs{$\mathcal{M}^{15}$}{$+20$}; \worstobs{$P_j(n)$}{$+20$}; \worstobs{$\mathcal{M}^{16}$}{$+20$} \\
B216L & \worstobs{$P_j(n)$}{$+122$} & \worstobs{$\mathcal{G}^{M}_{M}$}{$+150$} & & \worstobs{$\mathcal{G}^{M}_{M}$}{$+9.8$} & \worstobs{$P_j(n)$}{$+49$} \\
B72 & \worstobs{$P_j(n)$}{$+176$} & \worstobs{$\mathcal{G}^{M}_{M}$}{$+113$} & & \worstobs{$\mathcal{G}^{M}_{M}$}{$+5.4$} & \worstobs{$P_j(n)$}{$+15$} \\
B216H & \worstobs{$P_j(n)$}{$+340$} & \worstobs{$P_j(n)$}{$+93$}; \worstobs{$\mathcal{G}^{M}_{M}$}{$+91$} & \worstobs{$P_j(n)$}{$+209$} & \worstobs{---}{2.5} & \worstobs{$P_j(n)$}{$+33$} \\
B288 & \worstobs{$P_j(n)$}{$+422$} & \worstobs{$\mathcal{G}^{M}_{M}$}{$+105$} & \worstobs{$P_j(n)$}{$+147$} & \worstobs{---}{1.5} & \worstobs{$P_j(n)$}{$+38$} \\
\hline
\multicolumn{6}{l}{\emph{Corrected ground truth}} \\
B16 & \worstobs{$P_j(n)$}{$+109$} & \worstobs{$\mathcal{G}^{M}_{M}$}{$+96$} & & \worstobs{$P_j(n)$}{$+14$} & \worstobs{$P_j(n)$}{$+14$} \\
B216L & \worstobs{$P_j(n)$}{$+80$} & \worstobs{$\mathcal{G}^{M}_{M}$}{$+146$} & & \worstobs{$\mathcal{G}^{M}_{M}$}{$+6.2$} & \worstobs{$P_j(n)$}{$+43$} \\
B72 & \worstobs{$P_j(n)$}{$+133$} & \worstobs{$\mathcal{G}^{M}_{M}$}{$+93$} & & \worstobs{$\mathcal{G}^{M}_{M}$}{$+6.6$} & \worstobs{$P_j(n)$}{$+9.7$} \\
B216H & \worstobs{$P_j(n)$}{$+309$} & \worstobs{$\mathcal{G}^{M}_{M}$}{$+65$} & & \worstobs{---}{2.2} & \worstobs{$P_j(n)$}{$+21$} \\
B288 & \worstobs{$P_j(n)$}{$+310$} & \worstobs{$\mathcal{G}^{M}_{M}$}{$+78$} & & \worstobs{$\mathcal{M}^{8}$}{$-3.1$} & \worstobs{$P_j(n)$}{$+20$} \\
\hline
\end{tabular}}
\end{heretable}

\clearpage

\iflatexml\else
\noindent\begin{minipage}{\linewidth}
\centering
\includegraphics[width=\textwidth,height=0.42\textheight,keepaspectratio]{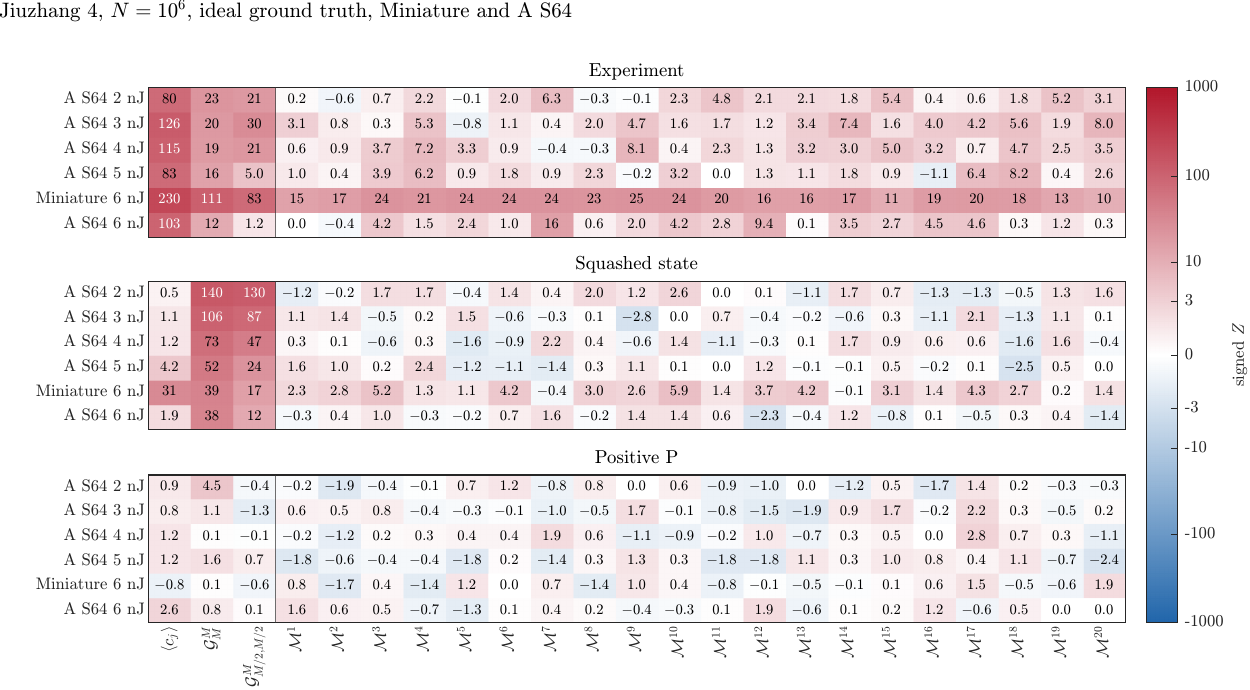}\par\medskip
\includegraphics[width=\textwidth,height=0.42\textheight,keepaspectratio]{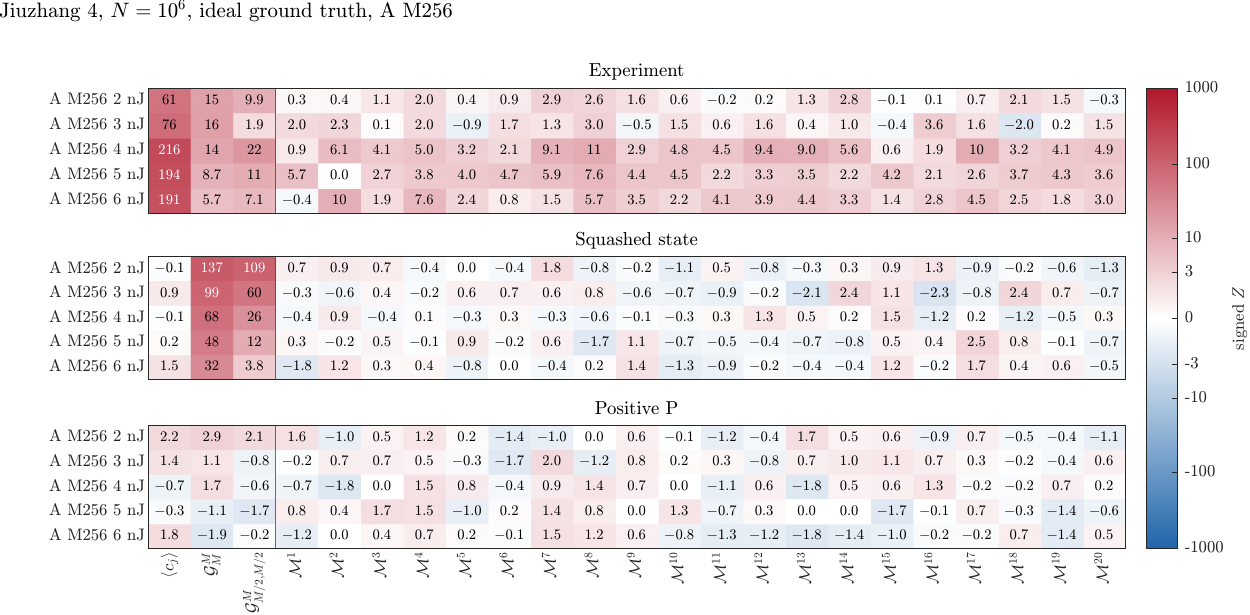}\par\smallskip
{\small Figure~\ref{fig:obs_J4}, first of two parts; the caption follows the second part.}
\end{minipage}\par

\medskip

\fi
\begin{herefigure}
\centering
\iflatexml
\includegraphics[width=\textwidth,height=0.42\textheight,keepaspectratio]{Figures/observables_J4_N1e6_ideal_A_S64.pdf}\par\medskip
\includegraphics[width=\textwidth,height=0.42\textheight,keepaspectratio]{Figures/observables_J4_N1e6_ideal_A_M256.pdf}\par\smallskip
\fi
\includegraphics[width=\textwidth,height=0.42\textheight,keepaspectratio]{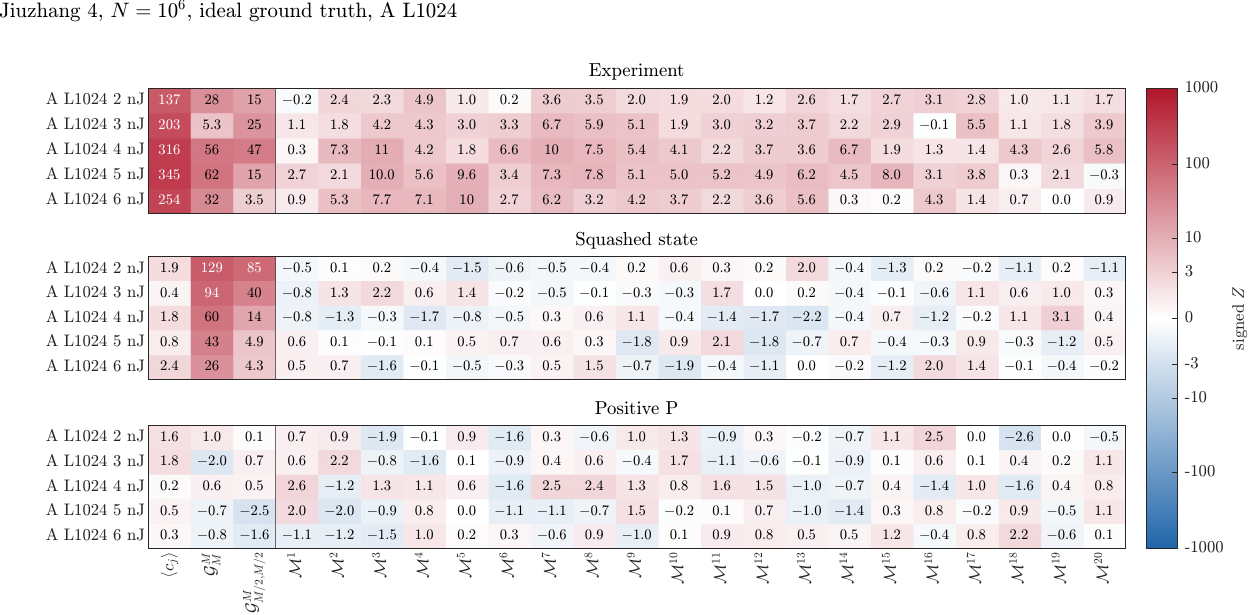}\par\medskip
\includegraphics[width=\textwidth,height=0.42\textheight,keepaspectratio]{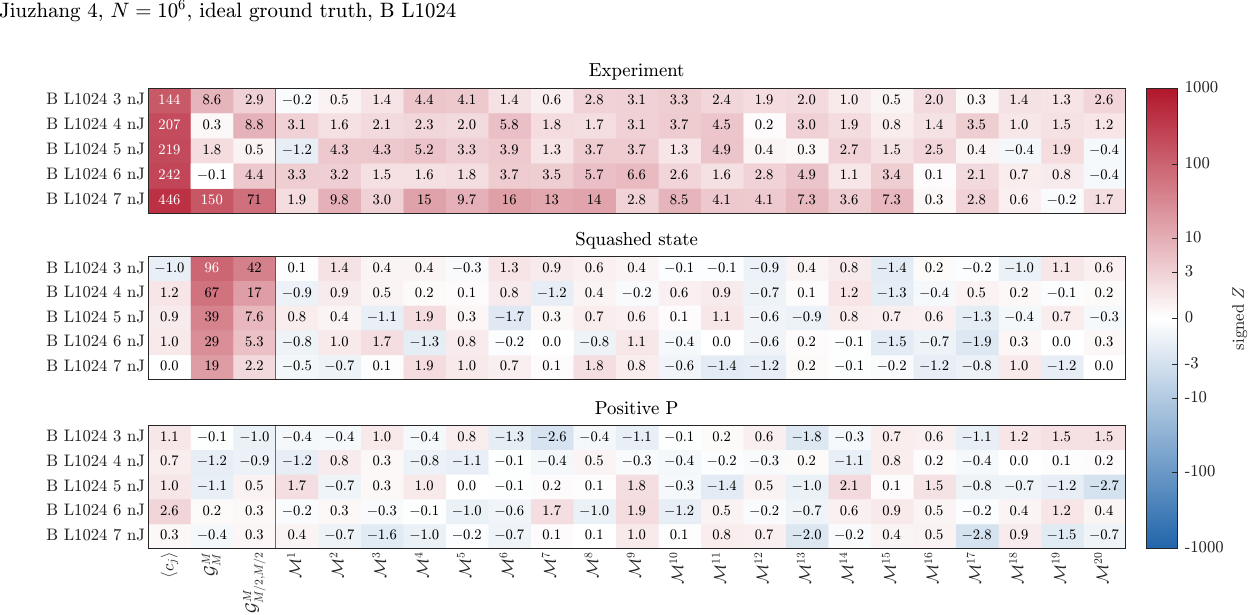}
\captionof{figure}{As Fig.~\ref{fig:obs_J2_1e7_ID}, for \textit{Jiuzhang~4}, with $N=10^{6}$. Each heat map is one configuration, named in its header (A S64, A M256, A L1024 and B L1024); the Miniature dataset is shown with A S64.\label{fig:obs_J4}}\par\kern0pt
\end{herefigure}

\medskip

\begin{heretable}
\centering
\captionof{table}{As Table~\ref{tab:worst_J2_1e7}, for \textit{Jiuzhang~4}, with $N=10^{6}$, against the ideal ground truth. Rows are in order of mean output photon number (Table~\ref{tab:datasets}). For B L1024 7~nJ the ideal squeezing parameters are inferred (Sec.~\ref{sec:datasets_params}); with the two alternatives given there, the experiment's $\langle c_j\rangle$ is $+387$ or $+364$.\label{tab:worst_J4}}\par\kern0pt
\renewcommand{\arraystretch}{1.25}
\begin{tabular}{lccc}
\hline
\textbf{Dataset} & \textbf{Experiment} & \textbf{Squashed} & \textbf{Positive-$P$} \\
\hline
A S64 2~nJ & \worstobs{$\langle c_j\rangle$}{$+80$} & \worstobs{$\mathcal{G}^{M}_{M}$}{$+140$} & \worstobs{$\mathcal{G}^{M}_{M}$}{$+4.5$} \\
A S64 3~nJ & \worstobs{$\langle c_j\rangle$}{$+126$} & \worstobs{$\mathcal{G}^{M}_{M}$}{$+106$} & \worstobs{---}{2.2} \\
A S64 4~nJ & \worstobs{$\langle c_j\rangle$}{$+115$} & \worstobs{$\mathcal{G}^{M}_{M}$}{$+73$} & \worstobs{---}{2.8} \\
A M256 2~nJ & \worstobs{$\langle c_j\rangle$}{$+61$} & \worstobs{$\mathcal{G}^{M}_{M}$}{$+137$} & \worstobs{---}{2.9} \\
A S64 5~nJ & \worstobs{$\langle c_j\rangle$}{$+83$} & \worstobs{$\mathcal{G}^{M}_{M}$}{$+52$} & \worstobs{---}{2.4} \\
Miniature 6~nJ & \worstobs{$\langle c_j\rangle$}{$+230$} & \worstobs{$\mathcal{G}^{M}_{M}$}{$+39$} & \worstobs{---}{1.9} \\
A S64 6~nJ & \worstobs{$\langle c_j\rangle$}{$+103$} & \worstobs{$\mathcal{G}^{M}_{M}$}{$+38$} & \worstobs{---}{2.6} \\
A M256 3~nJ & \worstobs{$\langle c_j\rangle$}{$+76$} & \worstobs{$\mathcal{G}^{M}_{M}$}{$+99$} & \worstobs{---}{2.0} \\
A M256 4~nJ & \worstobs{$\langle c_j\rangle$}{$+216$} & \worstobs{$\mathcal{G}^{M}_{M}$}{$+68$} & \worstobs{---}{1.8} \\
A L1024 2~nJ & \worstobs{$\langle c_j\rangle$}{$+137$} & \worstobs{$\mathcal{G}^{M}_{M}$}{$+129$} & \worstobs{---}{2.6} \\
A M256 5~nJ & \worstobs{$\langle c_j\rangle$}{$+194$} & \worstobs{$\mathcal{G}^{M}_{M}$}{$+48$} & \worstobs{---}{1.7} \\
A M256 6~nJ & \worstobs{$\langle c_j\rangle$}{$+191$} & \worstobs{$\mathcal{G}^{M}_{M}$}{$+32$} & \worstobs{---}{1.9} \\
B L1024 3~nJ & \worstobs{$\langle c_j\rangle$}{$+144$} & \worstobs{$\mathcal{G}^{M}_{M}$}{$+96$} & \worstobs{---}{2.6} \\
A L1024 3~nJ & \worstobs{$\langle c_j\rangle$}{$+203$} & \worstobs{$\mathcal{G}^{M}_{M}$}{$+94$} & \worstobs{---}{2.2} \\
B L1024 4~nJ & \worstobs{$\langle c_j\rangle$}{$+207$} & \worstobs{$\mathcal{G}^{M}_{M}$}{$+67$} & \worstobs{---}{1.2} \\
A L1024 4~nJ & \worstobs{$\langle c_j\rangle$}{$+316$} & \worstobs{$\mathcal{G}^{M}_{M}$}{$+60$} & \worstobs{---}{2.6} \\
B L1024 5~nJ & \worstobs{$\langle c_j\rangle$}{$+219$} & \worstobs{$\mathcal{G}^{M}_{M}$}{$+39$} & \worstobs{---}{2.7} \\
A L1024 5~nJ & \worstobs{$\langle c_j\rangle$}{$+345$} & \worstobs{$\mathcal{G}^{M}_{M}$}{$+43$} & \worstobs{---}{2.5} \\
B L1024 6~nJ & \worstobs{$\langle c_j\rangle$}{$+242$} & \worstobs{$\mathcal{G}^{M}_{M}$}{$+29$} & \worstobs{---}{2.6} \\
A L1024 6~nJ & \worstobs{$\langle c_j\rangle$}{$+254$} & \worstobs{$\mathcal{G}^{M}_{M}$}{$+26$} & \worstobs{---}{2.2} \\
B L1024 7~nJ & \worstobs{$\langle c_j\rangle$}{$+446$} & \worstobs{$\mathcal{G}^{M}_{M}$}{$+19$} & \worstobs{---}{2.8} \\
\hline
\end{tabular}
\end{heretable}

\begin{figurepair}
\begin{herefigure}
\centering
\includegraphics[width=\textwidth]{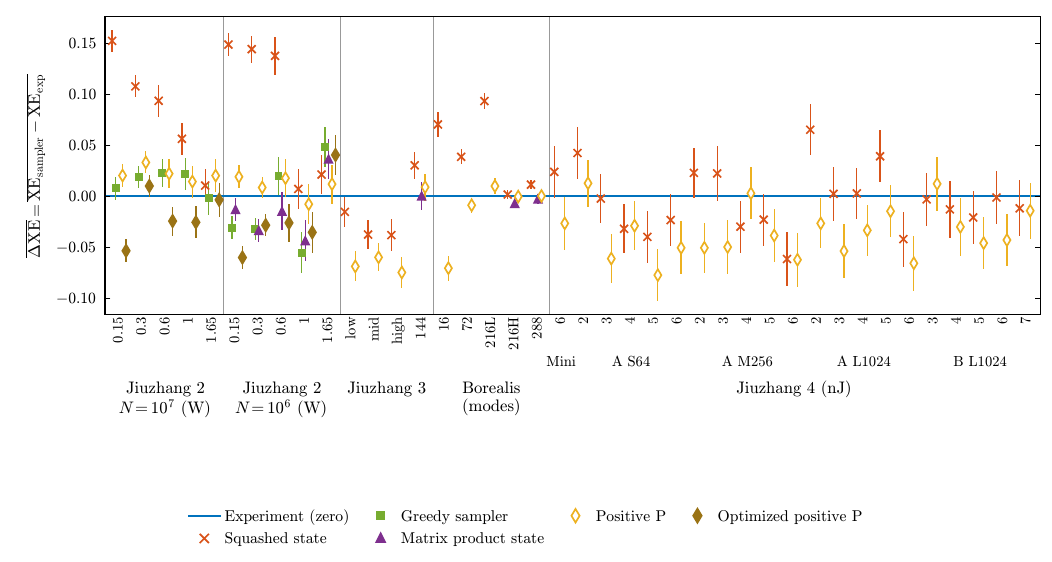}
\caption{Cross-entropy difference of each sampler relative to the experimental samples, ideal ground truth, averaged over fixed-count sectors with inverse-variance weights, Eq.~\eqref{eq:XE_mean}. Error bars are one propagated standard error. A positive value means the sampler's samples have lower mean log-probability under the ground truth than the experimental samples. \textit{Jiuzhang~2} is shown with $N=10^{7}$ and with $N=10^{6}$, the comparison that includes the MPS sampler; the others use $10^{6}$. Each eligible sector uses 1000 randomly selected samples. For Borealis the positive-$P$ sampler uses the same output step as in Fig.~1. The MPS sampler is shown for \textit{Jiuzhang~2} with $N=10^{6}$, the 144-mode \textit{Jiuzhang~3} dataset, and Borealis B216H and B288. \textit{Jiuzhang} sectors are click numbers $n\leq15$, with 256-bit Torontonians; Borealis sectors are total photon numbers $n=10$--$15$, with double-precision hafnians.\label{fig:xe_ID}}\par\kern0pt
\end{herefigure}
\begin{herefigure}
\centering
\includegraphics[width=\textwidth]{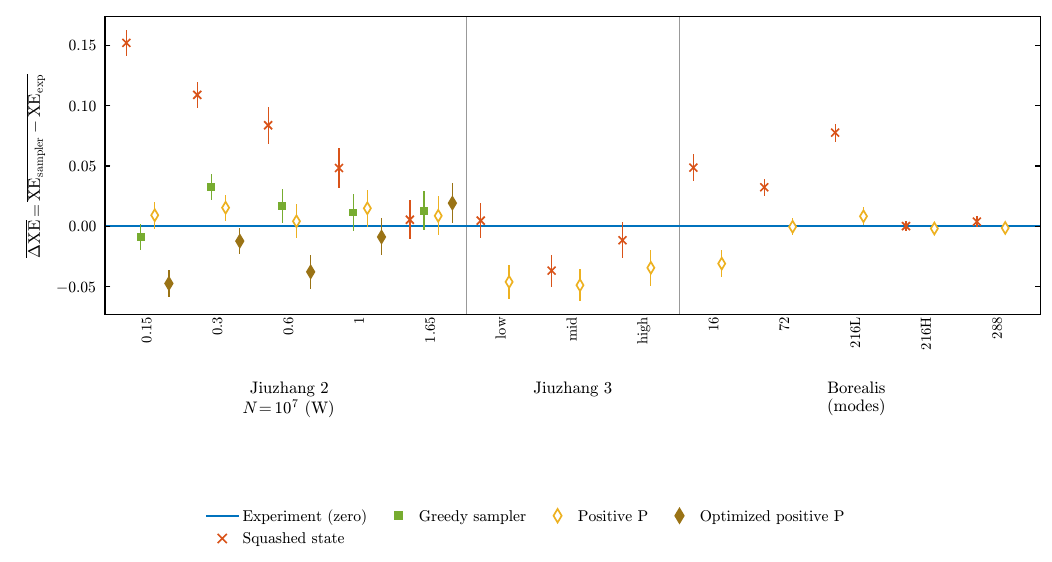}
\caption{As Fig.~\ref{fig:xe_ID}, against the corrected ground truth. \textit{Jiuzhang~4} has no corrected ground truth.\label{fig:xe_TD}}\par\kern0pt
\end{herefigure}
\end{figurepair}

\clearpage

\newcommand{\dKunresolved}{An unresolved one, $|K-1|<2\sigma_K$, meaning the slope cannot be distinguished from one, is drawn as a downward arrow, in the sampler's color, at the bottom of the panel.}
\begin{figure}[!ht]
\centering
\includegraphics[width=\textwidth]{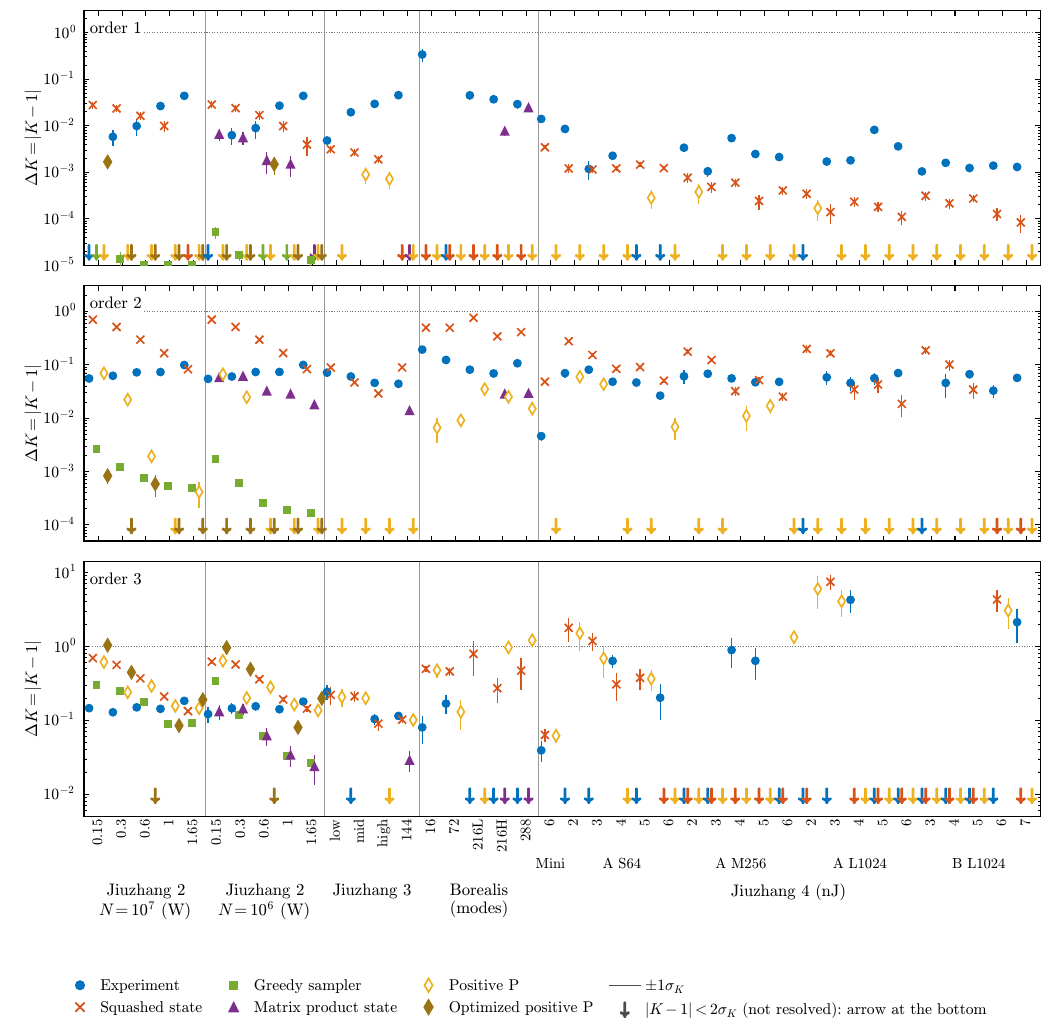}
\caption{Slope error $\Delta K=|K-1|$ of the sampled versus exact cumulants of orders one to three, ideal ground truth. $K$ is the least-squares slope over up to 10\,000 mode sets per order. \textit{Jiuzhang~2} is shown with $N=10^{7}$ and with $N=10^{6}$; the others use $10^{6}$. Borealis uses photon-number cumulants, with the positive-$P$ output step of Fig.~1. A resolved slope error, $|K-1|\ge2\sigma_K$, is shown with $\pm1\sigma_K$ bars. \dKunresolved{} $\sigma_K$ is the least-squares standard error from each sampler's residual scatter. A marker at the top edge with an arrow pointing up is off the scale.\label{fig:dK_ID}}
\end{figure}

\clearpage

\begin{figure}[!ht]
\centering
\includegraphics[width=\textwidth]{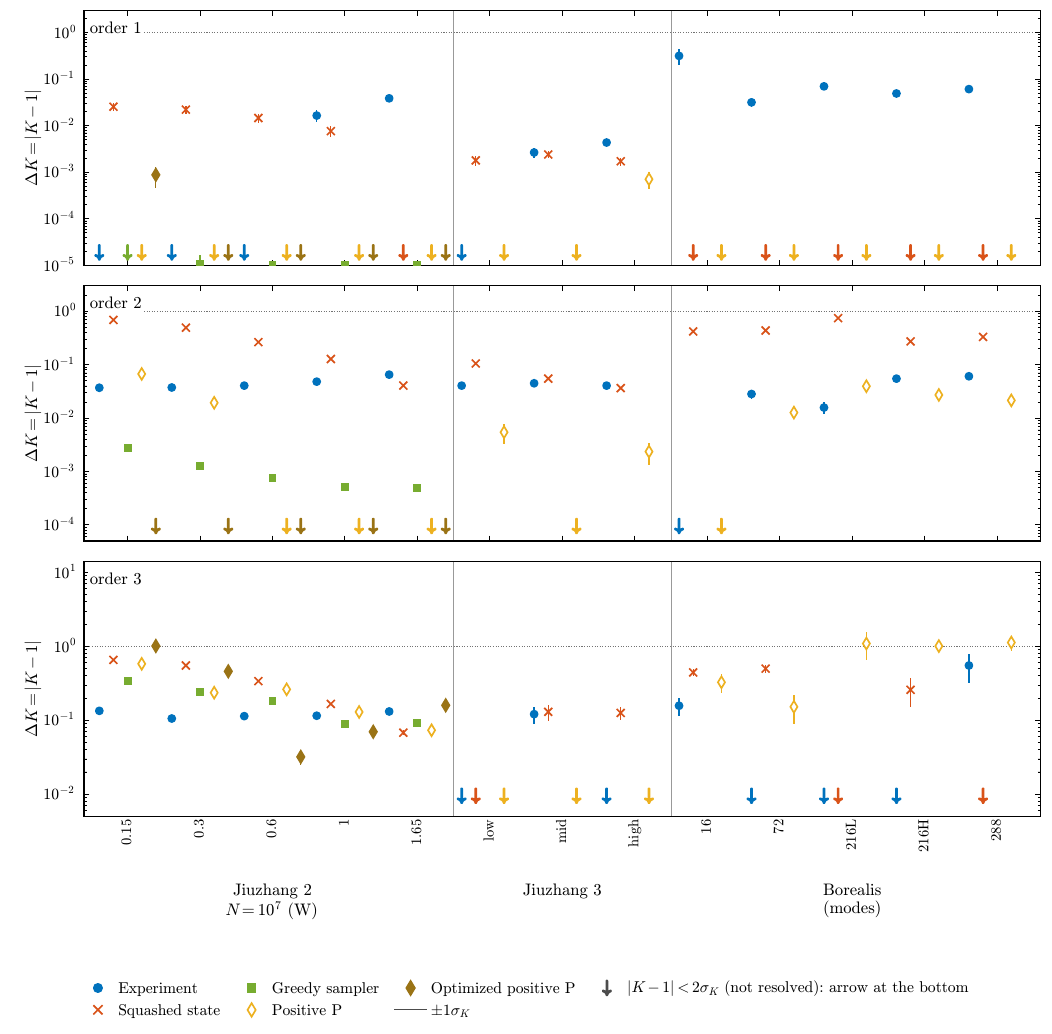}
\caption{As Fig.~\ref{fig:dK_ID}, against the corrected ground truth.\label{fig:dK_TD}}
\end{figure}

\clearpage

\begin{figure}[!ht]
\centering
\includegraphics[width=\textwidth]{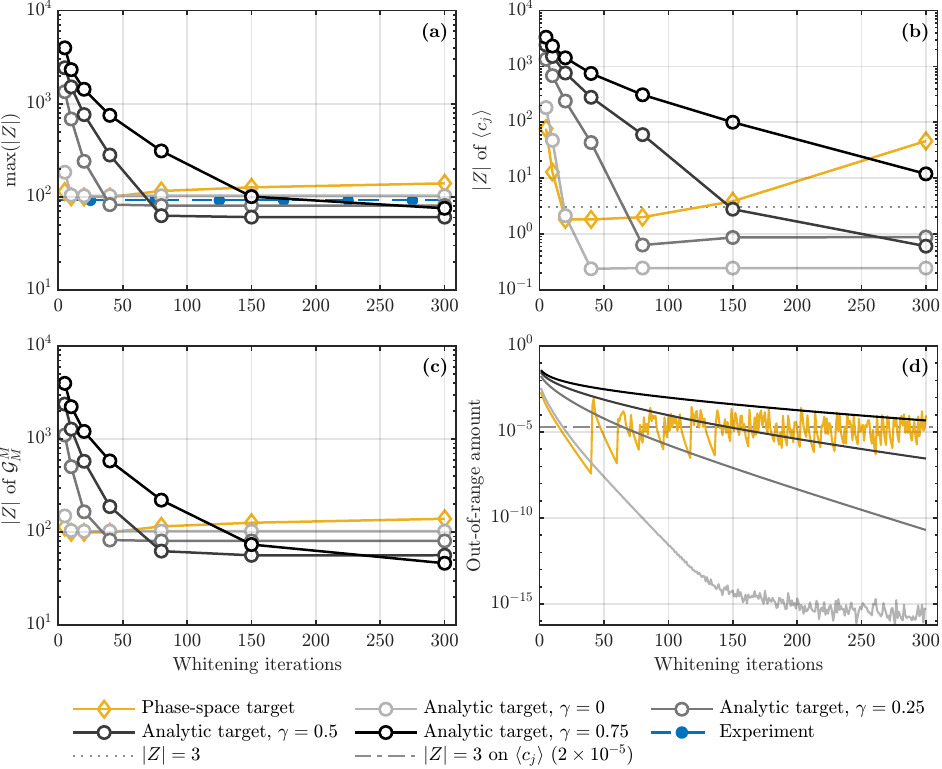}
\caption{Positive-$P$ sampling of \textit{Jiuzhang~2} at $150\,\mathrm{mW}$, with
$N = 10^{7}$ samples, tested against the ideal ground truth, as a function of
the number of whitening iterations. All curves except the experiment are the positive-$P$ sampler, with the whitening target as labeled. The phase-space target is estimated from each batch of trajectories. The analytic target takes the
mean and pairwise covariances of the no-click values $\pi_j$ from the Gaussian
state, with its diagonal set by $\gamma$ between the smallest positive semidefinite value ($\gamma = 0$) and $p_j(0)[1-p_j(0)]$ ($\gamma = 1$).
(a) {$\max|Z|$} over all observables;
(b) $|Z|$ of the per-mode click probabilities $\langle c_j\rangle$;
(c) $|Z|$ of the total click-count distribution $\mathcal{G}_{M}^{M}$;
in (a) to (c), blue is the experiment against the same ground truth.
(d) Out-of-range amount, the mean over samples and modes of the amount by which
$\operatorname{Re}\pi_j$ lies outside $[0,1]$ before the final clamp, averaged
over ten batches of $10^{6}$ trajectories. The dash-dot line marks the level
below which $\langle c_j\rangle$ agrees with the ground truth
($|Z| < 3$) at this $N$.\label{fig:alpha_iter}}
\end{figure}

\clearpage

\begin{heretable}
\centering
\captionof{table}{$\max|Z|$ over all 23 observables for \textit{Jiuzhang~2}, $N=10^{7}$, against the ideal and corrected ground truths, and the worst observable of the optimized positive-$P$ sampler. $w$ is the fraction of samples from the $\gamma=0$ run, chosen separately for each ground truth. The optimized sampler is scored against an independently seeded estimate of each ground truth. Against the corrected ground truth the experiment's $\mathcal{G}^{M}_{M}$ is fitted, so its column is not an independent test.\label{tab:optimised_pp}}\par\kern0pt
\renewcommand{\arraystretch}{1.2}
\setlength{\tabcolsep}{3pt}
\begin{tabular}{lcccccc}
\hline
\textbf{Power} & $w$ & \textbf{Exp.} & \textbf{Greedy} & \textbf{Baseline} & \textbf{Optimized} & \textbf{Worst} \\
\hline
\multicolumn{7}{l}{\emph{Ideal ground truth}} \\
150~mW & 0.20 & 91 & 86 & 101 & 57 & $\mathcal{M}^{15}$ \\
300~mW & 0.70 & 80 & 56 & 60 & 23 & $\mathcal{M}^{11}$ \\
600~mW & 0.85 & 196 & 23 & 31 & 8.0 & $\mathcal{M}^{13}$ \\
1000~mW & 0.85 & 420 & 13 & 18 & 3.1 & $\mathcal{M}^{5}$ \\
1650~mW & 0.80 & 796 & 8.7 & 7.7 & 1.8 & $\mathcal{G}^{M}_{M}$ \\
\hline
\multicolumn{7}{l}{\emph{Corrected ground truth}} \\
150~mW & 0.25 & 91 & 83 & 94 & 52 & $\mathcal{M}^{15}$ \\
300~mW & 0.70 & 74 & 55 & 56 & 21 & $\mathcal{M}^{11}$ \\
600~mW & 0.85 & 191 & 19 & 25 & 7.5 & $\mathcal{M}^{13}$ \\
1000~mW & 0.85 & 289 & 11 & 11 & 4.7 & $\mathcal{M}^{6}$ \\
1650~mW & 0.80 & 531 & 9.5 & 4.3 & 2.7 & $\mathcal{G}^{M}_{M}$ \\
\hline
\end{tabular}
\end{heretable}

\medskip

\begin{heretable}
\centering
\captionof{table}{Asymptotic cost of each sampler. $M$: output modes; $S$: input squeezed states; $n$: detected photons in a sample; $L$: whitening--coloring iterations; $B$: phase-space samples per batch; $\mathcal{K}$: order of the greedy sampler; $\chi$: MPS bond dimension; $d$: MPS local photon-number cutoff; $h$: cost of one hafnian in the MPS construction. Entries marked $\dagger$ follow from the algorithm as described in its paper but are not stated there.\label{tab:scaling}}\par\kern0pt
\renewcommand{\arraystretch}{1.3}
\begin{tabular}{llll}
\hline
\textbf{Sampler} & \textbf{Cost per sample} & \textbf{One-off cost} & \textbf{Memory} \\
\hline
Positive-$P$ (this work) & $O(MS+LM^{2})$ & analytic target only: $O(M^{2})$ and an SDP & $O(BM+M^{2})$ \\
Squashed state~\cite{MartinezCifuentes2023classicalmodelsmay} & $O(MS)^{\dagger}$ & --- & $O(MS)^{\dagger}$ \\
Greedy, order $\mathcal{K}$~\cite{villalonga2021efficient} & $O(2^{\mathcal{K}}M^{\mathcal{K}})$ & $O(M^{\mathcal{K}})$ marginals, each exponential in $\mathcal{K}$ & $O(2^{\mathcal{K}}M^{\mathcal{K}})^{\dagger}$ \\
MPS~\cite{oh2024classical} & $O(Md\chi^{2})^{\dagger}$ & $O(Md\chi^{2}h)$ construction & $O(Md\chi^{2})$ \\
Exact~\cite{madsenQuantumComputationalAdvantage2022,bulmerBoundaryQuantumAdvantage2022a} & $O(Mn^{3}2^{n/2})$ & --- & --- \\
\hline
\end{tabular}
\end{heretable}

\clearpage

\begin{herefigure}
\centering
\includegraphics[width=\textwidth]{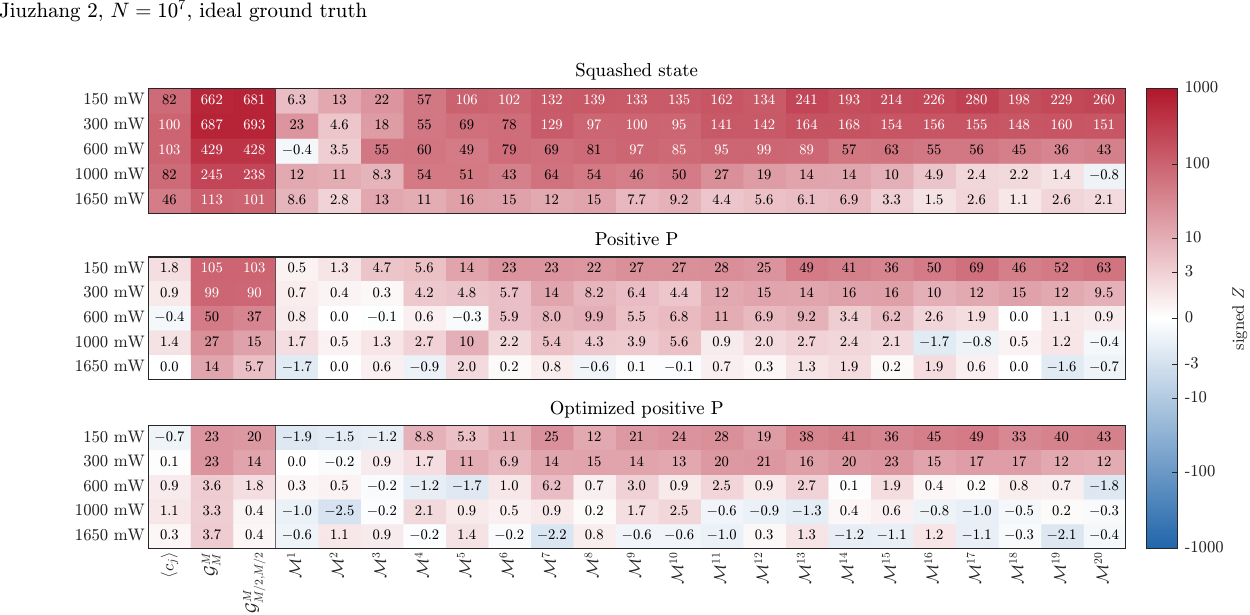}
\captionof{figure}{Signed $Z$-score of every observable for the lossless version of \textit{Jiuzhang~2} (Sec.~\ref{sec:lossless}), with $N=10^{7}$, against its own ideal ground truth. Observables, color scale and gray cells are as in Fig.~\ref{fig:obs_J2_1e7_ID}. There are no experimental samples for the lossless states. The bottom block is the optimized positive-$P$ sampler, rerun on the lossless datasets with the weights given in Sec.~\ref{sec:lossless}.\label{fig:obs_LL_J2}}\par\kern0pt
\end{herefigure}

\medskip

\begin{herefigure}
\centering
\includegraphics[width=\textwidth]{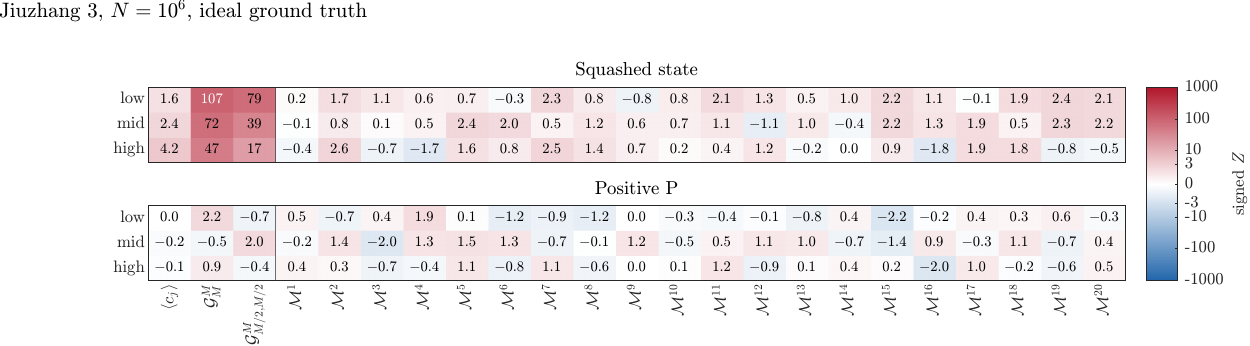}
\captionof{figure}{As Fig.~\ref{fig:obs_LL_J2}, for the lossless version of \textit{Jiuzhang~3}, with $N=10^{6}$.\label{fig:obs_LL_J3}}\par\kern0pt
\end{herefigure}

\medskip

\begin{herefigure}
\centering
\includegraphics[width=\textwidth]{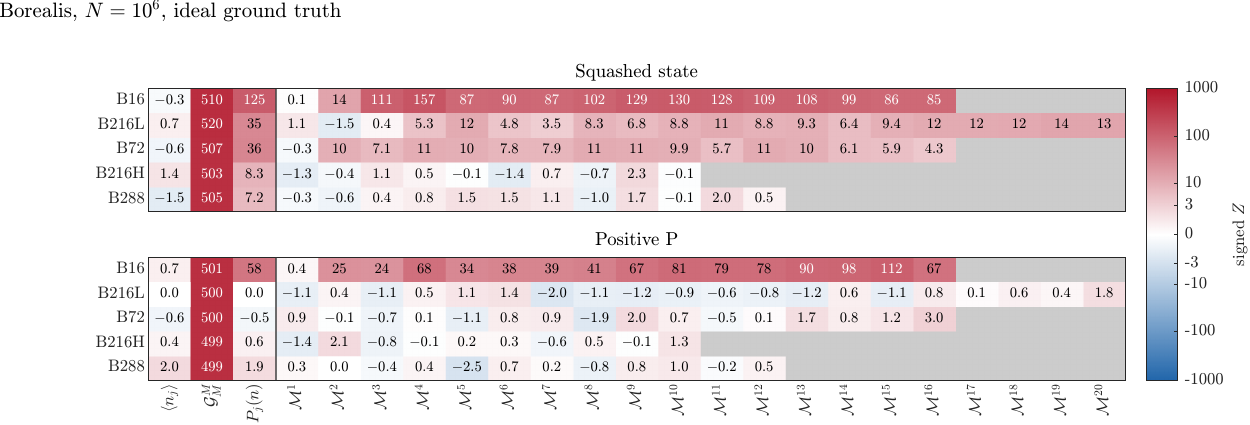}
\captionof{figure}{As Fig.~\ref{fig:obs_LL_J2}, for the lossless version of Borealis, with $N=10^{6}$. Observables are as in Fig.~\ref{fig:obs_B_ID}. The positive-$P$ sampler uses the Poisson output step for B16 and the categorical output step for the other datasets, as on the lossy versions. For these lossless datasets $\mathcal{G}^{M}_{M}$ is scored against its exact distribution, because the phase-space ground truth misses the even-total constraint (Sec.~\ref{sec:lossless}).\label{fig:obs_LL_B}}\par\kern0pt
\end{herefigure}

\bigskip

\iflatexml\else
\noindent\begin{minipage}{\linewidth}
\centering
\includegraphics[width=\textwidth,height=0.42\textheight,keepaspectratio]{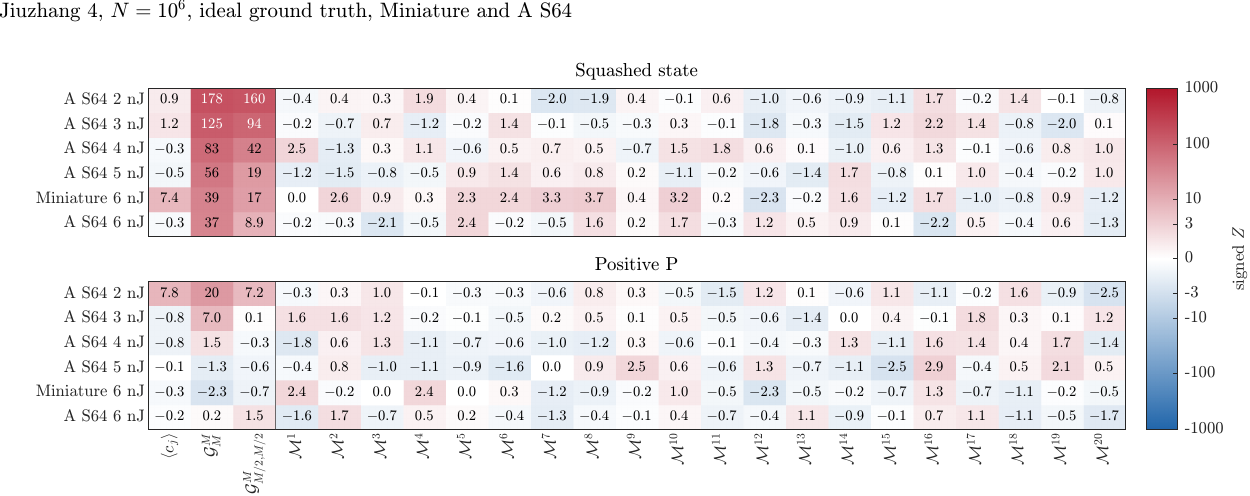}\par\medskip
\includegraphics[width=\textwidth,height=0.42\textheight,keepaspectratio]{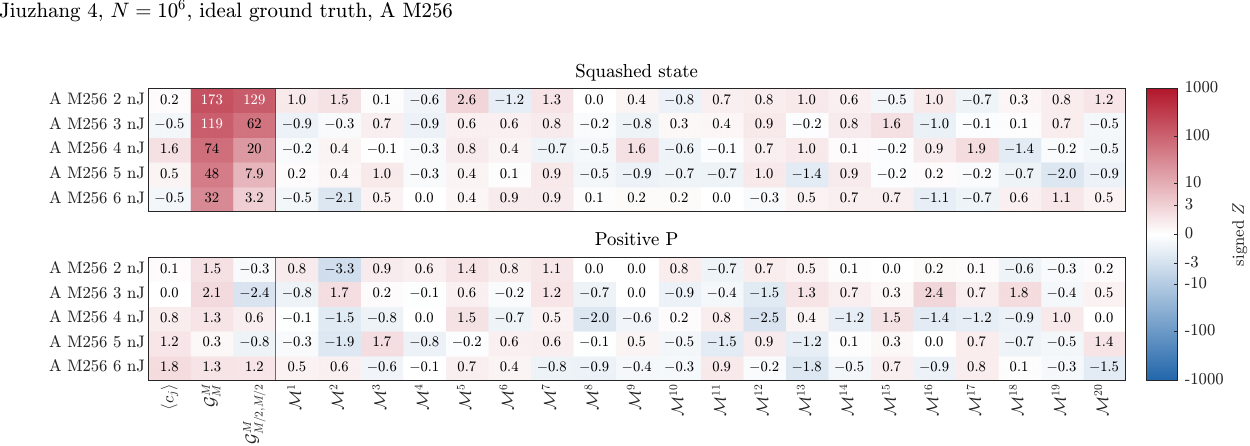}\par\smallskip
{\small Figure~\ref{fig:obs_LL_J4}, first of two parts; the caption follows the second part.}
\end{minipage}\par

\medskip

\fi
\begin{herefigure}
\centering
\iflatexml
\includegraphics[width=\textwidth,height=0.42\textheight,keepaspectratio]{Figures/observables_lossless_J4_N1e6_A_S64.pdf}\par\medskip
\includegraphics[width=\textwidth,height=0.42\textheight,keepaspectratio]{Figures/observables_lossless_J4_N1e6_A_M256.pdf}\par\smallskip
\fi
\includegraphics[width=\textwidth,height=0.42\textheight,keepaspectratio]{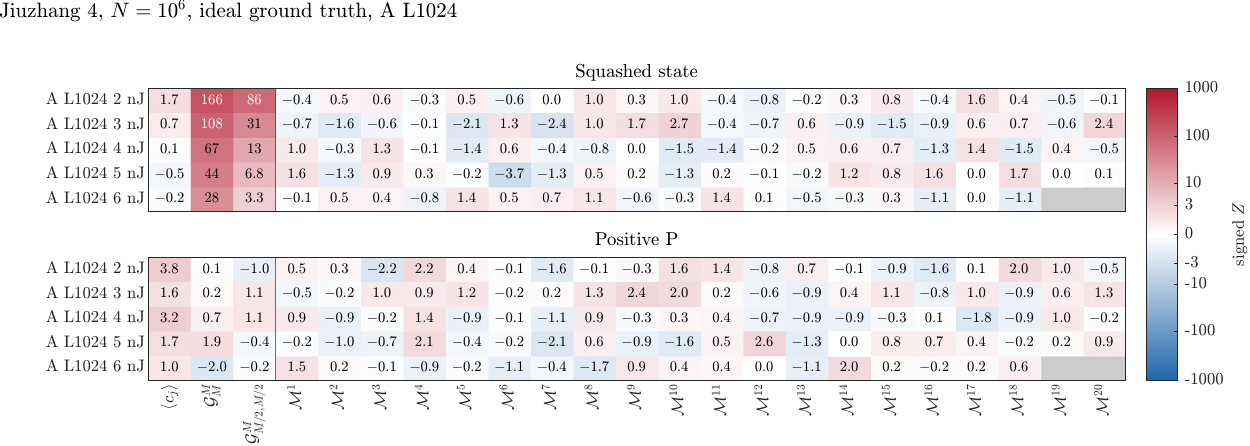}\par\medskip
\includegraphics[width=\textwidth,height=0.42\textheight,keepaspectratio]{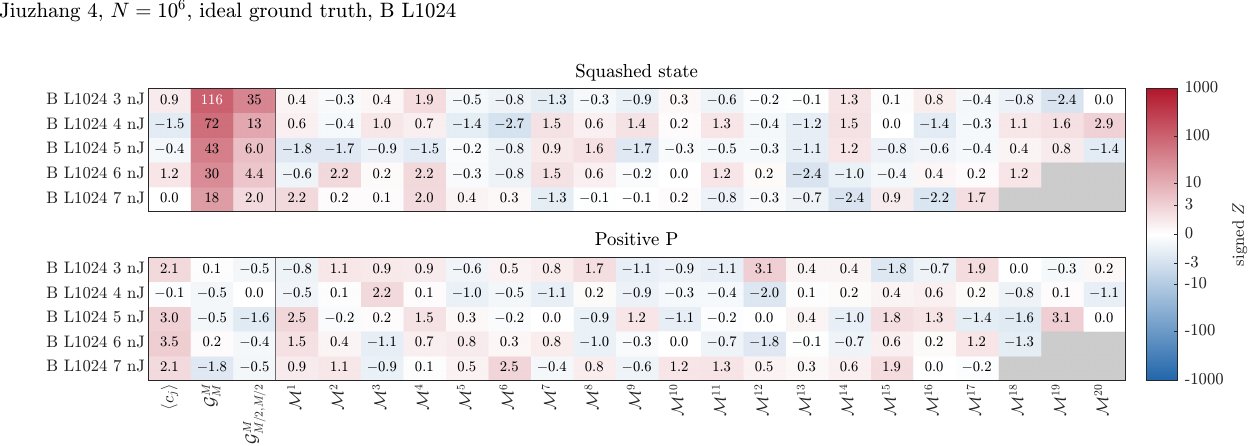}
\captionof{figure}{As Fig.~\ref{fig:obs_LL_J2}, for the lossless version of \textit{Jiuzhang~4}, with $N=10^{6}$. Each heat map is one configuration, named in its header (A S64, A M256, A L1024 and B L1024); the Miniature dataset is shown with A S64.\label{fig:obs_LL_J4}}\par\kern0pt
\end{herefigure}

\bigskip

\begin{heretable}
\centering
\captionof{table}{Evaluation time and peak memory for one grouped count
distribution, with $N_{\mathrm{traj}}=200$ trajectories. $\boldsymbol{M}$ is the partition,
$M=\sum_j M_j$ the total mode number, and $N_G=\prod_j(M_j+1)$ the number of
output probabilities. Memory is the size of the largest array each method allocates, calculated from its dimensions. Timings were measured on an AMD EPYC 7543 processor (2.8~GHz) of the Ngarrgu Tindebeek supercomputer, in double precision, with one computational thread in MATLAB R2023b, on synthetic positive-$P$ trajectories. Each time is the wall-clock time of the fastest of three runs, after one untimed run. Repeating the measurement on another node changed the times by up to a factor of three, but the speedups by less than 15\%.\label{tab:S-timing}}\par\kern0pt
\begin{tabular}{lrrrrrrr}
\hline
 & & & \multicolumn{2}{c}{\textbf{Time (ms)}} & & \multicolumn{2}{c}{\textbf{Memory (MB)}}\tabularnewline
\textbf{Partition $\boldsymbol{M}$} & \textbf{$M$} & \textbf{$N_G$} & \textbf{Fourier} & \textbf{Tree} & \textbf{Speedup} & \textbf{Fourier} & \textbf{Tree}\tabularnewline
\hline
$[500]$ & 500 & 501 & 357.2 & 47.1 & $7.6\times$ & 1.5 & 1.5\tabularnewline
$[50,50]$ & 100 & 2601 & 22.2 & 5.9 & $3.7\times$ & 7.9 & 0.4\tabularnewline
$[250,250]$ & 500 & 63\,001 & 435.1 & 30.9 & $14.1\times$ & 192 & 2.7\tabularnewline
$[10,10,10,10]$ & 40 & 14\,641 & 121.7 & 4.9 & $25.1\times$ & 45 & 1.7\tabularnewline
$[25,25,25,25]$ & 100 & 456\,976 & 5264.6 & 60.2 & $87\times$ & 1395 & 24\tabularnewline
$[100,100,100]$ & 300 & 1\,030\,301 & 9568.1 & 81.7 & $117\times$ & 3144 & 27\tabularnewline
\hline
\end{tabular}
\end{heretable}

\clearpage

%